\documentclass[twocolumn]{aastex701}
\usepackage{tikz}

\usepackage{graphicx}	
\usepackage{amsmath}	
\usepackage{ulem}
\usepackage{soul}
\usepackage{cancel} 

\newcommand{\grad}{\vec \nabla}
\newcommand{\dv}{\vec \nabla \cdot}
\newcommand{\curl}{\vec \nabla \times}
\newcommand{\lap}{\Delta}

\submitjournal{ApJ}
\shorttitle{3D MHD simulations of NS cores in the two-fluid approximation}
\shortauthors{A. Igoshev et al.}
\begin{document}
\title{Three-dimensional numerical simulations of neutron star cores in the two-fluid MHD approximation: simple configurations}

\correspondingauthor{Andrei Igoshev}
\author[orcid=0000-0003-2145-1022,gname='Andrei', sname='Igoshev']{Andrei Igoshev}
\altaffiliation{Royal Society University Research Fellow}
\affiliation{School of Mathematics, Statistics and Physics, Newcastle University, Newcastle upon Tyne,  NE1 7RU,  UK}
\affiliation{Department of Applied Mathematics, University of Leeds, LS2 9JT Leeds, UK}
\email[show]{andrei.igoshev@newcastle.ac.uk}

\author[orcid=0000-0002-9423-1170,sname='Moraga']{Nicol\'as A. Moraga}
\affiliation{Departamento de F\'{\i}sica, Facultad de Ciencias, Universidad de Chile, Las Palmeras 3425, \~Nu\~noa, Santiago, Chile}
\email{nicolas.moraga@ug.uchile.cl}

\author[orcid=0000-0003-4059-6796,sname='Reisenegger']{Andreas Reisenegger}
\affiliation{Departamento de F\'{\i}sica, Facultad de Ciencias B\'asicas, Universidad Metropolitana de Ciencias de la Educaci\'on, Av. Jos\'e Pedro Alessandri 774, \~Nu\~noa, Santiago, Chile}
\affiliation{Centro de Desarrollo de Investigaci\'on UMCE, Universidad Metropolitana de Ciencias de la Educaci\'on, Santiago, Chile}
\email{andreas.reisenegger@umce.cl}

\author[orcid=0000-0003-0994-2013,gname='Calum', sname='Skene']{Calum S. Skene}
\affiliation{Department of Applied Mathematics, University of Leeds, LS2 9JT Leeds, UK}
\affiliation{School of Physics and Astronomy, The University of Edinburgh, Edinburgh EH9 3FD, UK}
\email{cskene3@ed.ac.uk}

\author[orcid=0000-0001-8639-0967,gname='Rainer', sname='Hollerbach']{Rainer Hollerbach}
\affiliation{Department of Applied Mathematics, University of Leeds, LS2 9JT Leeds, UK}
\email[show]{r.hollerbach@leeds.ac.uk}

\date{Accepted XXX. Received YYY; in original form ZZZ}




\begin{abstract} 
Magnetic field evolution in neutron star cores is not fully understood. 
We 
describe the field evolution both for one barotropic fluid as well as two collisionally coupled barotropic fluids with different density profiles using 
the anelastic approximation and the Navier-Stokes equations to simulate the evolution in three dimensions. 
In the one-fluid case, a single fluid describes the motion of the charged particles.
In the two-fluid model, the neutral fluid is coupled to the electrically conductive fluid by collisions, the latter being dragged by the magnetic field. In this model, both fluids have distinct density profiles. This forces them to move at slightly different velocities, resulting in a relative motion between the two barotropic fluids -- ambipolar diffusion. 
We develop a code based on \texttt{Dedalus} and study the evolution of simple poloidal dipolar and toroidal magnetic fields. Previous 2D studies found that poloidal magnetic fields evolve towards a stable Grad-Shafranov equilibrium. In our 3D simulations we find an instability of the two-fluid system similar to the one in the barotropic fluid system. After the instability saturates, a highly non-linear Lorentz force introduces small-scale fluid motion that leads to turbulence, development of a cascade and significant, non-axially symmetric changes in the magnetic field configuration. Fluid viscosity plays an essential role in regularizing the small-scale fluid motion, providing an energy drain.  
\end{abstract}





\section{Introduction}

Magnetic fields determine multiple observational properties of isolated and accreting neutron stars \citep{Igoshev2021Univ}, including aspects of their  radio and X-ray emission. Structurally, a neutron star (NS) has a thin solid crust with a typical thickness of $\approx 1$~km and a liquid core of radius $\sim 10$~km. The core consists mostly of neutrons, with a small fraction of protons and electrons, as well as muons and potentially other more exotic species at increasing densities. In the crust, ions have very restricted mobility, so the evolution depends only on the motion of the electrons, and the magnetic evolution reduces to electron magnetohydrodynamics (eMHD).
In the past two decades, considerable progress has been achieved in eMHD modeling of the crust-confined field evolution governed by the Hall effect and Ohmic decay (for review see \citealt{Pons2019LRCA} and \citealt{Gourgouliatos2022Symm}). Significantly less is known about the structure and strength of magnetic fields in NS cores.

Unfortunately, the magnetic evolution in the core is far more complicated than in the crust, as the physics behind it is poorly understood. It is expected that the stellar core becomes superconducting and superfluid relatively early in the NS life \citep{migdal1959superfluidity}. This should influence the magnetic field evolution by modifying the kinetic coefficients that govern the field's dissipation, and by introducing macroscopic manifestation of the interaction between quantized neutron vortices and magnetic flux tubes \citep{glampedakis2011magnetohydrodynamics}. 
In this paper, we take a first step in modeling two fluids MHD in the NS core in 3D (modeled in 2D by \citealt{Castillo2020MNRAS,castillo2025AA} and \citealt{moraga2025magnetothermal}), but focus on a core composed of ‘normal’ (non-Cooper-paired) matter, meaning it is neither superconducting nor superfluid. This assumption is likely realistic for the high temperatures and strong magnetic fields found in young magnetars. 

At temperatures $\lesssim 5\times10^{8}\,\mathrm{K}$, the core is in the weak-coupling regime. Here, Urca reactions are essentially frozen, and the decreasing inter-particle collision rate weakens the coupling between neutrons and charged particles. Consequently, the electrically conductive fluid (electrons and protons) can move relative to the neutron fluid, giving rise to ambipolar diffusion. 




Initial attempts have been made to study ambipolar diffusion in the single-fluid approximation, including works by \cite{Castillo2017MNRAS,Passamonti2017MNRAS,Igoshev2023MNRAS}. In this approximation, the neutrons (or neutral fluid) are fixed and form a non-moving background. In this case, the only relevant speed is the ambipolar velocity, i.e., the velocity of the charged fluid. However, this simplified scenario cannot correctly capture the multi-fluid nature of the core, and over-estimates the typical evolutionary timescale of the magnetic field.  
Subsequent work by \citet{OfengeimGusakov2018PhRvD} (see also \citealt{castillo2025AA}) demonstrated that the bulk velocities of the neutron and charged fluids can be significantly larger than their relative (ambipolar) velocity. This result showed that the single-fluid approximation is incorrect and underscores the need to treat the neutron and charged components as two independent fluids. The first axially symmetric simulations in this approximation were performed by \cite{Castillo2020MNRAS,castillo2025AA,moraga2025magnetothermal}. In this simplified scenario, the magnetic evolution appears to converge toward a stable equilibrium state: the Grad-Shafranov (GS) equilibrium \citep{gradrubin54,shafranov66,Armaza_2015}. In this state, the Lorentz force is balanced by the pressure and gravity forces acting on the charged fluid, while the neutrons remain in diffusive equilibrium. Until now, no three-dimensional simulations have been performed in this approximation. 

Thus, in this paper we develop a new technique to model ambipolar diffusion in three spatial dimensions in the two-fluid approximation. Our approach relies on advances made in computational astrophysical and geophysical fluid dynamics over the past decades, e.g., the dynamo benchmark \citep{Jones2011Icar}, and understanding the crucial role played by viscous boundary layers despite their tiny extent, see, e.g., \cite{Livermore2016NatSR}. 

As a first step with this new technique, we focus on simple magnetic field configurations: (1) poloidal dipolar field, and (2) toroidal field. 
The stability of magnetic field configurations is a long-standing problem affecting massive stars, white dwarfs and neutron stars, see e.g. \cite{LanderJones2012MNRAS}. Pure poloidal and toroidal fields are known to be unstable in single-fluid three-dimensional MHD  \citep{Tayler1973MNRAS,MarkeyTayler1973MNRAS}, which was confirmed by numerical simulations \citep{LanderJones2012MNRAS,Mitchell2015MNRAS,Becerra2022MNRAS}. 
As explained above, axially symmetric two-fluid simulations 
evolve towards the 
GS-equilibrium \citep{Castillo2020MNRAS} which is stable. However, the stability of the GS-equilibrium under non-axisymmetric perturbations has not been studied. Earlier three-dimensional numerical simulations for ambipolar diffusion performed in the single-fluid limit showed that a poloidal magnetic field is unstable \citep{IgoshevHollerbach2023MNRAS}. 

 

Our paper is structured as follows. In Section~\ref{s:method} we describe the physical model and the methodology of our numerical simulations, including the simulation setup (Section~\ref{s:eos}) and relevant timescales of the problem (Section~\ref{s:timescales}). We introduce equations governing single fluid barotropic evolution (Section~\ref{s:baritrop_equations}) as well as two-fluid ambipolar diffusion (Section~\ref{s:twofluids_eq}) with initial conditions (Section~\ref{s:init_cond}) and diagnostic techniques (Section~\ref{s:diagnostic}). In Section~\ref{s:results} we present our results, first for the 1-barotropic-fluid (Sections~\ref{s:res_barotrop}) and later for two-fluid (Sections~\ref{s:res_twofluids}) simulations. We compare our two-fluid results with previous results in Section~\ref{s:previous}, and discuss the importance of dimensionless parameters in Section~\ref{s:dimensionless}, and future work in Section~\ref{s:future}. We conclude in Section~\ref{s:conclusion}.

\section{Methodology}
\label{s:method}

\subsection{Magnetic field evolution in the core}

The Eulerian equations describing the magnetic field evolution in the NS core in absence of superconductivity and superfluidity were derived in multiple publications \citep{IakovlevShalybkov1991ApSS,GoldreichReisenegger1992ApJ,Passamonti2017MNRAS, GusakovKantor2017PhRvD,OfengeimGusakov2018PhRvD}. Here we rewrite for completeness equations by \cite{OfengeimGusakov2018PhRvD} 
\begin{align}
& n_i \frac{\partial \vec u_i}{\partial t}  + n_i \left(\vec u_i \cdot \nabla \right) \left(\frac{\mu_i}{c^2} \vec u_i \right) =  \nonumber\\
& = - n_i \vec \nabla \mu_i  + e_i n_i \left( \vec E + \frac{1}{c} \vec u_i \times \vec B \right) \label{eq:Euler} \nonumber \\
& - \frac{\mu_i n_i}{c^2} \vec \nabla \phi - \sum_{j\neq i}\gamma_{ij} n_i n_j (\vec u_i - \vec u_j)\\
& \frac{\partial n_i}{\partial t} + \vec \nabla \cdot (n_i \vec u_i) = \Delta \Gamma_i \\
& \Delta \phi = \frac{4\pi G}{c^2} (P+\epsilon) \\
& \frac{\partial \vec B}{\partial t} = -c \vec \nabla \times \vec E \\
& \vec \nabla \times \vec B = \frac{4\pi}{c} \vec j = \frac{4\pi}{c} \sum_i e_i n_i \vec u_i\\
& \sum_i e_i n_i = 0\label{eq:Charged_neutrality}
\end{align}
In these equations the index $i$ corresponds to protons, electrons and neutrons. The first equation is momentum conservation equation, second is mass conservation equation, third is Poisson equation, fourth is Faraday equation, fifth Amp\`ere's law in the magnetohydrodynamic limit and the last is charge neutrality. Individual variables are: $\mu_i$ is the chemical potential, $n_i$ is the number density, $\vec u_i$ are velocities, $\phi$ is the gravitational potential, $\vec B$ is magnetic and $\vec E$ is electric field. The interaction between different species of particles is described with $\gamma_{ij}$ while $\Delta \Gamma_i$ are reaction rates for beta-decay reactions. Physical constants include speed of light  $c$, elementary charge $e_i$ (negative for electrons, positive for protons and zero for neutrons) and  gravitational constant $G$.

\subsection{Background stellar equilibrium and its perturbations}

\label{s:eos}
Simulations studying initially random magnetic fields suggest that the NS core reaches a global hydromagnetic equilibrium shortly after the proto NS phase within a few Alfvén times $t_\mathrm{Alf}\sim 1\, (B/10^{14}\,\mathrm{G})$~s (see e.g.~\citealt{Becerra2022random}).
On the other hand, as the ratio of the magnetic pressure to the characteristic fluid pressure 
$P$ inside the NS core is typically $B^{2}/8\pi P \sim 10^{-6}\,(B/10^{15}\,\mathrm{G})^{2}$ \citep{reisenegger2009A&A}, this initial equilibrium can be modeled as a non-magnetized, spherically symmetric background in chemical and hydrostatic equilibrium , i.e.
\begin{align}
    & \mu_{n}(r) = \mu_p(r) +\mu_e(r) \equiv \mu(r),\\
    & \vec \nabla \mu - \frac{\mu}{c^2} \vec\nabla \phi = 0,\label{eq: HSequilibrium}
\end{align}
(for more details see \citealt{moraga2025magnetothermal}), and the magnetic field only slightly perturbs it. The perturbations introduced by the magnetic field are quantified as
\begin{equation}
    \mu_{i}(\vec{r})\rightarrow \mu(r) +\delta\mu_{i}(\vec{r},t).
\end{equation}
Here, $\delta\mu_{i}(\vec{r},t)$ are the Eulerian chemical potential perturbations induced by the magnetic field. The dynamics of these perturbations is governed by the linearized version of the set of equations \eqref{eq:Euler}-\eqref{eq:Charged_neutrality}. These linearized equations can be further simplified.

First, we simplify the momentum equation by introducing electrically conductive fluid. This fluid moves with speed $\vec u_\mathrm{c}$:
\begin{equation}
\vec u_\mathrm{c} = \frac{\vec u_\mathrm{p} + \vec u_\mathrm{e}}{2}.    
\end{equation}
Assuming local charge neutrality, which means $n_\mathrm{e} = n_\mathrm{p} \equiv n_c$, the electric current is $\vec j = e n_c (\vec u_p - \vec u_e) $. Adding the momentum equations for the electron fluid and the proton fluid and dividing by two, we obtain:
\begin{align}
& n_\mathrm{c} \frac{\partial \vec u_\mathrm{c}  }{\partial t}   =  - n_c \vec \nabla \delta \mu_c  - \frac{\delta \mu_c n_c}{c^2}  \vec \nabla \phi  + \dfrac{\vec j}{c} \times \vec B  - \gamma_\mathrm{pn} n_c n_n (\vec u_c - \vec u_n). \label{eq:Linear_charged_particles}
\end{align}
Also, following previous publications we neglected $\gamma_{ne}$ because it is small. Here, we note that, with the aid of equation~\eqref{eq: HSequilibrium}, both the pressure plus the gravity terms can be absorbed into a single term of the form $-n_c\mu \grad(\delta \mu_c/\mu) $, or equivalently considering the redshifted perturbations $\mu^{\infty}_c \equiv \delta \mu \,e^{\phi/c^{2}}$, leads to the single term $e^{-\phi/c^{2}}\grad\delta \mu^{\infty}_c$. 
In this work we use the Newtonian gravity, i.e., assume $e^{-\phi/c^2} \equiv 1$. These corrections will be included in the future.
\begin{equation}
\frac{\partial \vec u_c}{\partial t} = - \vec \nabla \delta \mu_c + \dfrac{\vec{f}_{B}}{n_c} - n_n \gamma_\mathrm{np} (\vec u_c - \vec u_n). 
\label{eq:momentum_charged_fluid}
\end{equation}
Here, $\vec{f}_B = \vec{j}\times \vec{B}/c = (\vec \nabla \times \vec B)\times \vec B / (4\pi)$ is the classical Lorentz force. A similar equation can be written for neutrons following similar approximations about perturbations above the hydrostatic state:
\begin{equation}
\frac{\partial \vec u_n}{\partial t} = - \vec \nabla \delta \mu_n - n_c\gamma_\mathrm{np} (\vec u_n - \vec u_c)   
\label{eq:neutron_velocity}
\end{equation}
We add viscosity to both momentum equations. The reasons for this are as follows: (1) although small, viscosity is present in normal NS matter, (2) mathematically viscosity changes the nature of the partial differential equations from a first-order equation for $\delta \mu_i$ to a parabolic equation for $\vec u_i$ which allows us to impose proper boundary conditions for $\vec u_i$, (3) despite being small, it is well-known that fluid behavior for ${\nu\to 0}$ and $\nu\equiv0$ are different, e.g. D'Alembert's paradox.

We supplement two momentum equations with continuity equations written in the weak coupling limit, i.e. $\Delta \Gamma_i = 0$, where Urca reactions are frozen. We also use a soundproof approximation commonly applied in fluid dynamics. This anelastic approximation gives:  
\begin{equation}
\vec \nabla \cdot (n_i\vec u_i) = 0   
\label{eq:anelastic_constrain}
\end{equation}
The only remaining equation is the Faraday equation. We find the electric field by subtracting momentum equations for protons and electrons. It gives us
\begin{equation}
\frac{1}{en_c} \frac{\partial \vec j}{\partial t} = 2 e n_c E + \frac{2 e n_c}{c} \vec u_c \times \vec B + 2 n_c^2 \gamma_\mathrm{pe} (\vec u_p - \vec u_e )    
\end{equation}
Following standard MHD approach we assume that electric current adapts immediately i.e. $\partial \vec j / \partial t = 0$. It gives us:
\begin{equation}
\vec E = -\frac{1}{c} \vec u_c \times \vec B + \frac{\gamma_\mathrm{pe}}{e^2} \vec j.    
\end{equation}
If we substitute into the Faraday equation we obtain:
\begin{equation}
\frac{\partial\vec B}{\partial t} = \vec \nabla \times \left( \vec u_c \times \vec B \right) - \frac{\gamma_\mathrm{pe} c^2}{4\pi e^2} \; \vec \nabla \times \left(\vec \nabla \times \vec B \right)   
\end{equation}
In this equation $e^2 / \gamma_\mathrm{pe} = \sigma$, i.e. electric conductivity. We assume that it does not depend on the radial coordinate. We also assume that $\vec \nabla \cdot \vec B = 0$ which allows us to simplify the last term and obtain the standard induction equation:
\begin{equation}
\frac{\partial\vec B}{\partial t} = \vec \nabla \times \left( \vec u_c \times \vec B \right) + \frac{c^2}{4\pi \sigma} \; \Delta \vec B
\label{eq:induction_B}
\end{equation}

\subsection{NS equation of state}
\label{s:eos}

We use the 
HHJ equation of state 
\citep{Heiselberg_1999} to build a model of our neutron stars with mass $M_\mathrm{NS} = 1.4$~M$_\odot$ and radius $R_\mathrm{NS} = 12.2$~km. In this model the core radius is $R_\mathrm{c} = 11.2$~km. 
For our research, the most relevant radial profiles are 
number density of charged particles $n_\mathrm{c}(r)$ and  neutrons $n_\mathrm{n}(r)$, as well as the interaction between protons and neutrons $\gamma_\mathrm{np}(r)$. We make these numerical profiles dimensionless by fixing their value at the NS center. The relevant numerical values are summarised in Table~\ref{tab:units}. 

\begin{table}
    \caption{Physical values used to compute dimensionless quantities.}
    \label{tab:units}
    \centering
    \begin{tabular}{cllc}
    \hline
    Symbol                  & Value               & Units \\
    \hline
    $R_\mathrm{c}$                   & $1.12\times 10^{6}$ & cm \\
    $n_\mathrm{c}^0$                 & $4.2\times 10^{37}$ & cm$^{-3}$  \\
    $B_0$                   & $1.0\times 10^{15}$ & cm$^{-1/2}$~g$^{1/2}$~s$^{-1}$ \\
    $\gamma_\mathrm{np}^0$  & $1.1\times 10^{-46}$ & cm$^3$ g s$^{-1}$ \\
    \hline 
    \end{tabular}
    \tablecomments{We take $T = 10^{8}$~K when computing $\gamma^0_\mathrm{np}$.}
\end{table}

\begin{figure}
    \includegraphics[width=\columnwidth]{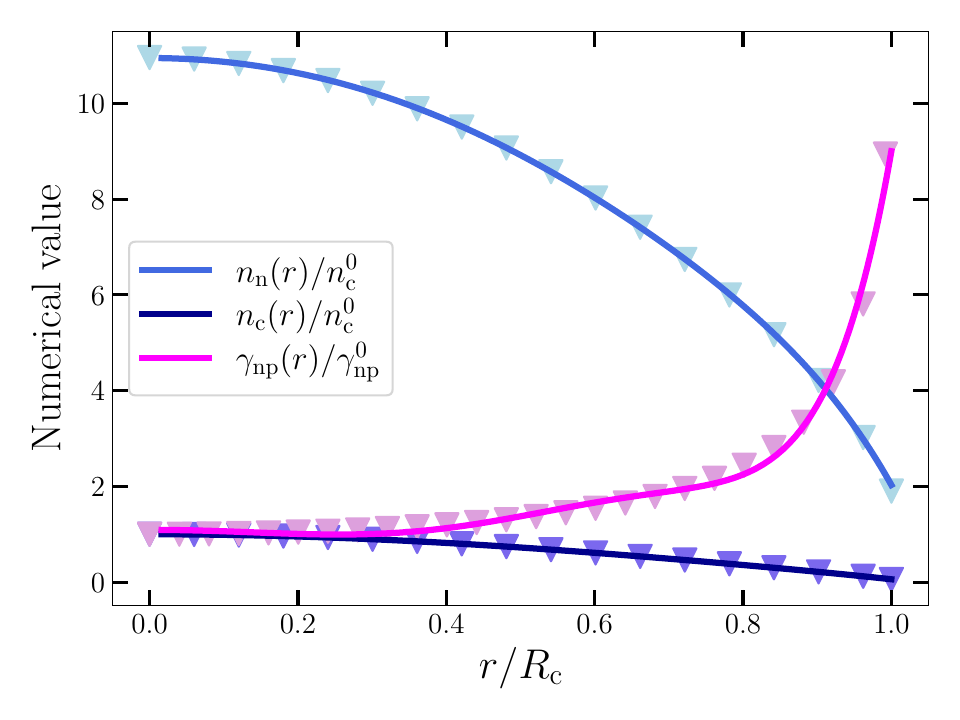}
    \caption{Dimensionless numerical radial profiles of number density for neutrons $n_\mathrm{n}$, charged particles $n_\mathrm{c}$, and interaction between protons and neutrons $\gamma_\mathrm{np}$. These curves correspond to polynomial fits to the HHJ equation of state. The solid lines show our actual numerical fit, while pale triangles are used to show the numerical values directly computed using the HHJ equation of state. Only $\gamma_\mathrm{np}$ deviates noticeably around $r/R_\mathrm{c} = 0.8$.}
    \label{fig:radial_profile}
\end{figure}

We show the dimensionless numerical profiles in Figure~\ref{fig:radial_profile}. These radial profiles are fitted using polynomials up to 8th degree

\begin{align} 
n_\mathrm{n} (r) / n_\mathrm{c}^0 & = a^\mathrm{n}_0 + a^\mathrm{n}_1 r^2 + a^\mathrm{n}_2 r^4 + a^\mathrm{n}_3 r^6 + a^\mathrm{n}_4 r^8, \\   
n_\mathrm{c} (r) / n_\mathrm{c}^0  & = a^\mathrm{c}_0 + a^\mathrm{c}_1 r^2 + a^\mathrm{c}_2 r^4 + a^\mathrm{c}_3 r^6 + a^\mathrm{c}_4 r^8, \\   
\gamma_\mathrm{np} (r) / \gamma_\mathrm{np}^0 & = a^\gamma_0 + a^\gamma_1 r^2 + a^\gamma_2 r^4 + a^\gamma_3 r^6 + a^\gamma_4 r^8.   
\end{align}

\begin{table*}
    \caption{Numerical coefficients used to represent radial profiles.}
    \label{t:radial}
    \centering
    \begin{tabular}{lrrrrrrc}
    \hline
    Type     & $a_0\quad$   & $a_1\quad$     & $a_2\quad$     & $a_3\quad$   & $a_4\quad$  & $\mathrm{max}\;|\Delta \epsilon|$\\  
    \hline
    Neutrons          & 10.9495  & -7.80889 & -2.53869 & 6.63592  & -5.21052   & 0.07\ \ \ \ \\
    Charged particles & 0.999744 & -1.16899 & 0.217623  & 0.158726 &  -0.147995 & 0.06\ \ \ \ \\
    Interactions      & 1.09179  & -3.25125 & 30.9124   & -64.7779 & 45.0334    & 0.17\ \ \ \ \\
    \hline
    \end{tabular}
    \tablecomments{The last column shows the maximum fractional error $\mathrm{max}\; |\Delta \epsilon|$ of our polynomial fit with respect to tabulated values.}
\end{table*}

We summarize the numerical coefficients for these fits in Table~\ref{t:radial}. With this approach, we enforce that the final radial profiles satisfy the regularity conditions near the NS center. By limiting these polynomials to degree 8, we remove an excessive numerical noise which occurs when functions are expanded in the radial basis. The numerical function $\gamma_\mathrm{np} (r)$ deviates more strongly from its polynomial fit in comparison to the number densities. Still, the deviations are within 10~\% for number densities and within 20~\% for interactions. We do not study here the sensitivity of magnetic field evolution to the exact equation of state. We use the same equation of state to model magnetic field evolution in the case of 1-barotropic-fluid MHD system as well as in the case of the complete two-fluid system.

\subsection{Timescales}
\label{s:timescales}

The problem includes a range of timescales which need to be chosen in such a way as to correctly represent the physics.
The fundamental ambipolar diffusion timescale is $t_\mathrm{ad,0}$
\begin{equation}
t_\mathrm{ad,0} = \frac{40\pi \gamma_\mathrm{np}^0 (n_\mathrm{c}^0)^2 R_\mathrm{c}^2}{B_0^2} \approx 10^6 \; \mathrm{yrs}  ,
\label{eq:tau_ad}
\end{equation}
computed for $B_0 = 10^{15}$~G and temperature $T = 10^8$~K.
We measure time in these units for both 1-barotropic fluid and two-fluid simulations.
This is basically the same as eq.~(20) in \cite{IgoshevHollerbach2023MNRAS}, with the substitution
\begin{equation}
\frac{m_\mathrm{p}^*}{t_\mathrm{pn}}\quad{\rm there}\quad
\longrightarrow\quad 10\,\gamma_{\mathrm{np}}^0\,n_\mathrm{c}^0\quad{\rm here}, 
\end{equation}
where the factor of $10$ appears because $n_\mathrm{n}^0 \approx 10 n_\mathrm{c}^0$. 
The estimate eq.~(\ref{eq:tau_ad}) translates into the velocity unit $u_0 = R_\mathrm{c} / t_\mathrm{ad, 0}$. In actual simulations, we observe charged particle and neutron velocities which are larger than unity. This means that the evolution proceeds faster than our initial expectation $1 \times t_\mathrm{ad, 0}$. This behavior is characteristic of the two-fluid model. In the single-fluid model, the magnetic field evolution is driven solely by advection generated by charged fluid motion. However, in the two-fluid case, the advection velocity also includes the contribution from the bulk neutron flow, leading to a faster dynamical evolution.

The shortest timescale in our problem is the Alfv\'en timescale. The physical Alfv\'en timescale in a NS is too short (tens of seconds), so instead we scale it up, by introducing a numerical coefficient Am which fixes the Alfv\'en timescale with respect to the ambipolar diffusion timescale. The exact meaning of the coefficient is
\begin{equation}
\mathrm{Am} = \frac{4\pi m^* n_\mathrm{c}^0 R_\mathrm{c}^2}{t_\mathrm{ad,0}^2 B_0^2} = \left( \frac{t_\mathrm{Alf}}{t_\mathrm{ad,0}} \right)^2,
\end{equation}
where $m^*$ is an effective mass\footnote{which we increase by a factor of $10^{25}$ in comparison to the actual mass of a proton to track evolution on Alfv\'en timescale} and $t_\mathrm{Alf} = R_\mathrm{c} / u_\mathrm{Alf} = \sqrt{4\pi m^* n_\mathrm{c}^0} \; (R_\mathrm{c} / B)$. Here $u_\mathrm{Alf}$ is the classical Alfv\'en velocity. 
If a simulation is started with no initial fluid velocities, but only with a magnetic field, fluid velocities are growing until an equilibrium is achieved on a timescale $t_\mathrm{Alf}$, which is $t_\mathrm{Alf}\approx t_\mathrm{ad,0} \sqrt{\mathrm{Am}}$.
When choosing our simulations and selecting various numerical coefficients we try to reach the following hierarchy of timescales
\begin{equation}
t_\mathrm{Alf} \ll t_\mathrm{ad, 0} \ll t_\mathrm{Re},\; t_\mathrm{Ohm}    ,
\end{equation}
where $t_\mathrm{Re}$ and $t_\mathrm{Ohm}$ correspond to decay timescales due to viscosity and Ohmic decay, respectively. The initial Ohmic timescale is
\begin{equation}
t_\mathrm{Ohm} = \mathrm{Rm} \; t_\mathrm{ad,0}   .
\end{equation}
It becomes much shorter when the magnetic field's spatial scale becomes smaller than $R_\mathrm{c}$. 
If a simulation is started with initial velocities but no magnetic field, these velocities decay due to the fluid viscosity on a timescale of
\begin{equation}
t_\mathrm{Re} = \mathrm{Re}  \; t_\mathrm{ad,0}    .
\end{equation}
If $t_\mathrm{ad,0} \approx 1$, it means that we should choose $\mathrm{Am}\ll 1$ and $\mathrm{Rm}\gg 1$ as well as $\mathrm{Re} > 1$. We summarize all numerical values for these coefficients in Table~\ref{tab:parameters}.

Besides the time evolution, fluid and magnetic viscosities (Reynolds and magnetic Reynolds numbers) play a very important role in determining the size of boundary layers. In order to simulate the NS correctly it is necessary to resolve the boundary layers forming at the crust-core interface.


\subsection{1-barotropic-fluid MHD system}
\label{s:baritrop_equations}

\begin{table*}
    \caption{Dimensionless numbers used in the simulations.}
    \label{tab:parameters}
    \centering
    \begin{tabular}{cllccc}
    \hline
      & Number & \ \ \ \ \ Ratio of ... & Equation & Simulation  & Range of physical values \\
    \hline
    Am         & Ambipolar  & inertial to magnetic forces &$\left(t_\mathrm{Alf} / t_\mathrm{ad, 0}\right)^2$ & $10^{-5}$ & $\sim10^{-30}$\\ 
    Re         & Reynolds   & inertial to viscous forces & $uR_\mathrm{c} / \nu$  & $2$~--~$0.2$ & $10^{-4}$~--~$10^{-5}$\\
    Rm         & Mag.~Reynolds  & induction to diffusion & $uR_\mathrm{c} / \eta$ & $100$ & $10^4$~--~$10^9$ \\
    \hline
    \end{tabular}
    \tablecomments{For values sensitive to temperature, we provide a range of values corresponding to temperatures $10^6$~--~$10^9$~K. Values used in the simulation (Am, Re and Rm) define the simulation properties. Values $u$ corresponds to typical values inside NS for fluid velocity, $\nu$ to viscosity and $\eta$ to magnetic diffusivity. }
\end{table*}

In order to test our code and understand how two-fluid ambipolar diffusion is different from 1-barotropic-fluid MHD, we start with the latter one. This model does not intend to represent NS physics correctly, but instead informs us about behavior of MHD system in same conditions.

Thus, the MHD system of equations is written using the anelastic approximation as follows
\begin{equation}
\mathrm{Am}\; n_\mathrm{c} \; \frac{D \vec u_\mathrm{c}}{D t} = - n_\mathrm{c} \grad \delta \mu_\mathrm{c} + \vec f_B + \vec f_\nu (\vec u_\mathrm{c}) ,
\label{eq:uc_barotrop}
\end{equation}
\begin{equation}
\dv (n_\mathrm{c} \vec u_\mathrm{c}) = 0 ,
\label{eq:anelastic}
\end{equation}
\begin{equation}
\frac{\partial \vec B}{\partial t} = \curl (\vec u_\mathrm{c} \times \vec B) + \frac{\lap \vec B}{\mathrm{Rm}}  ,
\label{eq:induction}
\end{equation}
\begin{equation}
\dv \vec B = 0  .
\label{eq:gauss}
\end{equation}
These are essentially the momentum equation for an electrically conducting fluid eq. (\ref{eq:momentum_charged_fluid}) with anelastic constraint eq. (\ref{eq:anelastic_constrain}) as well as the induction equation (\ref{eq:induction_B}). Here we neglect neutron fluid as well as a drag from interaction with neutron fluid completely.
We combine these equations with the integral condition:
\begin{equation}
\int \delta \mu_\mathrm{c} \; dV = 0 .   
\label{eq:integ_cond}
\end{equation}
Eq.\ (\ref{eq:uc_barotrop}) does not change if we add a constant to $\delta \mu_c$ i.e. $\delta \mu_c' = \delta \mu_c + C$. Moreover, at different times we can add different constants without any change for
eq.\ (\ref{eq:uc_barotrop}). In order to set a constraint on the deviation from chemical equilibrium we add the integral condition eq.\ (\ref{eq:integ_cond}). In our research we are not interested in the exact value of $\delta \mu_c$; thus, the integral condition eq.\ (\ref{eq:integ_cond}) is sufficient. \cite{castillo2025AA} showed a physical form of the integral condition (their eqs.\ 14-15), which is not important in our research since we never use the exact value of $\delta \mu_c$ and are only interested in its gradient which is not affected by the integral condition.  
The viscous force is written as
\begin{equation}
\vec f_\nu(\vec u_\mathrm{c}) = \frac{\mathrm{Am}}{\mathrm{Re}} \vec\nabla \cdot \left[2n_c\left(\overleftrightarrow{e}-\frac{1}{3} (\vec\nabla\cdot \vec{u})\overleftrightarrow{I}\right)\right]
\end{equation}
where $\overleftrightarrow{I}$ is the identity tensor and $\overleftrightarrow{e}=((\vec\nabla\vec u)^T+\vec\nabla\vec u)/2$ is the strain rate tensor.
The way the viscous force is written assumes uniform kinematic viscosity and zero bulk viscosity, similar to the anelastic dynamo benchmark \citep{Jones2011Icar}, for example. 



In the Navier-Stokes equation we use the Lagrangian derivative
\begin{equation}
\frac{D\vec u}{Dt} \equiv \frac{\partial \vec u}{\partial t} + (\vec u\cdot \vec \nabla ) \vec u \equiv \frac{\partial \vec u}{\partial t} + \frac{1}{2} \nabla (\vec u)^2 - \vec u \times (\curl \vec u) .   
\end{equation}

This system as well as the complete two-fluid system described in the next section lacks an explicit gravity force which would be proportional to $n_\mathrm{c}\delta \mu_\mathrm{c} \nabla \phi / c^2$, where $\phi$ is the gravitational potential. The motivation for its removal is that this force should be small in comparison to the pressure term $n_\mathrm{c}\vec \nabla \delta \mu_\mathrm{c}$, similar to considerations by \cite{OfengeimGusakov2018PhRvD}. 

The absence of this force has two significant advantages: (1) the anelastic approach is fully justified in this case with no energy conservation problems arising, unlike the solar tachocline problem \citep{Moss2022PhRvF,Wilczynski2022JFM}, and (2) an explicit equation of state is not required (besides the integral condition) because eq.~(\ref{eq:uc_barotrop}) splits into a divergence-free part for velocities and a curl-free part for pressures (chemical potential perturbations). The chemical potential perturbations simply need to balance pressure-like component of the Lorentz force. Moreover, the anelastic approximation is further justified because the sound speed in the NS core is significantly faster than the Alfv\'en speed and $\ln \left( n_\mathrm{c}^0 / n_\mathrm{c} (r=1)  \right) \approx 2.8 > 1 $. Thus, incompressible condition is insufficient. 

Our boundary conditions are as follows. We use stress-free boundary conditions for the charged fluid at the crust-core interface
\begin{align} 
\vec{u}_\mathrm{c} \cdot \hat r \; \Big|_{r=R_\mathrm{c}} &= 0 ,\\
\left. \frac{\partial}{\partial r} \left[ \vec u_\mathrm{c} \cdot \hat \theta \right] \; \right|_{r=R_\mathrm{c}} &= 0,\\
\left. \frac{\partial}{\partial r} \left[ \vec u_\mathrm{c} \cdot \hat \phi \right] \; \right|_{r=R_\mathrm{c}} &= 0 .
\end{align}

While the real boundary conditions in a NS core are unknown, stress-free ones are plausible. They describe a case where the magnetic field lines can move freely at the surface following the charged particle fluid. Phrasing it differently, these boundary conditions assume that fluid shear from motion in the core is not transferred to the crust. A possible alternative is slip-free condition. In this case magnetic field lines are frozen in the crust. We plan to study the effect of boundary conditions in our future work.

We use a potential boundary condition for the magnetic field. We assume that the field inside matches a potential magnetic field outside, which 
corresponds to the case of a vacuum outside
\begin{align}
\left.\frac{\partial  B_{r,\ell}}{\partial r} + \frac{(\ell+2)}{r} B_{r,\ell} \; \right|_{r = R_\mathrm{c}} &= 0   \label{eq:boundary_vacuum1} ,\\ 
(\curl \vec B) \cdot \vec r \; \Big|_{r = R_\mathrm{c}} &= 0  .
\label{eq:boundary_vacuum2}
\end{align}

Here $\vec B_{r,\ell}$ is the $\ell$-component of the spectral expansion of the radial component of the magnetic field over spherical harmonics. This boundary condition does not exclude the possible formation of surface currents (at the crust-core interface) which could have $\theta$ and $\phi$ components.

We primarily include the magnetic $\Delta \vec B / \mathrm{Rm}$ and fluid $\vec{f}_\nu (\vec u_\mathrm{c})$ diffusivities to satisfy the boundary conditions and properly resolve the formation of boundary layers. In a real non-superconductive neutron star core, magnetic diffusion is relatively large in comparison to expected ambipolar timescale, so our Rm is just two orders of magnitude smaller\footnote{Assuming $B_0 = 10^{15}$~G.}. The real Am/Re value is much smaller, and we have no hope to properly model boundary layers for velocity. 

\subsection{Two-fluid ambipolar diffusion MHD system}
\label{s:twofluids_eq}

In the case of two-fluid ambipolar diffusion we supplement our system eqs.~(\ref{eq:uc_barotrop}~--~\ref{eq:gauss}) with an additional Navier-Stokes equation for the neutron fluid,
i.e.\ eq.\ (\ref{eq:neutron_velocity}). 
Similarly, we add another anelastic constraint for the neutron fluid and another integral condition. The complete system of equations is as follows\footnote{We note that the integral condition presented here is a simplified one. The most physically accurate formulation corresponds to the separate conservation of charged particles and neutrons when weak interactions are neglected (see Eqs. 16–17 in \citealt{moraga2025magnetothermal}).}
\begin{align}
\mathrm{Am}\; n_\mathrm{c} \; \frac{\partial \vec u_\mathrm{c}}{\partial t} &= -n_\mathrm{c}\grad \delta \mu_\mathrm{c} + \vec f_B \nonumber\\
& + \gamma_{\mathrm{np}} n_\mathrm{n} n_\mathrm{c} (\vec u_\mathrm{n} - \vec u_\mathrm{c})  +  \vec{f}_\nu (\vec u_\mathrm{c})
\label{eq:uc},\\
\mathrm{Am}\; n_\mathrm{n}\; \frac{\partial \vec u_\mathrm{n}}{\partial t} &= -n_\mathrm{n} \grad \delta \mu_\mathrm{n} \nonumber\\
&+ \gamma_{\mathrm{np}} n_\mathrm{c} n_\mathrm{n} (\vec u_\mathrm{c} - \vec u_\mathrm{n})  + \vec{f}_\nu (\vec u_\mathrm{n}), \label{eq:un}
\end{align}
\begin{align}
\dv (n_\mathrm{c} \vec u_\mathrm{c}) &= 0,  \\  
\dv (n_\mathrm{n} \vec u_\mathrm{n}) &= 0 ,   
\end{align}
\begin{equation}
\frac{\partial \vec B}{\partial t} = \curl (\vec u_\mathrm{c} \times \vec B) + \frac{\lap \vec B}{\mathrm{Rm}} ,
\end{equation}
\begin{equation}
\dv \vec B  = 0, \label{eq:gauss2}   
\end{equation}
\begin{align}
\int \delta \mu_\mathrm{c} \; dV &= 0 , \\  
\int \delta \mu_\mathrm{n} \; dV &= 0  .  
\end{align}
The viscous force for neutrons $\vec{f}_\nu (\vec u_\mathrm{n})$ depends on the neutron density profile. Unlike the 1-barotropic-fluid case, our equations for two-fluid evolution do not include $\left(\vec u_\mathrm{c} \cdot \vec \nabla\right) \vec u_\mathrm{c}$. It follows the assumptions made in Section 2.1 and 2.2.
This term has a stabilizing effect for 1-barotropic-fluid simulations, helping the system to remain stable during the saturation phase with less viscosity. In the case of two fluids, the simulation is already more stable and this term can be removed. Physically this term corresponds to transport of vorticity. Vorticity is created at the boundary and can be transported inside the core volume if this term is present. If the term is absent, vorticity needs to be dispersed by viscous interaction at the boundary. When we added this term to two-fluid simulations, we did not notice any significant difference. Our motivation to neglect $\left(\vec u_\mathrm{c} \cdot \vec \nabla\right) \vec u_\mathrm{c}$ term is that inertial forces are small for both fluids. 

Similarly to the 1-barotropic-fluid MHD system, we use stress-free boundary condition for velocities at the crust-core interface
\begin{align}
\left.\vec u_\mathrm{c} \cdot \hat{r} \;\right|_{r = R_\mathrm{c}} &= 0, \\
\left.\vec u_\mathrm{n} \cdot \hat{r} \;\right|_{r = R_\mathrm{c}} &= 0 ,\\
\left.\frac{\partial }{\partial r} \left[ u^{\theta, \phi}_\mathrm{c} \right]\; \right|_{r = R_\mathrm{c}} &= 0, \\  
\left.\frac{\partial }{\partial r} \left[ u^{\theta, \phi}_\mathrm{n} \right]\; \right|_{r = R_\mathrm{c}} &= 0  .   
\end{align}

We use the same potential boundary conditions eqs.~(\ref{eq:boundary_vacuum1}, \ref{eq:boundary_vacuum2}) for the magnetic field. It is important to highlight that all previous studies \citep{Castillo2017MNRAS,Passamonti2017MNRAS,OfengeimGusakov2018PhRvD,Castillo2020MNRAS,IgoshevHollerbach2023MNRAS,Moraga2024MNRAS} modeled the core using the Euler equations, where the only applicable boundary condition was the \textit{non-penetration condition}, i.~e. $\vec{u}_\mathrm{c} \cdot \hat{r} \,|_{r=R_\mathrm{c}} = \vec{u}_\mathrm{n} \cdot \hat{r} \,|_{r=R_\mathrm{c}} =0$. Here, by employing the Navier-Stokes equations, we solve a set of parabolic equations, which allows for the implementation of additional and more diverse boundary conditions at the crust-core interface.

In a stable state, if we neglect the viscosity, the eqs. (\ref{eq:uc} and \ref{eq:un}) correspond to force balance:
\begin{equation}
-\vec f_B = \vec f_\mathrm{c} + \vec f_\mathrm{n} \equiv -n_\mathrm{c}\grad \delta \mu_\mathrm{c} - n_\mathrm{n}\grad \delta \mu_\mathrm{n}. 
\label{eq:two_fluid_force_balance}
\end{equation}

\subsection{Initial conditions}
\label{s:init_cond}

\begin{figure}
    \begin{minipage}{0.49\linewidth}
    \includegraphics[width=\columnwidth]{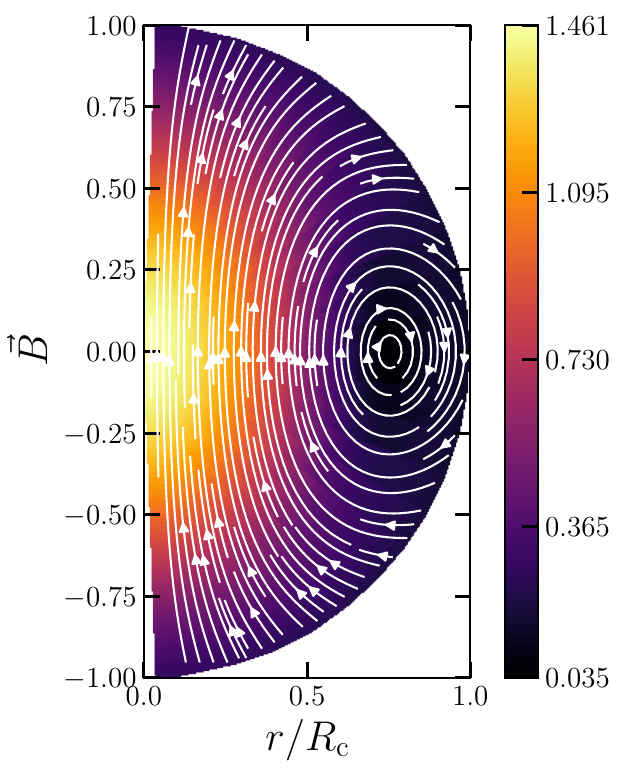}
    \end{minipage}
    \begin{minipage}{0.49\linewidth}
    \includegraphics[width=\columnwidth]{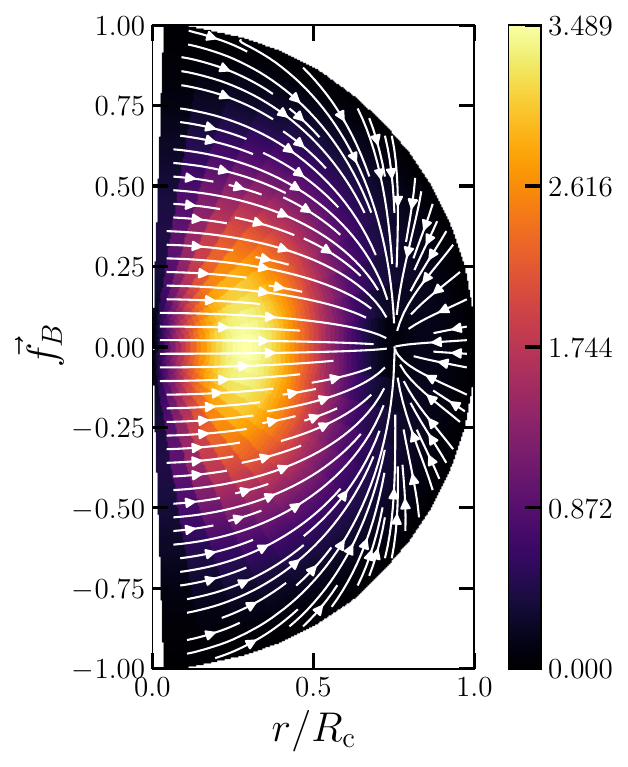}
    \end{minipage}
    \begin{minipage}{0.49\linewidth}
    \includegraphics[width=\columnwidth]{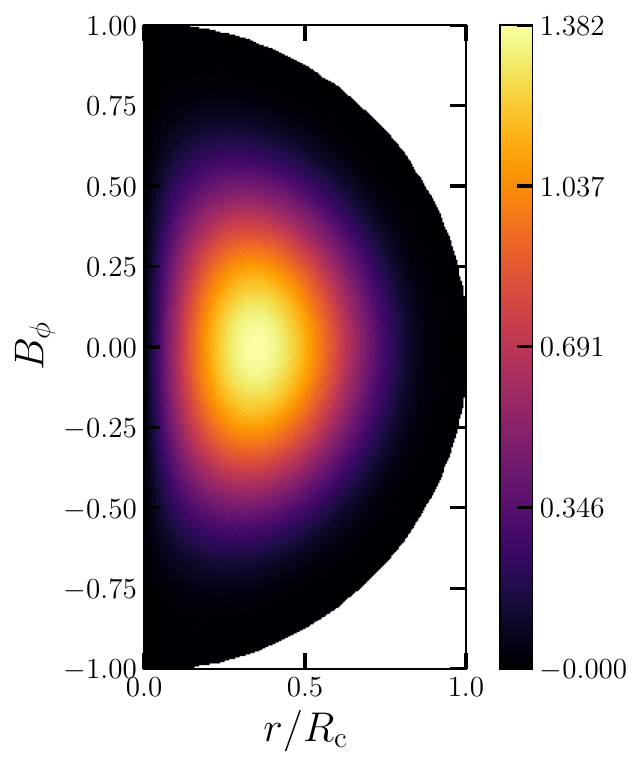}
    \end{minipage}
    \begin{minipage}{0.49\linewidth}
    \includegraphics[width=\columnwidth]{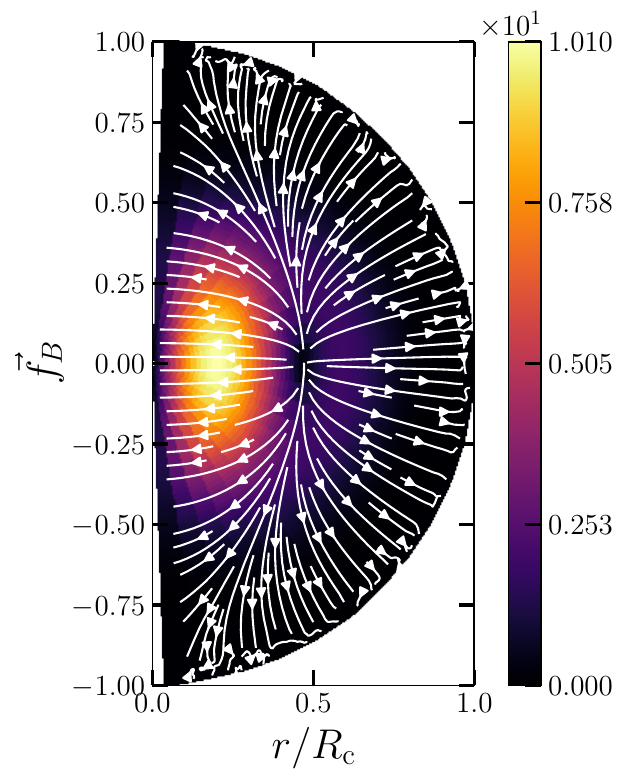}
    \end{minipage}
    \caption{Initial conditions for the magnetic field. Upper panels are for poloidal field, lower panels for toroidal field. We show the magnetic field (left column) as well as its Lorentz force (right column).}
    \label{fig:init_cond}
\end{figure}

As mentioned in the introduction, in this article we choose simple initial conditions. 
We model the purely poloidal dipolar field as
\begin{align}
B_r      & = 2\sqrt\frac{3}{4\pi}\left(a_0^\mathrm{p} + a_2^\mathrm{p} r^2 + a_4^\mathrm{p} r^4 + a_6^\mathrm{p} r^6 + a_{8}^\mathrm{p} r^{8}\right) \cos(\theta),\\
B_\theta & = -2\sqrt\frac{3}{4\pi}\left(a_0^\mathrm{p} + 2 a_2^\mathrm{p} r^2 + 3 a_4^\mathrm{p} r^4 + 4 a_6^\mathrm{p} r^6 + 5 a^\mathrm{p}_8 r^8\right) \sin(\theta) ,\\
B_\phi & = 0,
\end{align}
with numerical coefficients summarized in Table~\ref{t:init_cond}. This initial condition guarantees $\vec u_\mathrm{n} \cdot \hat r = 0$ and $\vec u_\mathrm{c} \cdot \hat r = 0$ at the crust-core interface, in the absence of viscosity at the beginning of our simulations.

For a toroidal magnetic field we use the following radial dependence
\begin{align}
B_r       &= 0, \\
B_\theta  &= 0, \\
B_\phi    &= 2.5568\;r\left(a_0^\mathrm{t} +a_2^\mathrm{t} r^2 + a_4^\mathrm{t} r^4 +a_6^\mathrm{t} r^6 + a_8^\mathrm{t} r^8\right) \sin(\theta).
\end{align}


\begin{table}
    \caption{Numerical values for polynomials describing the radial dependence of initial magnetic field configurations.}
    \label{t:init_cond}
    \centering
    \begin{tabular}{cccccc}
    \hline
    Field    & $a_0$ & $a_2$ & $a_4$ & $a_6$ & $a_8$ \\
    \hline
    Poloidal & 1.5564 & -2.6170 & 1.9775 & -0.7897 & 0.15275 \\
    Toroidal & 2.4880 & -8.8394 & 11.590 & -6.6142 & 1.37541 \\
    \hline
    \end{tabular}

\end{table}

\begin{table}
    \caption{Summary of models together with parameters used in numerical simulations.}
    \label{t:simulation_summary}
    \centering
    \begin{tabular}{lcccccccc}
     \hline
      & Type & Am & Re & Rm & $N_r$ & $L_\mathrm{m}$ & $M_\mathrm{m}$ & Physics \\
     \hline
     A   & Pol & $10^{-5}$  & 0.1 & $10^2$ & 96 & 24 & 48 &  Barotropic\\
     B   & Pol & $10^{-5}$  & 0.1 & $10^2$ & 128 & 48 & 96 & Barotropic  \\
     C   & Pol & $10^{-5}$  & 1  & $10^2$ & 196 & 24 & 48 &  Barotropic  \\
     \hline
     D   & Pol & $10^{-5}$ & 2 & $10^2$ & 96 & 24 & 48  &  Two-fluids \\
     E   & Pol & $10^{-5}$ & 2 & $10^2$ & 128 & 48 & 96  &  Two-fluids   \\
     F   & Tor & $10^{-5}$ & 2 & $10^2$ & 96 & 24 & 48 &  Two-fluids  \\
     G   & Pol & $10^{-5}$ & 4 & $10^2$ & 128 & 48 & 96 & Two-fluids  \\
    \hline
    \end{tabular}
    \tablecomments{Pol corresponds to purely poloidal while Tor corresponds to purely toroidal initial magnetic fields. Barotropic refers to the 1-barotropic-fluid system. }
\end{table}

We run numerical simulations for two magnetic field configurations which are initially axially symmetric. To each of these configurations we add small non-axisymmetric perturbations in all the components of the magnetic field which 
are normally distributed with zero mean and a standard deviation of $10^{-4}$ of our $B_0$. 
We list all the configurations with their respective resolutions in Table~\ref{t:simulation_summary}.
We show the initial magnetic field configurations in Figure~\ref{fig:init_cond}. All initial velocities are chosen to be zero.

\subsection{Simulations}

We solve the MHD systems of eqs.~(\ref{eq:uc_barotrop}~--~\ref{eq:gauss}) and eqs.~(\ref{eq:uc}~--~\ref{eq:gauss2}) in spherical geometry using the \texttt{Dedalus} v.3 code \citep{Vasil2019JCPX,Lecoanet2019,Burns2020PhRvR}. 
\texttt{Dedalus} is a pseudo-spectral code, i.e.\ the numerical solution is written as an expansion over a set of basis functions. These basis functions are spherical harmonics $Y_\ell^m (\theta, \phi)$ over angular coordinates and  Jacobi polynomials $P_n^{\alpha,\beta} (r)$ in the radial direction. For example, the chemical potential perturbation $\delta \mu$ is expanded as:
\begin{equation}
\delta \mu \; (r,\theta, \phi) = \sum_{k,\ell,m} a_{\ell,m,k} P_k^{\alpha,\beta} (r) Y_\ell^m (\theta, \phi),  
\label{eq:expansion}
\end{equation}
where $a_{\ell,m,k}$ are numerical coefficients. For vectors, \texttt{Dedalus} uses spin-weighted spherical harmonics; all mathematical details can be found in \cite{Vasil2019JCPX}. The series eq.~(\ref{eq:expansion}) is truncated at $N_r$ for maximum power of radial polynomials and at $L_\mathrm{m}$ and $M_\mathrm{m}$ for maximum power of spherical harmonics. These numbers determine the numerical resolution. We summarize these values for each run in Table~\ref{t:simulation_summary}. After solutions are expanded into a series, the collocation principle is used, i.e. equations are required to be exactly satisfied at all points of the numerical mesh. This requirement transforms a system of partial differential equations into an algebraic system which is further solved using LU-decomposition. The time propagation is done using the Runge-Kutta method.

We solve equations in a full sphere domain, so the NS center is also included. We have to use significantly finer resolution in the radial direction to resolve the boundary layers.

\subsection{Diagnostic of the simulations}
\label{s:diagnostic}
In order to understand the simulations and check their validity, we track a number of different numerical quantities and store them every 20 iterations. We summarize all these values in this section.
First, we compute the total magnetic energy
\begin{equation}
E_\mathrm{m} = \frac{1}{2}\int \vec{B}^2 dV    .
\end{equation}
Second, we store the kinetic energies for charged particles and neutrons
\begin{equation}
E^\mathrm{c}_\mathrm{k} = \frac{\mathrm{Am}}{2} \int n_\mathrm{c} \vec{u}_\mathrm{c}^2 dV,    
\end{equation}
\begin{equation}
E^\mathrm{n}_\mathrm{k} = \frac{\mathrm{Am}}{2} \int n_\mathrm{n} \vec{u}_\mathrm{n}^2 dV.    
\end{equation}
Third, in order to have a better understanding for the structure of the flow and magnetic field, we define an axially symmetric part of the magnetic field by averaging each of its spherical components over the azimuthal angle,
\begin{equation}
    \langle B_j\rangle(r,\theta)=\frac{1}{2\pi}\int_0^{2\pi}d\phi\,B_j(r,\theta,\phi)
\end{equation}
for $j=r,\theta,\phi$, with the associated 
axially symmetric part of the magnetic energy,
\begin{equation}
E_\mathrm{m}^\mathrm{s} = \frac{1}{2} \int \langle \vec B \rangle^2 dV  .
\end{equation}
We are specifically interested in the non-axisymmetric part of the energy, which is defined as
\begin{equation}
E_\mathrm{m}^{\mathrm{ns}} = E_\mathrm{m} - E_\mathrm{m}^\mathrm{s}  .  
\end{equation}
Fourth, we store contributions of individual channels to energy decay. The Ohmic losses are computed as
\begin{equation}
L_\mathrm{Ohm} = \int \frac{(\curl \vec B)^2}{\mathrm{Rm}} dV  .  
\end{equation}
The viscous losses are computed as
\begin{equation}
L_\mathrm{Re} = - \int \left[ n_\mathrm{n} \; \vec u_\mathrm{n} \cdot \vec f_\nu (\vec u_\mathrm{n}) + n_\mathrm{c} \; \vec u_\mathrm{c} \cdot \vec f_\nu (\vec u_c) \right] \; dV   .
\end{equation}
Losses due to the ambipolar diffusion are computed as
\begin{equation}
L_\mathrm{ad} = \int n_\mathrm{c} n_\mathrm{n} \gamma_\mathrm{np} (\vec u_\mathrm{c} - \vec u_\mathrm{n})^2 dV    .
\end{equation}
The Poynting flux is computed as
\begin{equation}
L_\mathrm{p} = \int \dv \left\{ \vec B \times \left[ \vec u_\mathrm{c} \times \vec B \right] \right\} dV   .
\end{equation}

Thus, the total energy dissipation reads (see e.g. \citealt{Moraga2024MNRAS,moraga2025magnetothermal})
\begin{equation}
\mathcal{\dot E} = -L,    
\end{equation}
where 
\begin{equation}
    \mathcal{\dot E} \equiv \dot E_\mathrm{m} + \dot E^\mathrm{c}_\mathrm{k} + \dot E^\mathrm{n}_\mathrm{k},
\end{equation}
and
\begin{equation}
    L \equiv - L_\mathrm{ad} - L_\mathrm{p} - L_\mathrm{Re} - L_\mathrm{Ohm}.
\end{equation}


We examine forces in the simulation using the root-mean-square value computed for the Lorentz force as
\begin{equation}
f_{B,\mathrm{rms}} =\sqrt{ \int \left((\vec \nabla \times \vec B)\times \vec B\right)^2 dV}    .
\end{equation}

We store individual flow snapshots every 2000-8000 iterations, depending on the speed of the simulation.
Using \texttt{Dedalus} expansion of vector field into spherical harmonics, we use $B_{l,m}$ components of the expansion to calculate 
the energy contained in each $\ell$
\begin{equation}
    \mathcal{E}_\ell = \sum_m \frac{1}{2}\int \vec{B}_{\ell,m}^2dV,
\end{equation}
and each $m$
\begin{equation}
    \mathcal{E}_m = \sum_\ell \frac{1}{2}\int \vec{B}_{\ell,m}^2dV.
\end{equation}
These energy spectra allow us to understand the simulation results at later stages.

We finish this subsection by remarking that in our simulations we would like to minimize the contribution of $L_\mathrm{Ohm}$ and $L_\mathrm{Re}$ to the total energy decay.

\section{Results}
\label{s:results}

\subsection{1-barotropic-fluid MHD}
\label{s:res_barotrop}

We run simulations for $180\; t_\mathrm{Alf}$ with an initial condition which consists of an axially symmetric poloidal magnetic field with added noise. We show the total energy of the magnetic field and charged fluid as well as the non-axisymmetric part of these energies in Figure~\ref{f:energies_barotrop}.

Based on the noticeable difference in energy evolution, we identify five separate stages. Initially, we describe the extent of these stages and general evolution, and we will later examine each stage in detail.

\subsubsection{Stages of simulations based on energy evolution}

\begin{figure*}
    \begin{minipage}{0.49\linewidth}
    \includegraphics[width=\columnwidth]{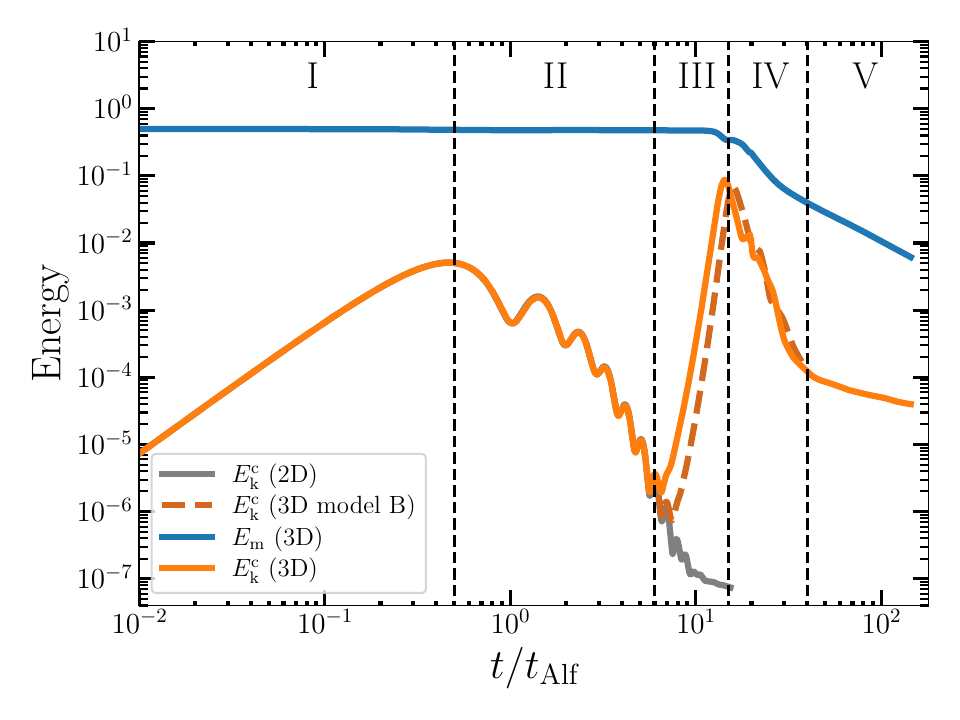}
    \end{minipage}
    \begin{minipage}{0.49\linewidth}
    \includegraphics[width=\columnwidth]{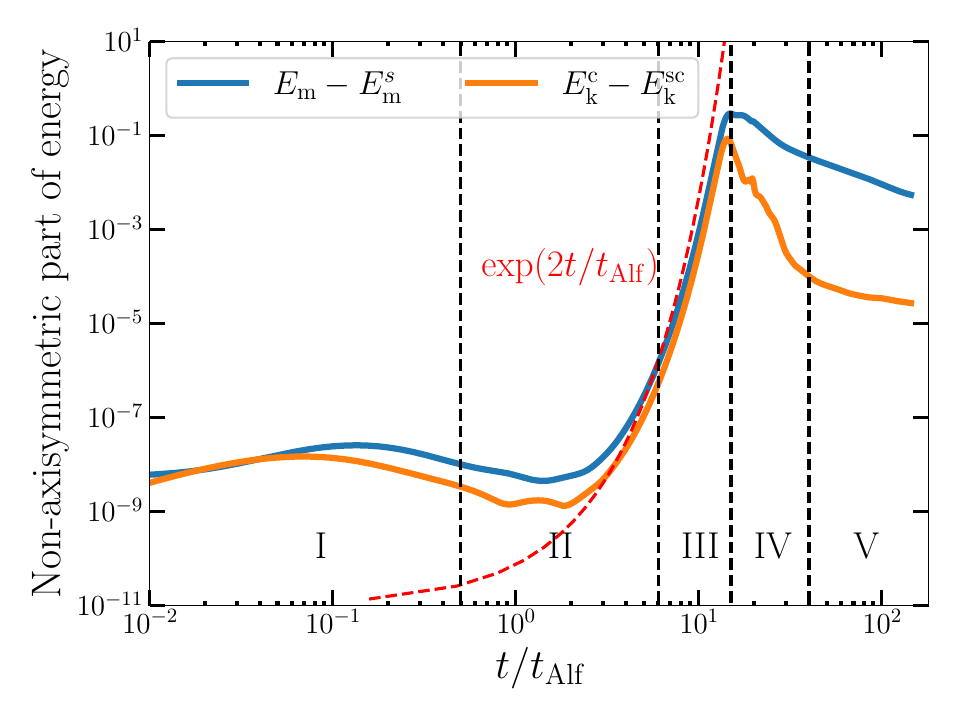}
    \end{minipage}
    \caption{Evolution of kinetic and magnetic energies for 1-barotropic-fluid MHD simulation. Left panel: total magnetic and kinetic energies. Right panel: non-axisymmetric part of magnetic ($E_\mathrm{m} - E_\mathrm{m}^\mathrm{s}$) and kinetic ($E_\mathrm{k} - E_\mathrm{k}^\mathrm{s}$) energies. We mark the following stages: I is the acceleration stage, II is evolution towards 2D force balance, III is  instability, IV is magnetic turbulence, V is resistive decay. The red dashed line in the right panel shows exponential growth with timescale $t_\mathrm{Alf}/2$. We show in grey the results of axially symmetric simulations. }
    \label{f:energies_barotrop}
\end{figure*}

(I) \textit{Acceleration stage:} This stage ends at $\approx 0.5\;t_\mathrm{Alf}$. The charged fluid is initially at rest, thus, it gets accelerated by the Lorentz force. At this stage, the magnetic field stays nearly constant, so the total magnetic energy decays by $\approx 2$\% by the end of this stage. The kinetic energy of charged particles grows until reaching a maximum.  At this stage the simulation stays approximately axially symmetric; the non-axisymmetric part of the magnetic and kinetic energy stays close to the level of the initial perturbations.
Thanks to our choice of Am, the maximum kinetic energy of charged particles is  significantly smaller than the magnetic energy. 

(II) \textit{Evolution towards 2D force equilibrium:} This stage ends at $\approx 6\;t_\mathrm{Alf}$. The kinetic energy decays because the magnetic field configuration evolves towards an axisymmetric force balance. In this equilibrium a substantial part of the Lorentz force is canceled by the term describing chemical potential perturbations. At this stage, we also see oscillations in kinetic energy and similar oscillations in magnetic energy. These oscillations represent Alfv\'en waves and are slowly damped by fluid viscosity.  

(III) \textit{Instability:} This stage ends at $\approx 15\;t_\mathrm{Alf}$. The non-axisymmetric part of both kinetic and magnetic energy grows from $\approx 10^{-7}$ (comparable to the level of initial perturbations) to $\approx 10^{-1}$, which is comparable to the initial axisymmetric part of magnetic energy, see right panel of Figure~\ref{f:energies_barotrop}. The growth follows $\exp(2 t /t_\mathrm{Alf})$ quite closely which strongly supports instability. 
 
(IV) \textit{Magnetic turbulence:} this stage ends at $\approx 40\;t_\mathrm{Alf}$. This stage is characterized by a significant contribution of viscosity which helps the system to evolve.
We discuss the dominant energy loss channel in the following section.

(V) \textit{Resistive decay and cascade:} after a new equilibrium is reached, the energy decays mostly due to Ohmic losses. Small scales decay faster than the large scales. 

Below we investigate each stage in detail.

\subsubsection{Acceleration stage}

\begin{figure}
    \begin{minipage}{0.99\linewidth}
    \includegraphics[width=\columnwidth]{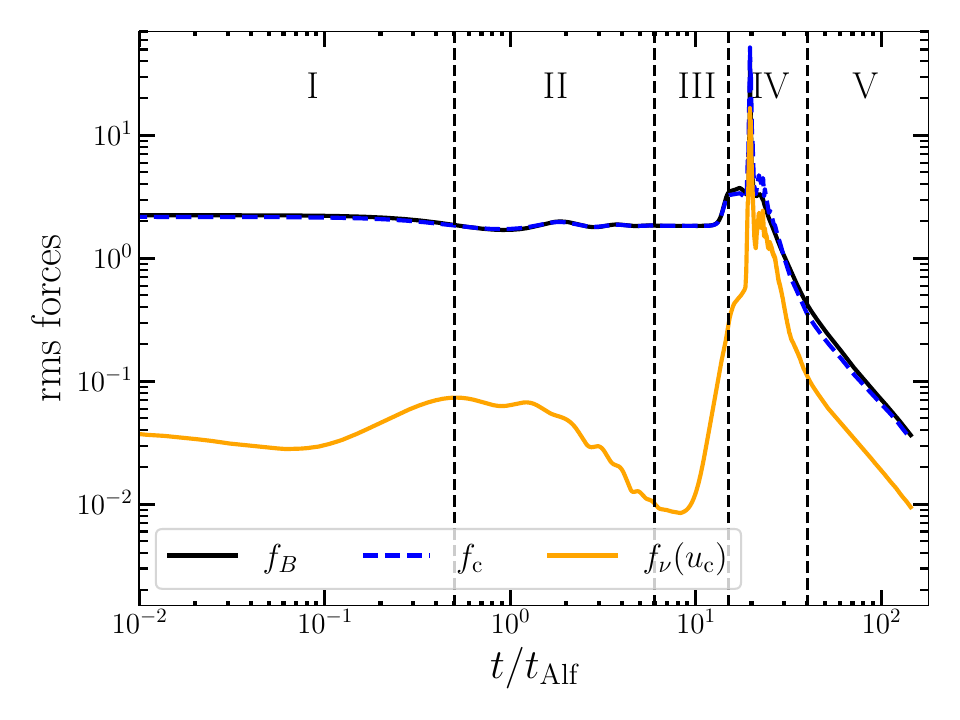}
    \end{minipage}
    \caption{Root-mean square forces during 1-barotropic-fluid simulation A.}
    \label{f:rms_forces_K}
\end{figure}

\begin{figure}
    \includegraphics[width=\columnwidth]{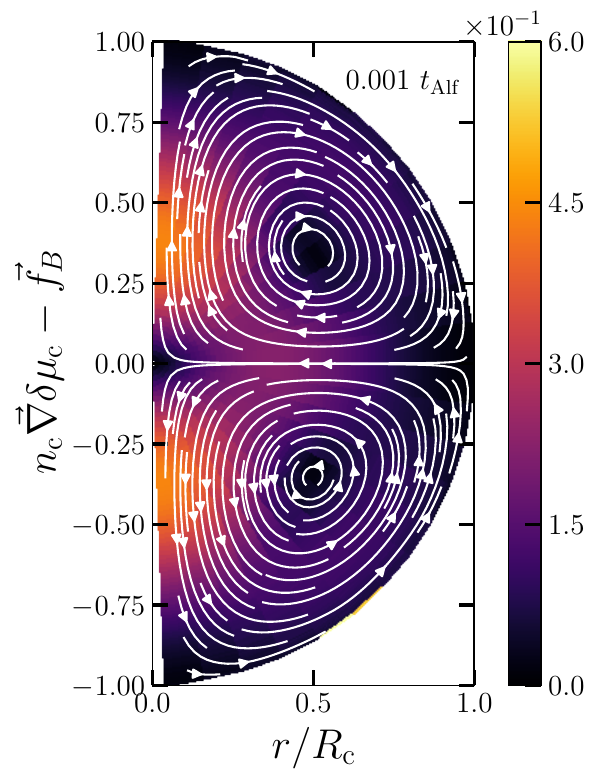}
    \caption{Meridional cuts showing force balance during the acceleration stage for 1-barotropic fluid. 
    }
    \label{f:forces_K}
\end{figure}

\begin{figure*}
    \begin{minipage}{0.32\linewidth}
    \includegraphics[width=\columnwidth]{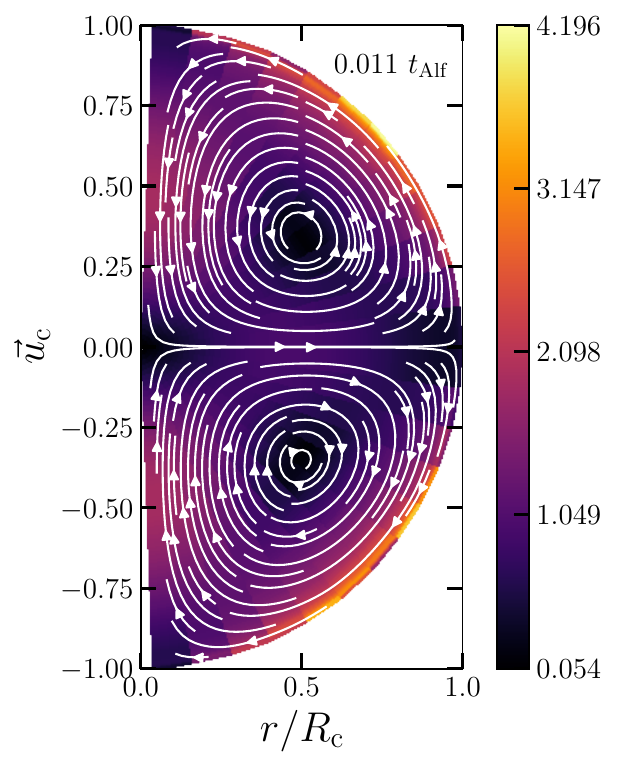}
    \end{minipage}
    \begin{minipage}{0.32\linewidth}
    \includegraphics[width=\columnwidth]{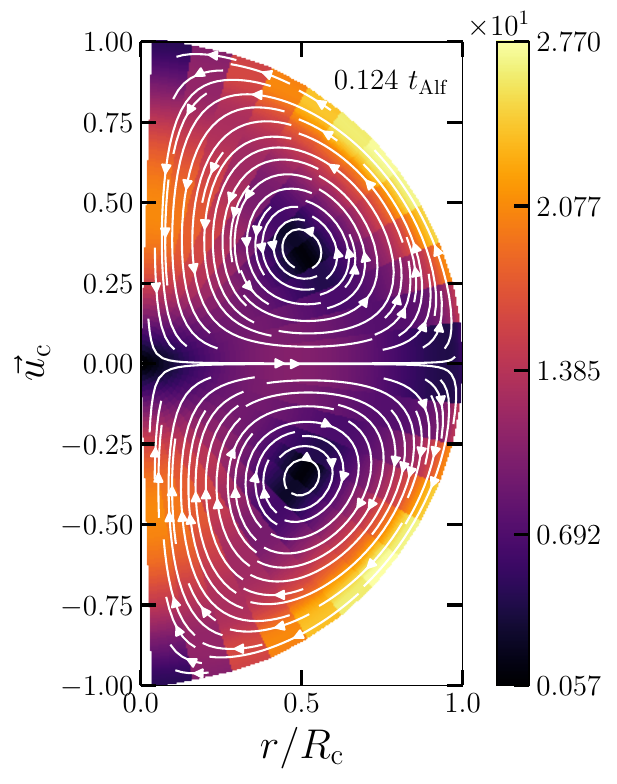}
    \end{minipage}
   \begin{minipage}{0.32\linewidth}
   \includegraphics[width=\columnwidth]{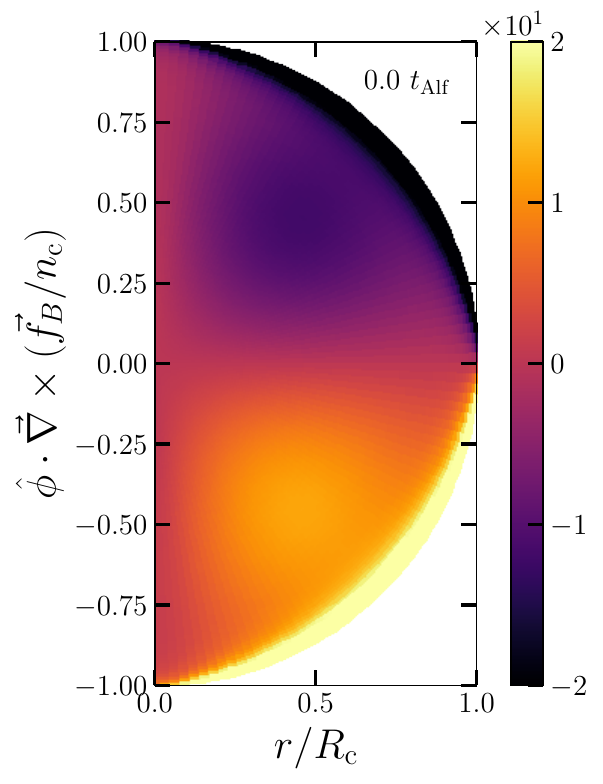}
   \end{minipage}
    \caption{The first two panels show meridional cuts of
    $\vec{u}_\mathrm{c}$ during the acceleration stage at $t=0.011\;t_\mathrm{Alf}$ and $0.124\;t_\mathrm{Alf}$.
    The third panel shows the curl of the Lorentz force divided by number density of charged particles. In this case, we did not include the initial noise.}
    \label{fig:curlfB}
\end{figure*}

We show rms as well as detailed force balances in Figures~\ref{f:rms_forces_K} and \ref{f:forces_K}. 
It is clear from these figures that already during the acceleration stage the pressure forces from charged particles can balance a significant part of the Lorentz force. 


We show the meridional cut for charged particle velocities in the left panel of Figure~\ref{fig:curlfB} at two distinct times during the acceleration stage. The velocity profile looks similar and consists of two vortices rotating in opposite directions in the northern and southern hemispheres. 

The velocity structure at this stage is explained by the solenoidal part of the Lorentz force. In order to demonstrate it, we divide 
eq.~(\ref{eq:uc_barotrop}) by charged particles number density and compute its curl, we obtain
\begin{equation}
\mathrm{Am} \frac{D \vec{\omega}_\mathrm{c} }{Dt} = \curl \left( \frac{\vec f_B}{n_\mathrm{c}} \right)  + \curl \left( \frac{ \vec{f}_\nu (\vec{u}_\mathrm{c}) }{n_\mathrm{c}}\right),
\label{eq:vorticity}
\end{equation}
where $\vec{\omega}_\mathrm{c} = \curl  \vec{u}_\mathrm{c}$. 

We show the $\phi$-component of the curl of the Lorentz force divided by charged particles number density in Figure~\ref{fig:curlfB}. We clearly see the source of northern and southern circulations as the curl of Lorentz force changes sign across the equator, which agrees well with circulations of opposite direction seen in the same figure. The vorticity is negative in the northern hemisphere and positive in the southern. 
The density profiles makes vorticity grows significantly in a small layer closer to the surface. Thus, it is expected that the largest velocities will be near the crust-core interface at this stage.

\subsubsection{Evolution toward 2D force balance}
\begin{figure*}
    \begin{minipage}{0.32\linewidth}
    \includegraphics[width=\columnwidth]{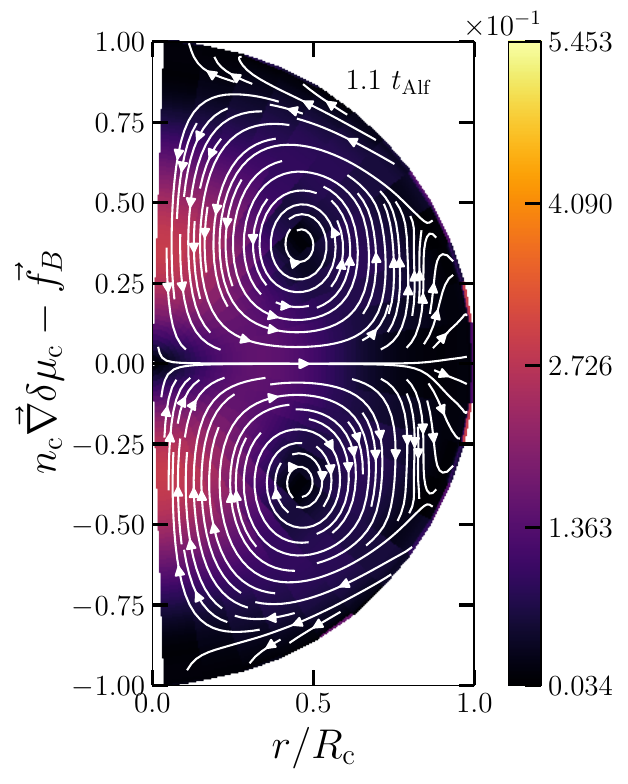}
    \end{minipage}
    \begin{minipage}{0.32\linewidth}
    \includegraphics[width=\columnwidth]{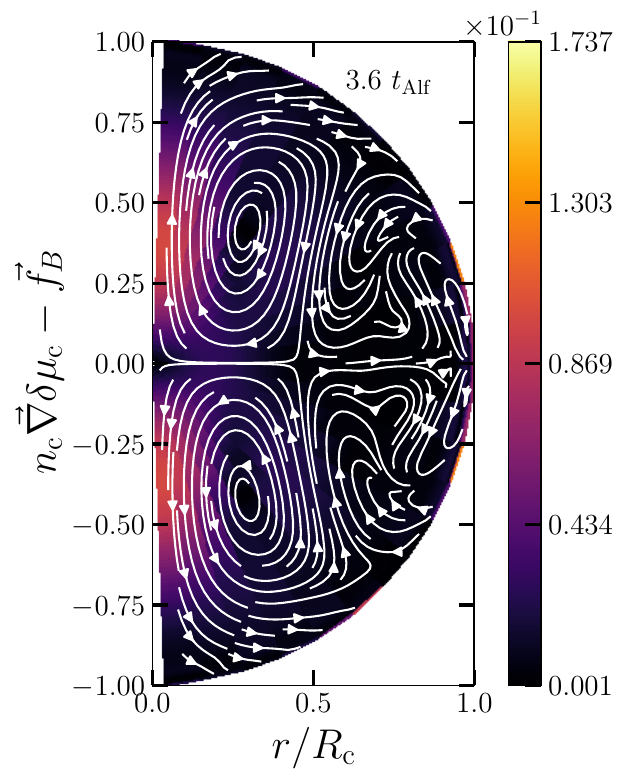}
    \end{minipage}
    \begin{minipage}{0.32\linewidth}
    \includegraphics[width=\columnwidth]{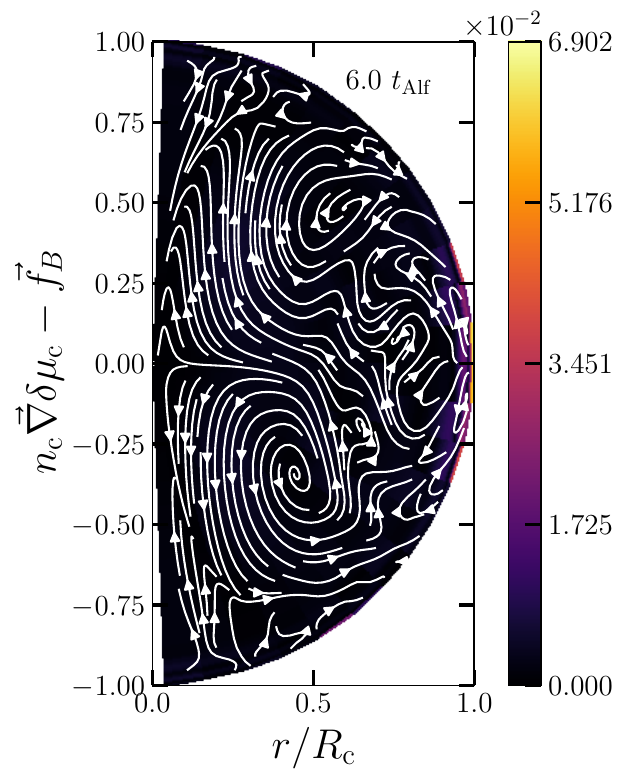}
    \end{minipage}
    \caption{Meridional cuts for 1-barotropic-fluid simulations. We show the deviation from the force balance state $n_\mathrm{c}\vec \nabla \delta \mu_\mathrm{c} - \vec f_B$ over the course of the second stage of the simulations. }
    \label{f:diff}
\end{figure*}

During the previous stage, the Lorentz force stayed nearly constant. At this stage the system evolves towards satisfying the condition
\begin{equation}
\vec f_B - n_\mathrm{c} \vec \nabla \delta \mu_\mathrm{c} \approx 0.   
\end{equation}
i.e. role of viscosity decreases.
We illustrate this tendency in Figure~\ref{f:diff}. The difference between the Lorentz force and the chemical potential gradient decays an order of magnitude during six Alfv\'en times, suggesting the approach to the GS-equilibrium \citep{gradrubin54,shafranov66}. At $6\;t_\mathrm{Alf}$ the difference concentrates near the crust-core interface where the magnetic field must satisfy its particular boundary condition. 

In order to check whether the instability arises exclusively due to 3D perturbations, we run shorter simulations which are designed to be axially symmetric with no added noise\footnote{Axially symmetric with added noise behaves the same. This is also true for fully three-dimensional without any noise where we start with exact axially symmetric initial conditions.}. We show the energy evolution in Figure~\ref{f:energies_barotrop} with the grey line, which initially overlaps with the orange line. It is clear that kinetic energy decays even further during stage III as well. 

\begin{figure*}
    \begin{minipage}{0.32\linewidth}
    \includegraphics[width=\columnwidth]{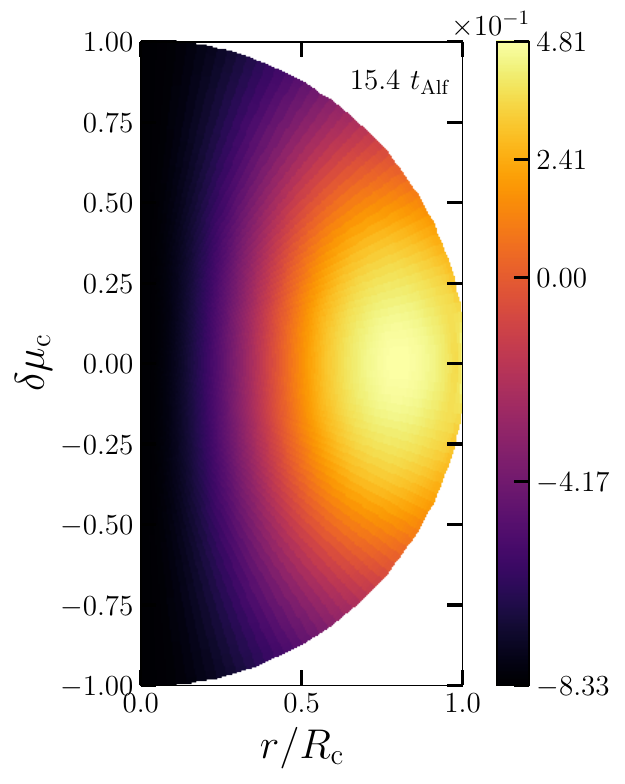}
    \end{minipage}
    \begin{minipage}{0.32\linewidth}
    \includegraphics[width=\columnwidth]{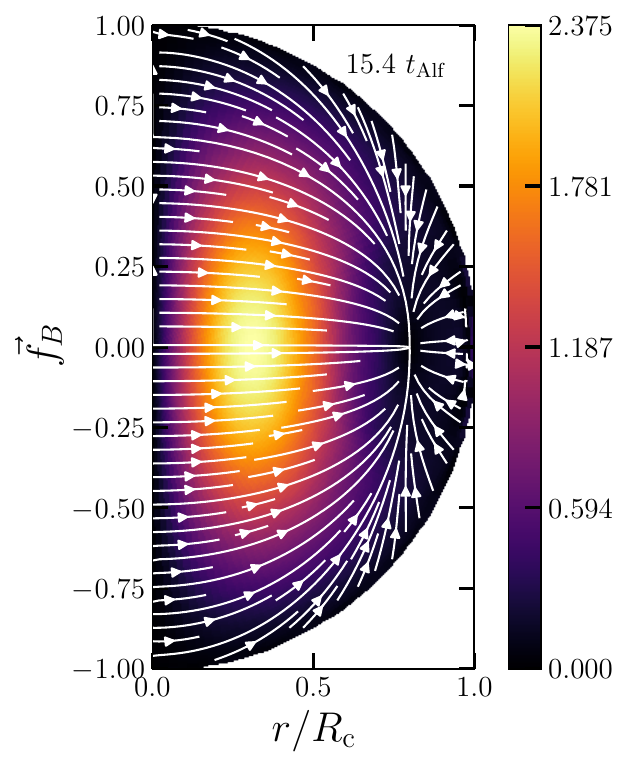}
    \end{minipage}
    \begin{minipage}{0.32\linewidth}
    \includegraphics[width=\columnwidth]{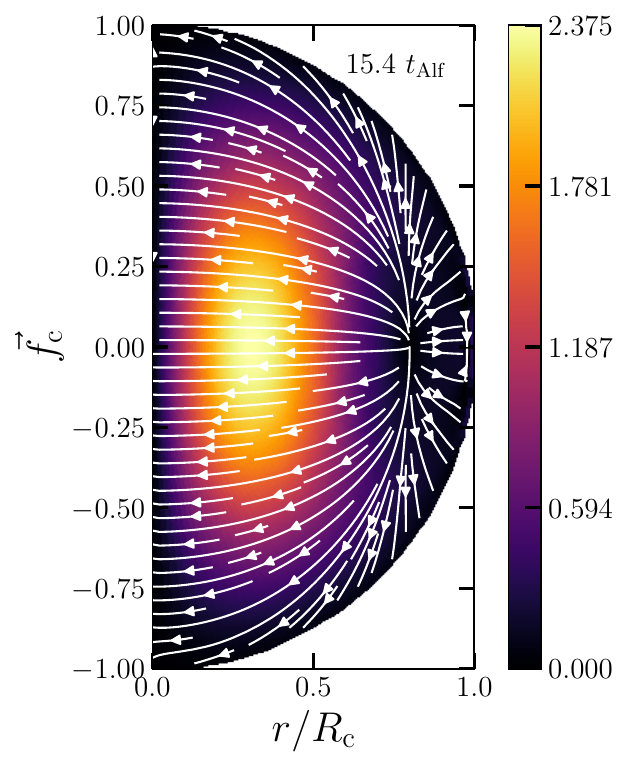}
    \end{minipage}
    \caption{Results of 1-barotropic-fluid axially symmetric simulations at $\approx 15\; t_\mathrm{Alf}$.}
    \label{f:stable}
\end{figure*}

We show the hydro-magnetic equilibrium obtained in longer axisymmetric simulations in Figure~\ref{f:stable}. Simple visual comparison between $\vec f_B$ and $\vec f_\mathrm{c} = -n_\mathrm{c}\vec \nabla \delta \mu_\mathrm{c}$ reveals that these are very similar forces. While the magnetic field has changed only slightly from its initial configuration, see Figure~\ref{fig:init_cond}, the Lorentz force, which is the non-linear product of magnetic field and its curl, evolves significantly.

Our choice for Am and Re makes Alfv\'en waves possible. They are not immediately damped and could exist in the volume for up to $\approx 10\; t_\mathrm{Alf}$. If Am is chosen as $\mathrm{Am}\approx 10^{-7}$, the Alfv\'en waves are damped faster and no visible oscillations occur in energy.


\subsubsection{Instability}

\begin{figure*}
    \begin{minipage}{0.49\linewidth}
    \includegraphics[width=\columnwidth]{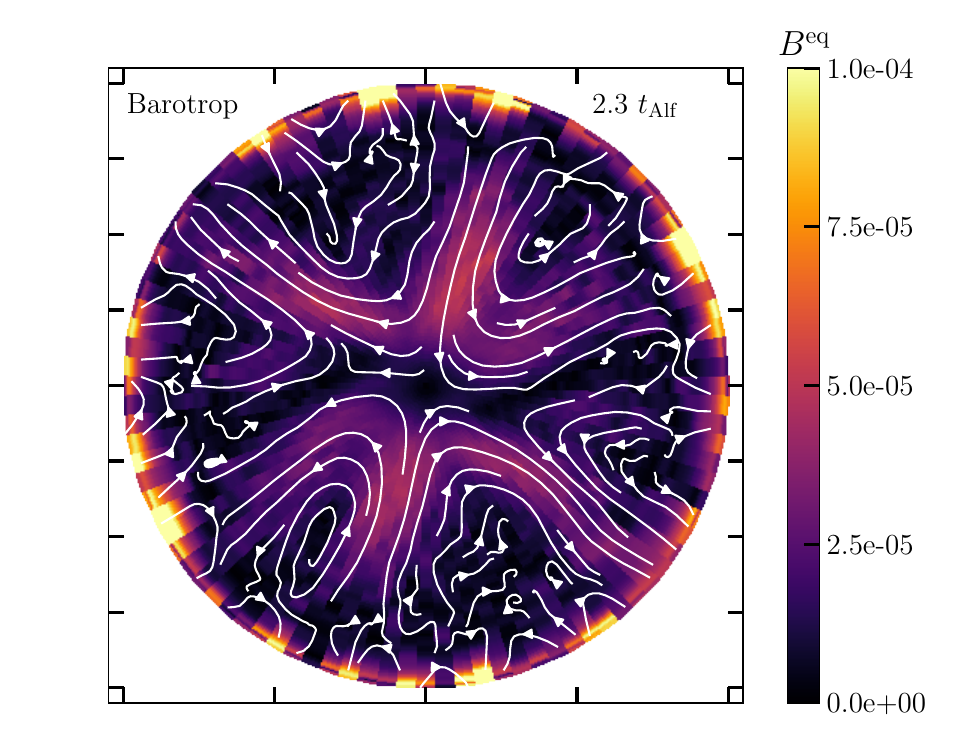}
    \end{minipage}
    \begin{minipage}{0.49\linewidth}
    \includegraphics[width=\columnwidth]{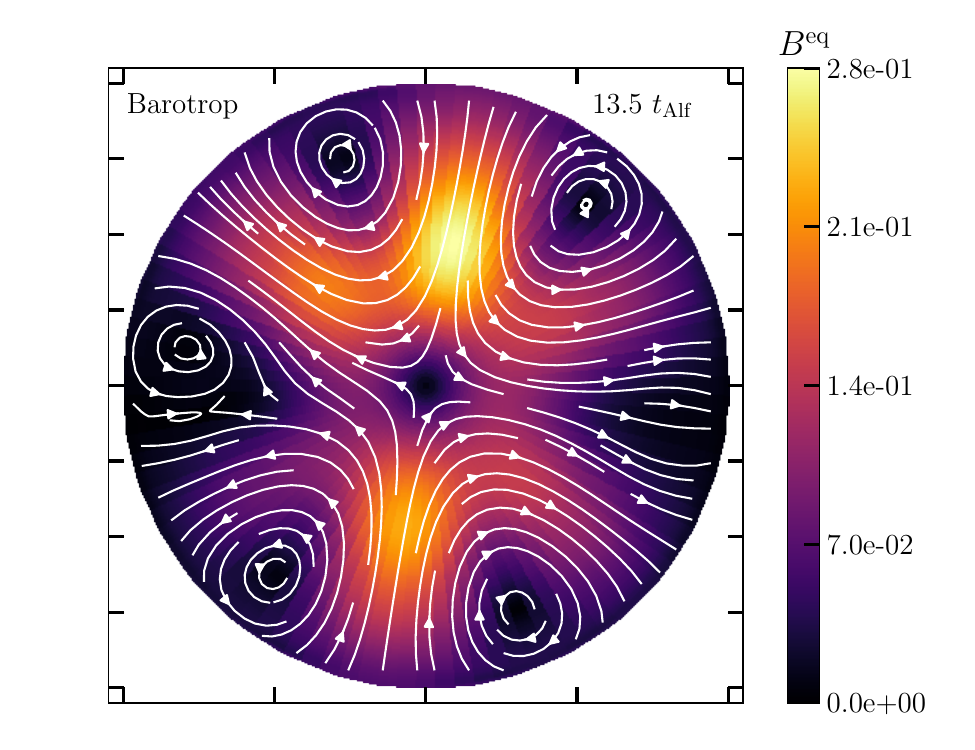}
    \end{minipage}
    \caption{Equatorial cuts for 1-barotropic-fluid simulations. We show magnetic field in the equatorial plane, i.e. only $B_r$ and $B_\phi$ components of the magnetic field. }
    \label{f:B_eq}
\end{figure*}

\begin{figure}
    \includegraphics[width=\columnwidth]{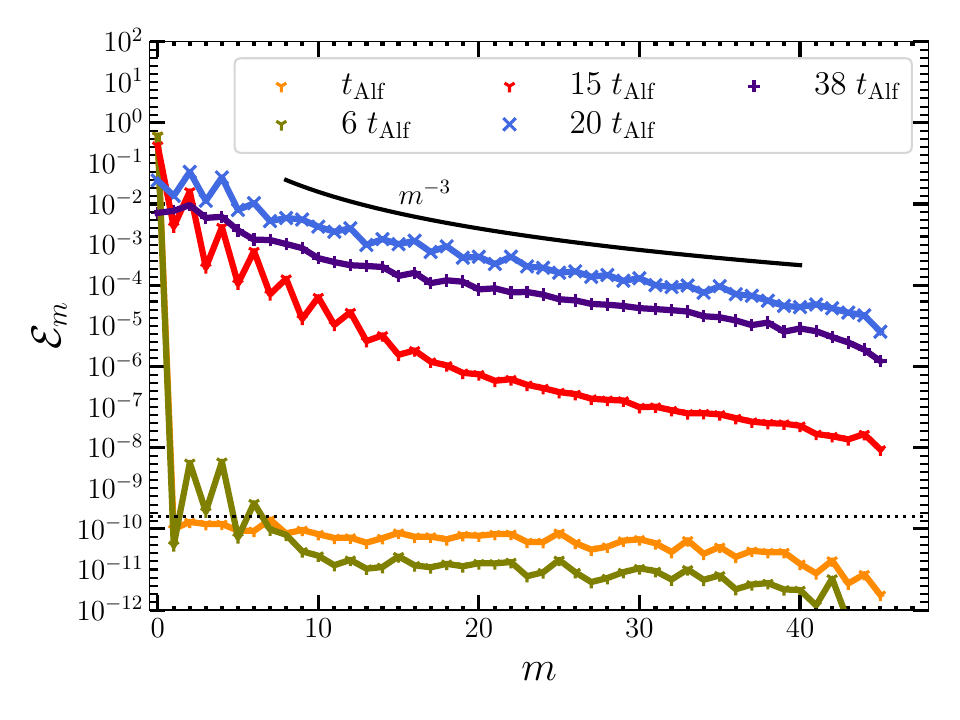}
    \caption{Magnetic energy spectra at different stages of the barotropic simulation. }
    \label{f:magnetic_energy_barotrop}
\end{figure}

\begin{figure}
    \includegraphics[width=\columnwidth]{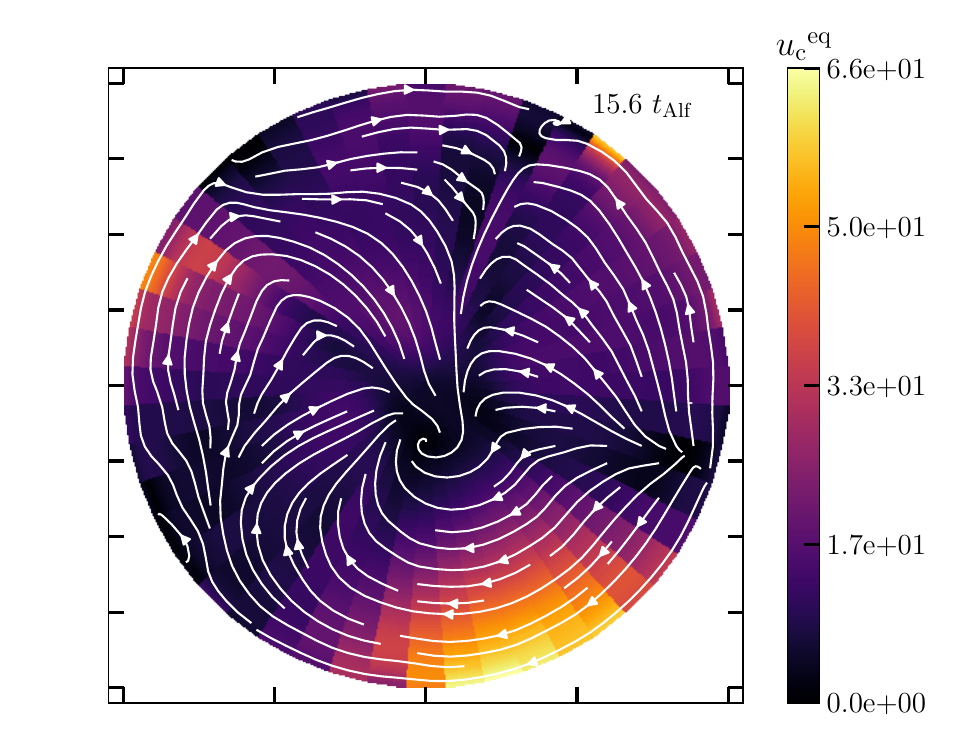}
    \caption{Charged fluid velocity for simulation A shown in equatorial cut.  }
    \label{f:velocity_field_K_instability}
\end{figure}

As mentioned above, the instability does not develop in axially symmetric simulations, and is primarily seen as growth in the non-axisymmetric parts of kinetic and magnetic energy, as seen from the right panel in Figure~\ref{f:energies_barotrop}. Thus, to trace the instability better, we examine the equatorial slices of magnetic fields in Figure~\ref{f:B_eq}. These magnetic fields are initialized as Gaussian noise in our initial conditions. Indeed, the magnetic field lines of our initial dipolar field cross the equatorial cut at right angles with $B_r = 0$ and $B_\phi=0$ in the slice. When we examine these equatorial slices during the simulation, we see a growth of regular structure. We show the magnetic energy spectra in Figure~\ref{f:magnetic_energy_barotrop}. In this plot we see that $m=2$ is growing rapidly, as well as $m=4$. We additionally show the velocity field near the saturation of instability in Figure~\ref{f:velocity_field_K_instability}.
Earlier, \cite{Braithwaite2009MNRAS} found that mode $m=2$ is the most unstable mode for an axially symmetric poloidal magnetic field. Modes $m\geq 2$ grew in their simulations.  The instability found by \cite{Braithwaite2009MNRAS} is seen as loops of magnetic fields move in direction parallel to the NS axis of symmetry.  

\subsubsection{Magnetic Turbulence}

The end of stage III is the first moment when the total magnetic energy decays noticeably. During the interval ranging from $14\; t_\mathrm{Alf}$ to $20\; t_\mathrm{Alf}$ the magnetic energy decreases from 0.45 to 0.3, i.e. by a factor of 1.5. It is the moment when simulations similar to \cite{LanderJones2012MNRAS} become invalid due to their intrinsic linear nature. 

\begin{figure}
    \begin{minipage}{0.49\linewidth}
    \includegraphics[width=\columnwidth]{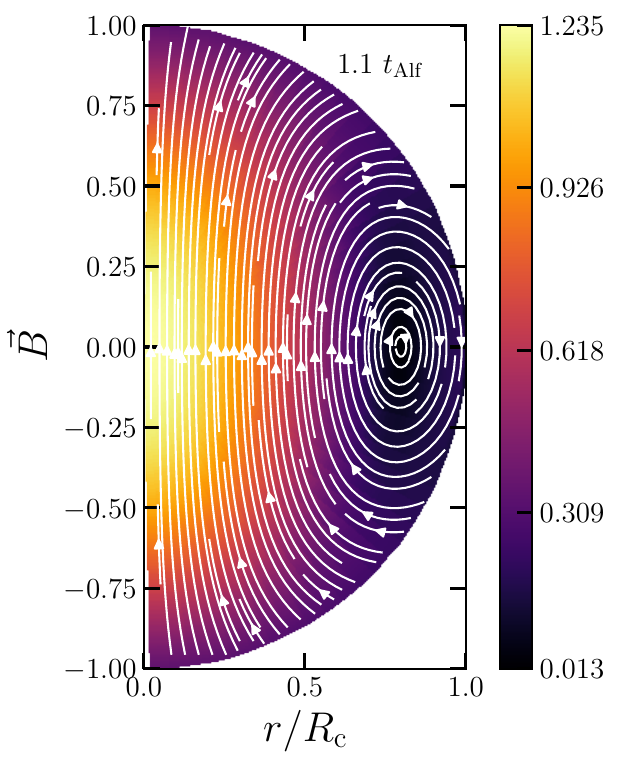}
    \end{minipage}
    \begin{minipage}{0.49\linewidth}
    \includegraphics[width=\columnwidth]{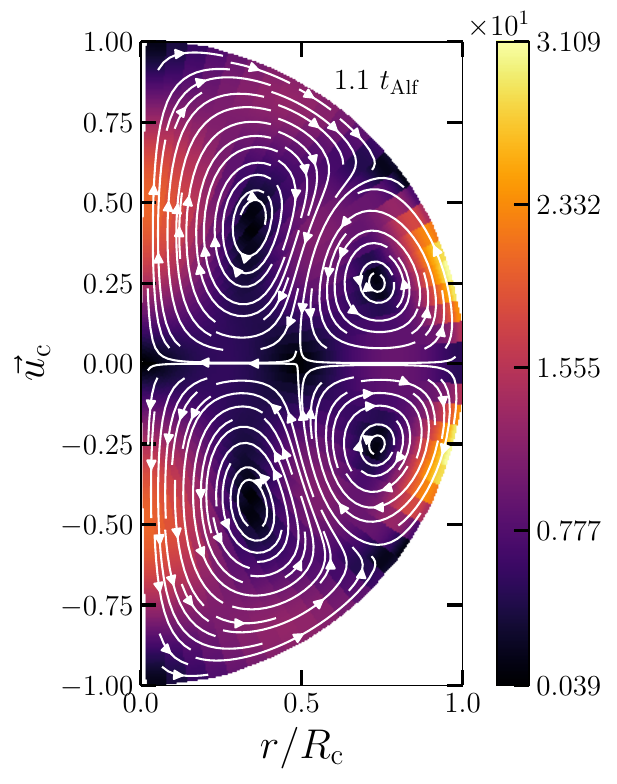}
    \end{minipage}
    \begin{minipage}{0.49\linewidth}
    \includegraphics[width=\columnwidth]{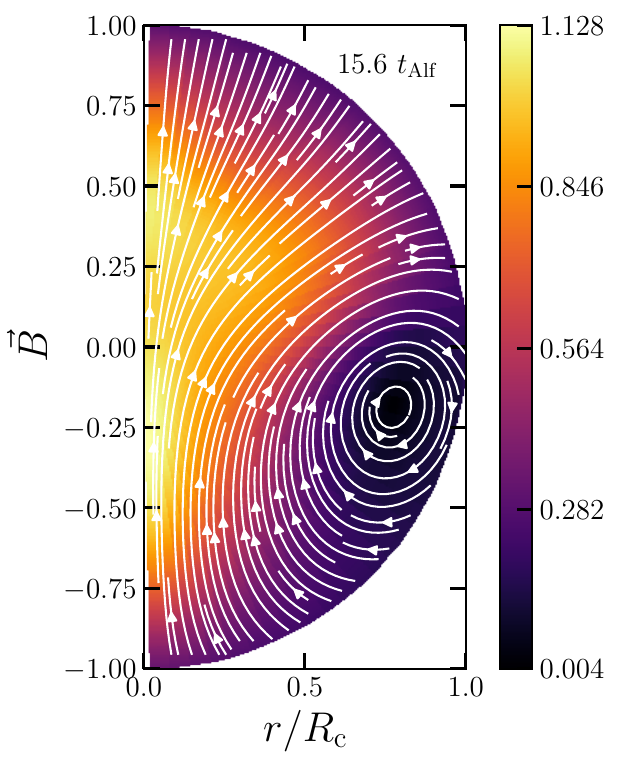}
    \end{minipage}
    \begin{minipage}{0.49\linewidth}
    \includegraphics[width=\columnwidth]{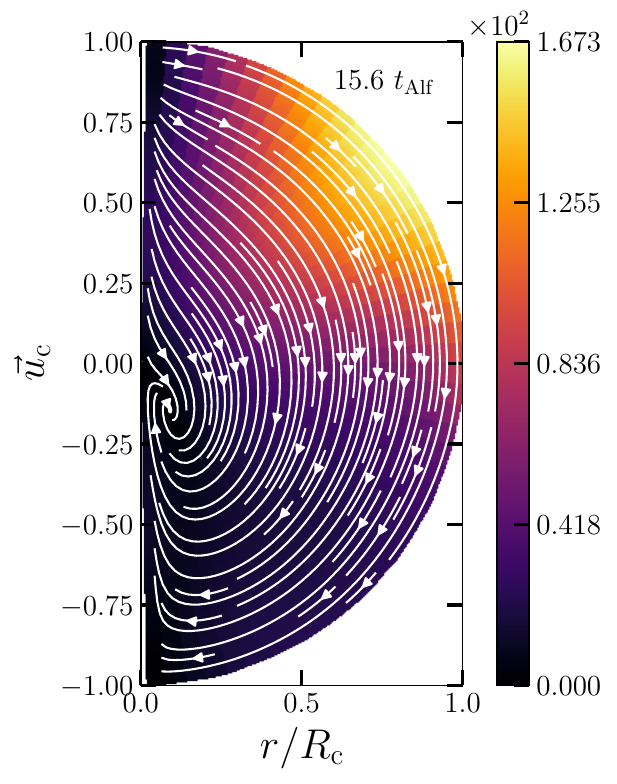}
    \end{minipage}
    \begin{minipage}{0.49\linewidth}
    \includegraphics[width=\columnwidth]{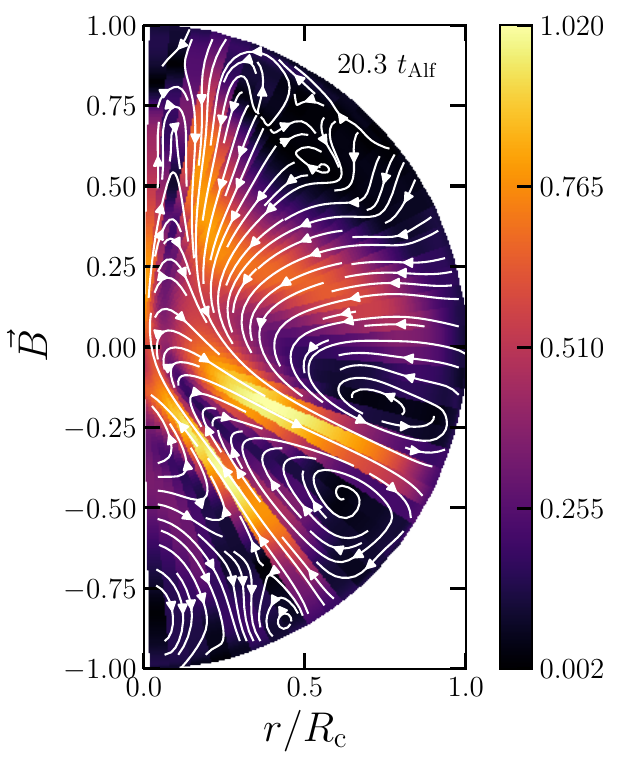}
    \end{minipage}
    \begin{minipage}{0.49\linewidth}
    \includegraphics[width=\columnwidth]{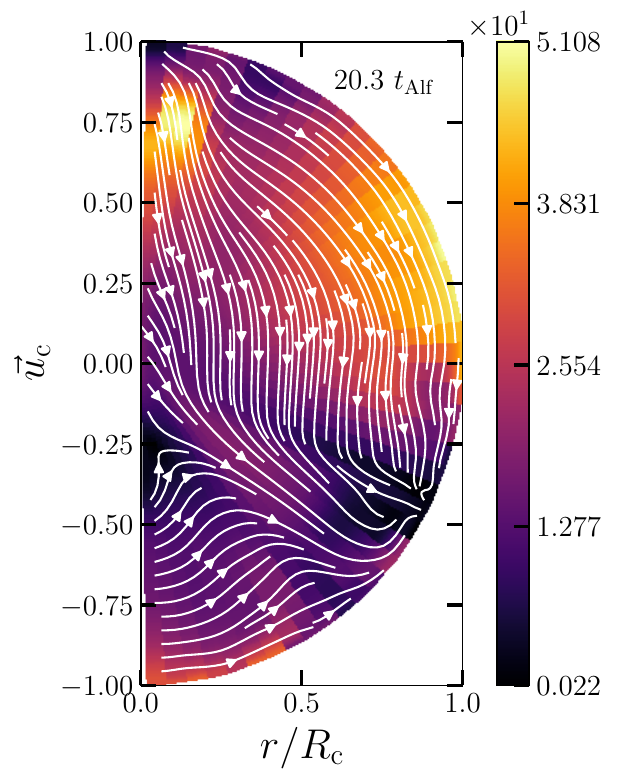}
    \end{minipage}
    \begin{minipage}{0.49\linewidth}
    \includegraphics[width=\columnwidth]{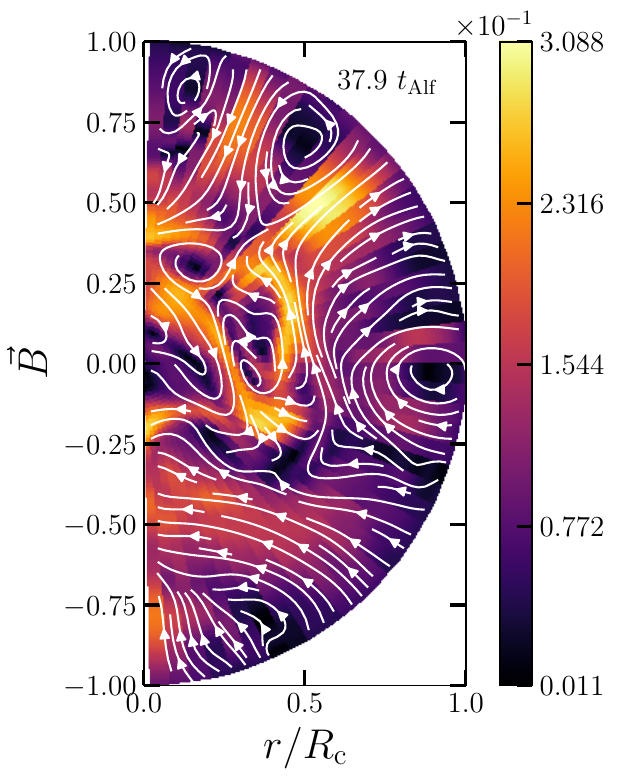}
    \end{minipage}
    \begin{minipage}{0.49\linewidth}
    \includegraphics[width=\columnwidth]{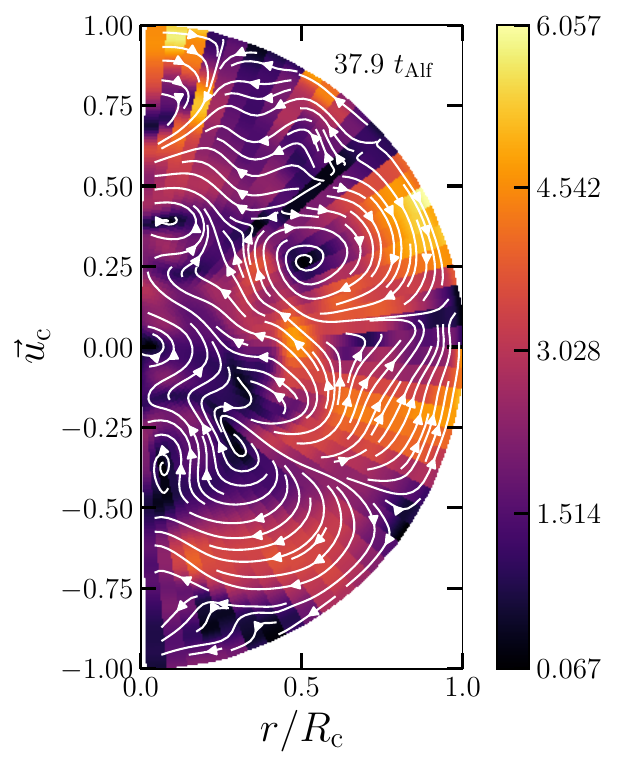}
    \end{minipage}
    \caption{Evolution of magnetic field and charged particle velocities at different stages in 1-barotropic-fluid simulation B. 
    }
    \label{fig:stable}
\end{figure}

In our simulations we see a significant transformation in the structure of magnetic and velocity fields around $20\; t_\mathrm{Alf}$; see a few examples for meridional cuts in Figure~\ref{fig:stable}.
While the magnetic field preserves its large-scale structure at $15\;t_\mathrm{Alf}$, it is split into separate magnetic layers around $20\; t_\mathrm{Alf}$. This process progresses further producing small-scale structure of the field by $38\; t_\mathrm{Alf}$.

The viscous contribution to the force balance peaks at a similar time. In Figure~\ref{f:rms_forces_K} we see that rms Lorentz, viscous and chemical potential perturbations forces grows rapidly and reach a peak. The viscous force stays subdominant through most of the simulation, but increases nearly three orders of magnitude during this transition. This growth in amplitude similarly indicates that a large number of small-scale velocity structures develop during this stage. 

\begin{figure}
    \begin{minipage}{0.99\linewidth}
    \includegraphics[width=\columnwidth]{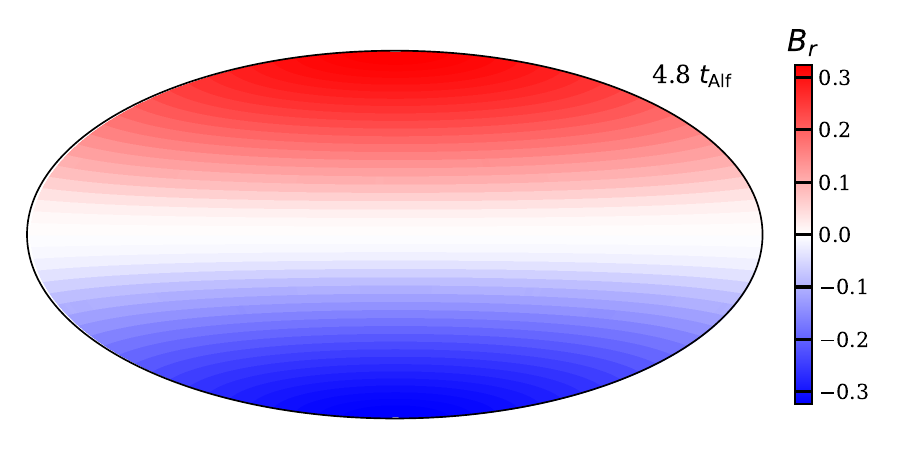}
    \end{minipage}
    \begin{minipage}{0.99\linewidth}
    \includegraphics[width=\columnwidth]{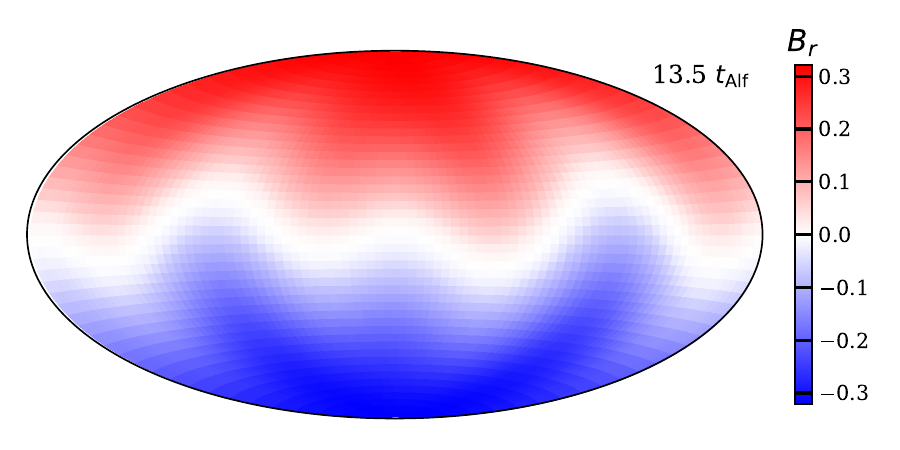}
    \end{minipage}
    \begin{minipage}{0.99\linewidth}
    \includegraphics[width=\columnwidth]{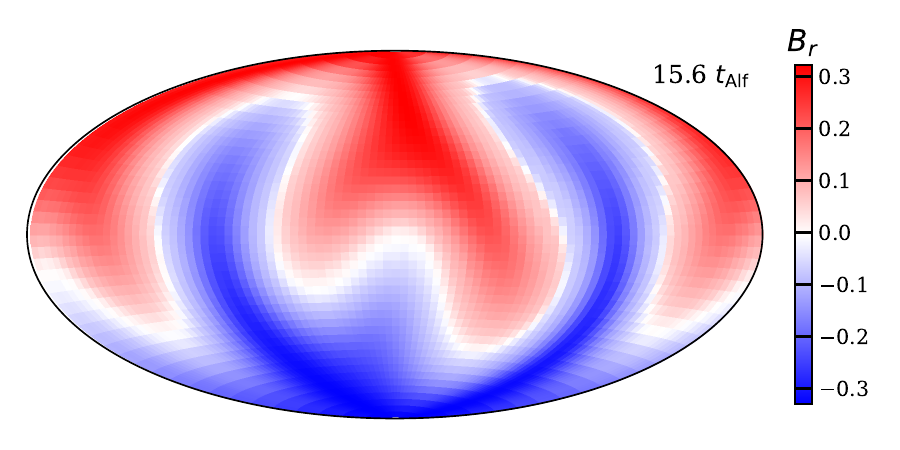}
    \end{minipage}
    \begin{minipage}{0.99\linewidth}
    \includegraphics[width=\columnwidth]{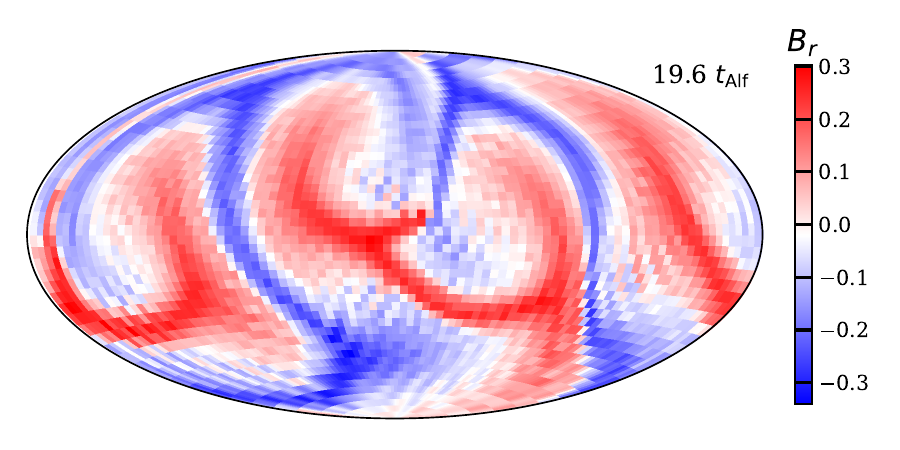}
    \end{minipage}
    \begin{minipage}{0.99\linewidth}
    \includegraphics[width=\columnwidth]{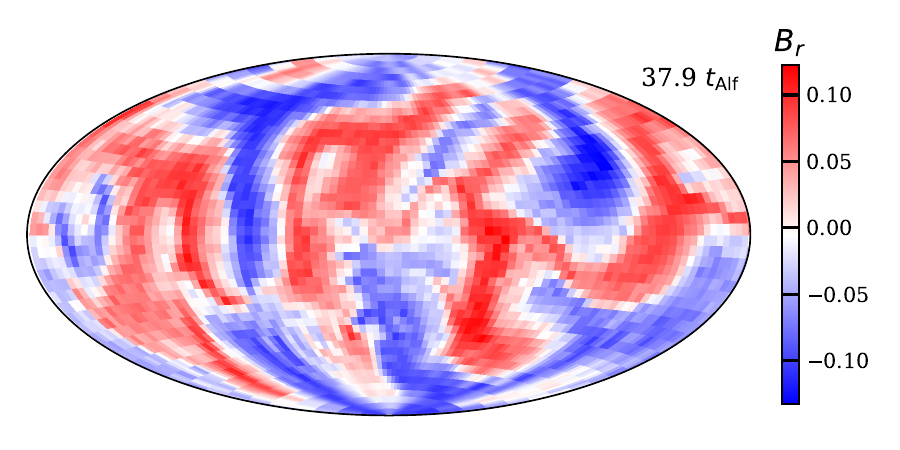}
    \end{minipage}
    \caption{Time evolution of surface radial magnetic field in higher-resolution 1-barotropic-fluid simulations. }
    \label{fig:Br_surface}
\end{figure}

This is further supported by spectral analysis. In Figure~\ref{f:magnetic_energy_barotrop} we see that the magnetic energy spectrum at $15-20\; t_\mathrm{Alf}$ times is dramatically different from its initial spectrum. While the initial spectrum was dominated by $m=0$, the new energy spectrum follows a cascade with $\mathcal{E}_\mathrm{m}\propto m^{-3}$. When the instability is developing (at earlier times), it has a limited number of well-defined harmonics which grow exponentially, including $m=2$, $m=4$ and $m=6$, with $m=2$ growing the fastest. This is not the case during stage IV. The energy in $m=0$ decays quickly and becomes comparable to $m=2$. We plot the surface map of radial magnetic fields in Figure~\ref{fig:Br_surface}. 

In order to better understand this stage we run another simulation with significantly better angular resolution -- simulation B. The onset of three-dimensional instability occurs slightly later in this simulation, see Figure~\ref{f:energies_barotrop}. It is probably related to the fact that initial perturbations have smaller physical size in this simulation due to the way they are initialized, so it takes slightly longer to grow the instability. Otherwise, the simulation behaves very similarly. It develops  instability which becomes turbulent and dissipates the magnetic energy.

\subsubsection{Resistive decay and stable configuration}

The evolution at stage V is remarkable because the magnetic field configuration does not change significantly from $\approx 40\;t_\mathrm{Alf}$ to the end of the simulation at $180\;t_\mathrm{Alf}$. Even if we compute a new Alfv\'en timescale rescaled for stage V magnetic field $t'_\mathrm{Alf} \approx 4\; t_\mathrm{Alf}$ corresponding to a magnetic field strength 0.25 of its original value, the simulation stays stable for at least $140/4 = 35\;t'_\mathrm{Alf}$, while the original third stage corresponding to development of instability lasts from $6\;t_\mathrm{Alf}$ to $15\;t_\mathrm{Alf}$.
As mentioned in the introduction, there are no known stable magnetic field configurations in a barotropic star. Any solution which stays stable on tens of Alfv\'en timescales and longer is thus especially interesting and worth studying. 

\begin{figure*}
\begin{minipage}{0.49\linewidth}
\includegraphics[width=\columnwidth]{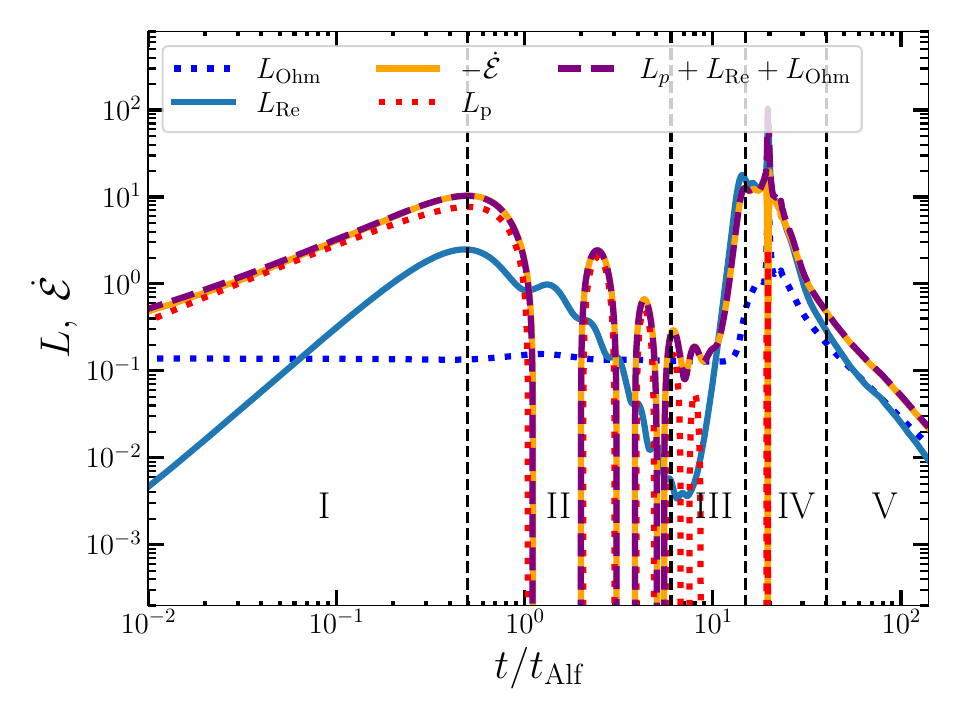}
\end{minipage}
\begin{minipage}{0.49\linewidth}
\includegraphics[width=\columnwidth]{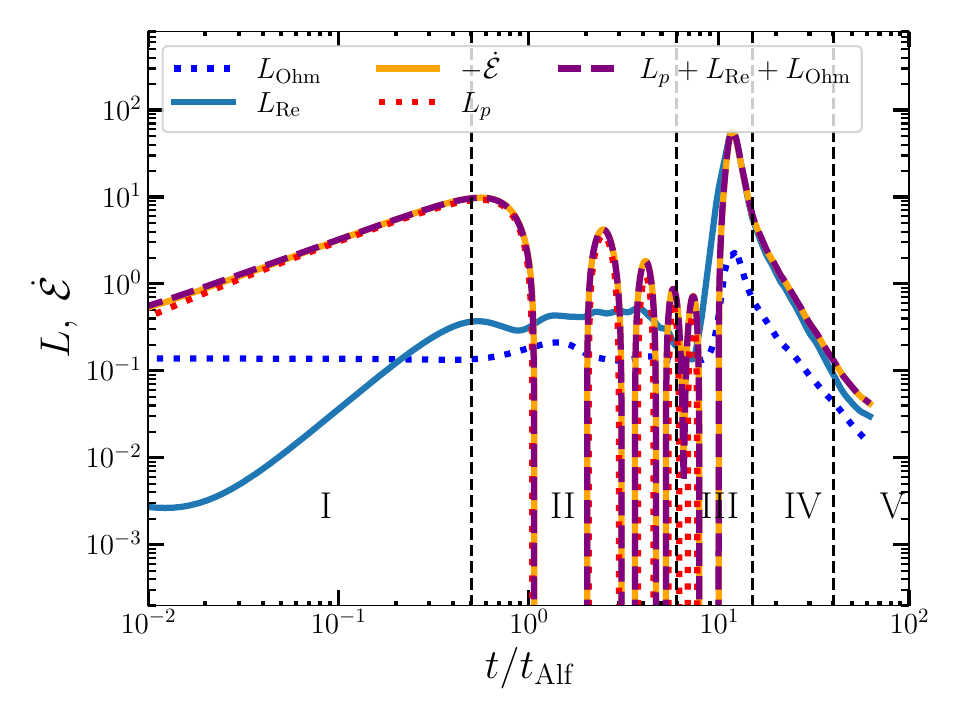}
\end{minipage}
\caption{Evolution of total energy and relative contribution of different decay terms for 1-barotropic-fluid simulations A (left panel) and C (right panel). The coincidence between the orange solid and purple dashed lines indicates well-conserved total energy. Evolution between $1\; t_\mathrm{Alf}$ and $\approx 10\; t_\mathrm{Alf}$ is dominated by Alfv\'en waves.  }
    \label{fig:energy_stable}
\end{figure*}

We show the magnetic field configuration at $38\;t_\mathrm{Alf}$ in Figures~\ref{fig:stable} and \ref{fig:Br_surface}. We see that the field is predominantly small-scale and non-axially symmetric. As measured by the energy, we see that the field does decay in this stage, Figure~\ref{fig:energy_stable}. When we examine the individual components responsible for decay, we see that the decay rate is governed by viscous and resistive losses. It is interesting to note that while magnetic diffusion is not that far away from its realistic value, the viscosity has been significantly amplified. It is unclear then if a more realistic, lower viscosity would change the result at this stage. To test this we perform an additional simulation where we decrease the viscosity by one order of magnitude. We present the result in the following section.

At later times, Ohmic losses become comparable to viscous losses. They cause small-scale magnetic fields to decay first. Thus, the magnetic field configuration simplifies at later stages.

\subsubsection{Role of fluid viscosity}

\begin{figure*}
    \begin{minipage}{0.49\linewidth}
    \includegraphics[width=\columnwidth]{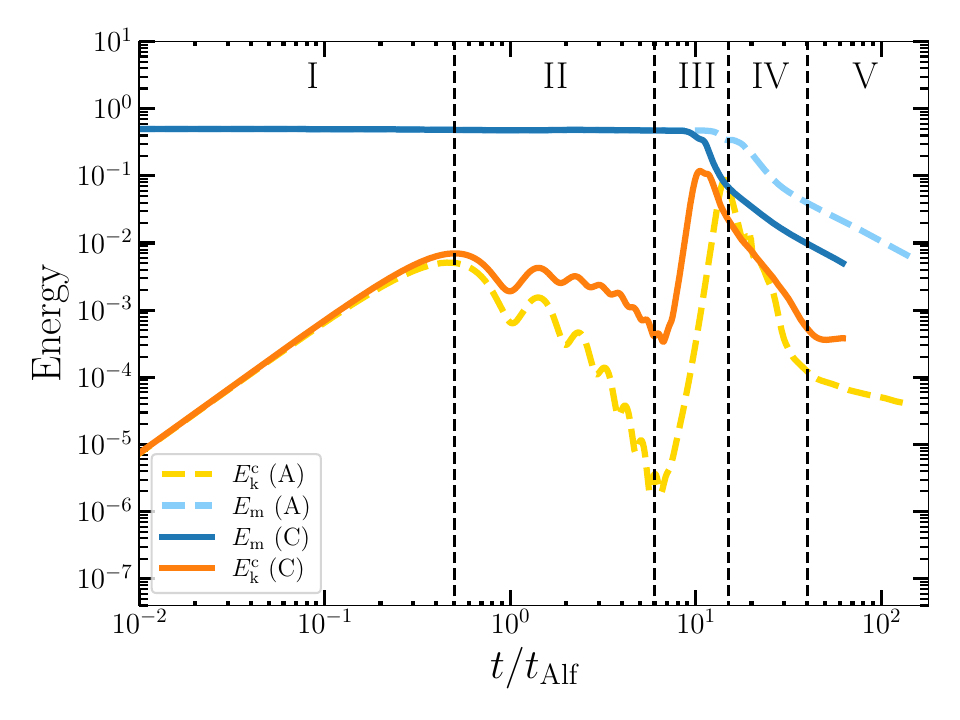}
    \end{minipage}
    \begin{minipage}{0.49\linewidth}
    \includegraphics[width=\columnwidth]{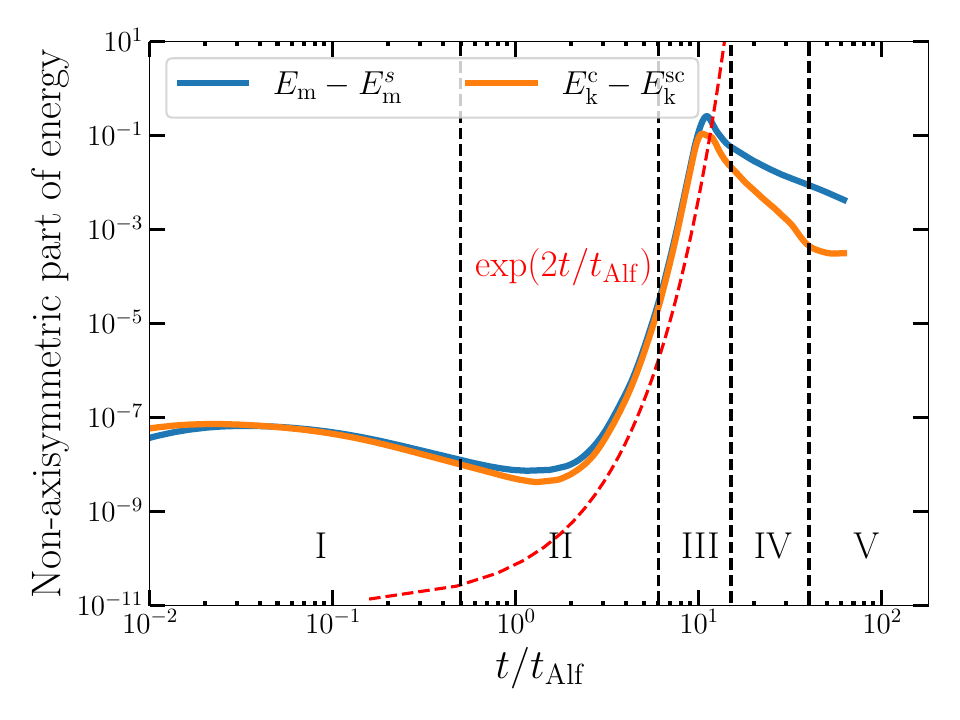}
    \end{minipage}
    \caption{Evolution of kinetic and magnetic energies for 1-barotropic-fluid MHD simulations A and C. Left panel: total magnetic and kinetic energies. Right panel: non-axisymmetric part of magnetic and kinetic energies in simulation C. We mark the following stages: I is the acceleration stage, II is evolution towards 2D force balance, III is  instability, IV is magnetic turbulence, V is resistive decay. The red dashed line in the right panel shows exponential growth with timescale $t_\mathrm{Alf}/2$.  }
    \label{f:energies_barotrop_diff_Re}
\end{figure*}

In order to understand the role of fluid viscosity for our simulations, we run an additional simulation C, where the viscosity is ten times smaller, and comparable to what it is in the two-fluid simulations. To successfully run this we had to increase the radial resolution significantly. Overall, the evolution proceeds through similar five stages to our standard simulation. We show the energy evolution in Figure~\ref{f:energies_barotrop_diff_Re}. 

Three aspects differ in the energy evolution. First, the kinetic energy of the charged fluid does not decay as much during stage II in simulation C, as it did in simulation A. Second, growth of  instability happens slightly earlier in simulation C. Third, the kinetic energy decay is slower at stage V. The first and the last differences are directly related to viscosity. Since the viscosity is much smaller in simulation C, fluid motion takes longer to decay while the system evolves towards the Grad-Shafranov equilibrium. This explains why kinetic energy is much larger at the end of stage II.  Similarly, fluid motion decays more slowly at stage V. Some shift in growth of  instability is probably associated to how noise is added to initial conditions, and depends on exact numerical resolution. 

It is important to note here that the viscous dissipation similarly plays a major role during the turbulence stage of simulation C, i.e. during $10-30\; t_\mathrm{Alf}$, see right panel of Figure~\ref{fig:energy_stable}. This means that the scale of the velocity field becomes smaller, but the total dissipation still reaches comparable values to simulation A. This is expected in the case of turbulence, because energy is transferred down the cascade until it reaches the scale where the viscous dissipation dominates. So, the exact scale where the dissipation occurs is not important and would just constrain the scales of velocities.

\subsubsection{Quality of energy conservation}

\begin{figure}
    \includegraphics[width=\columnwidth]{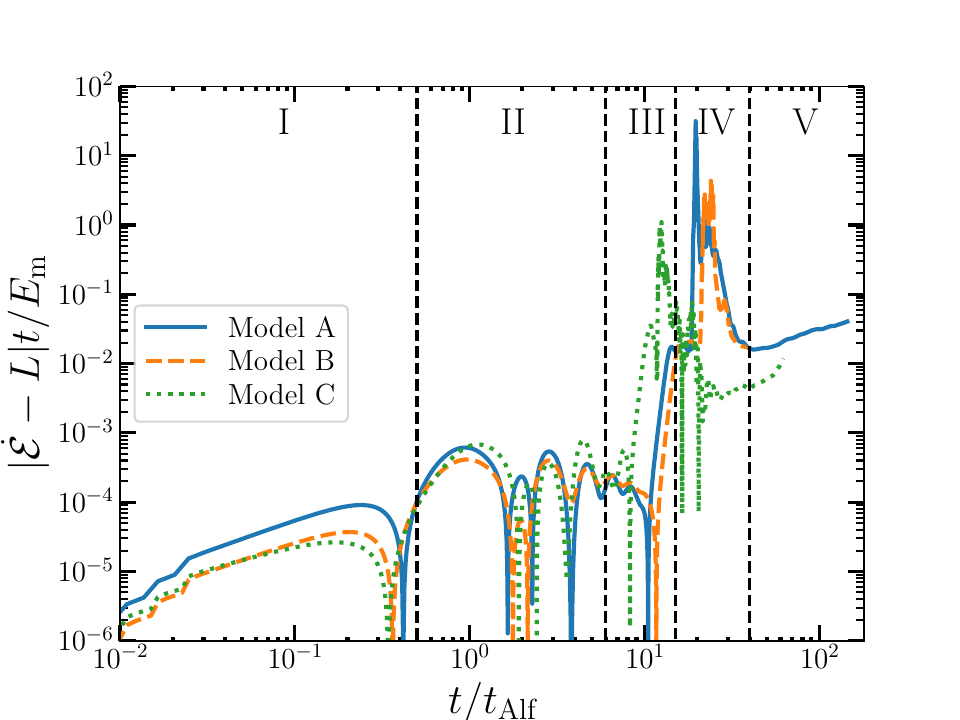}
\caption{Energy conservation in simulations A, B and C. We illustrate it as a difference between change in total energy $\dot{\mathcal{E}}$ and sum of all losses $L = -L_\mathrm{Ohm} - L_\mathrm{Re} - L_\mathrm{p}$. We compare this difference with total magnetic energy $E_\mathrm{m}$. }
\label{f:energy_conservation_quality_AC}
\end{figure}

We show the quality of energy conservation in Figure~\ref{f:energy_conservation_quality_AC}. The quality of energy conservation is typically better than $0.1$~\% except for the transition from instability to turbulence. When the resolution is increased, the quality of the energy conservation is increased as well. From the plot, it seems that increasing the resolution and viscosity would allow to increase the level of energy conservation even during the transition and keep it below 10\% if necessary. In this paper, we run 1-barotropic-fluid simulations to contrast and compare their evolution with the two-fluid model, and not for their own sake. We are therefore not excessively worried by deviation from energy conservation above 10 \%, as long as similar problems do not occur in our basic two-fluid simulations. After the saturation of instability, the errors related to energy conservation become smaller again and in the case of simulation C, these decay below 1 \% level.

\subsection{Two-fluid MHD}
\label{s:res_twofluids}

Now that we better understand how the simple 1-barotropic-fluid MHD system evolves in a spherical domain with our initial conditions, we examine a more complicated system of two-fluid equations. Our basic simulation is D. In order to make the description simpler, we follow the same structure as in the previous section.

\subsubsection{Stages of simulations based on energy evolution}

\begin{figure}
    \includegraphics[width=\columnwidth]{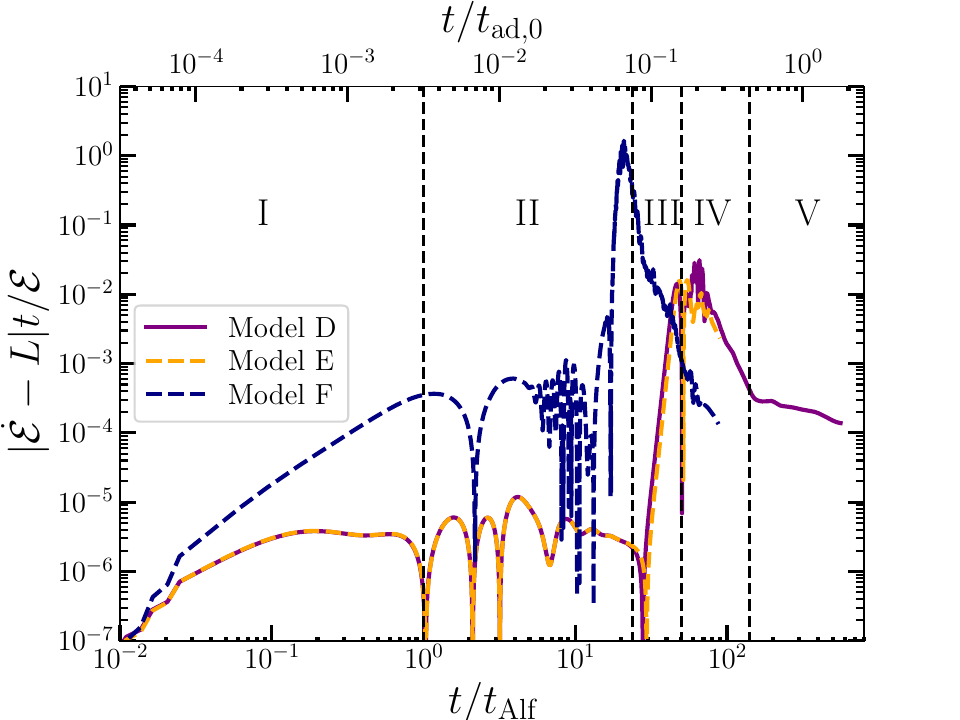}
\caption{Energy conservation in simulations D, E and F. We illustrate it as a difference between change in total energy 
$\dot{\mathcal{E}}$ and sum of all losses $L \equiv - L_\mathrm{ad} - L_\mathrm{p} - L_\mathrm{Re} - L_\mathrm{Ohm}$. We compare this difference with the total energy $\mathcal{E}$. }
\label{f:energy_conservation_quality_H}
\end{figure}

We show the quality of energy conservation in Figure~\ref{f:energy_conservation_quality_H}. We show the difference between magnetic and kinetic energy changes versus the dissipative terms. In order to make it comparable to total energy we multiply fluxes by time i.e. computing $|\dot{\mathcal{E}} - L|t$. Overall, the simulations D and E preserve energy better than $0.05$~\% except for a period of turbulence when the quality is only at the $\approx 2-3$~\% level, which is probably related to formation of small-scale structures. The model F, which includes toroidal magnetic field, conserves energy significantly worse, possibly because much larger velocities develop in this model, see also Section~\ref{s:toroidal}.

\begin{figure*}
    \begin{minipage}{0.49\linewidth}
    \includegraphics[width=\columnwidth]{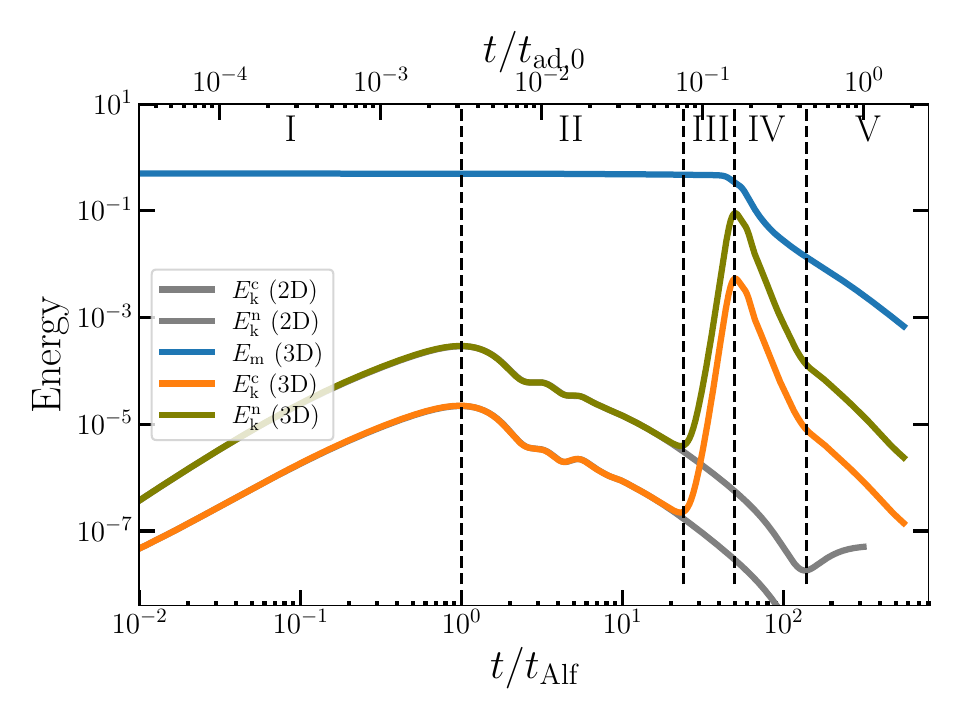}
    \end{minipage}
    \begin{minipage}{0.49\linewidth}
    \includegraphics[width=\columnwidth]{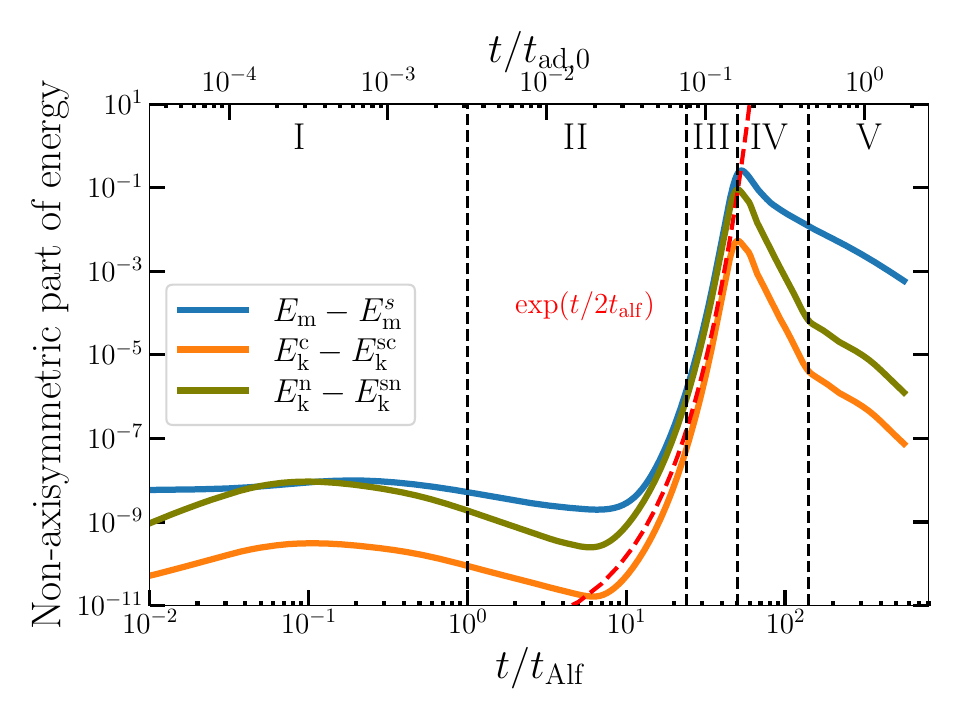}
    \end{minipage}
    \caption{Evolution of kinetic and magnetic energies for the two-fluid MHD simulation D. Left panel: total magnetic and kinetic energies. Right panel: non-axisymmetric part of magnetic and kinetic energies. We mark the following stages: I is the acceleration stage, II is evolution towards 2D force balance, III is  instability, IV is magnetic turbulence, V is resistive decay. The red dashed line in the right panel shows exponential growth with timescale $2t_\mathrm{Alf}$. We show in grey the results of axially symmetric simulations. }
    \label{f:energies_H}
\end{figure*}

Comparing Figures~\ref{f:energies_barotrop} and \ref{f:energies_H} we immediately notice that both simulations go through very similar stages. This behavior is not entirely surprising, because both fluids in a two-fluid simulation are coupled. 
It is worth noting that the two-fluid simulation is significantly slower. The acceleration stage takes at least twice the time (in units of $t_\mathrm{Alf}$ computed for charged particles only) 
that it took in 1-barotropic-fluid simulations. The growth of  instability occurs later in the simulation.

The extent of the individual stages is as follows: (I) the acceleration stage continues until $\approx 1\; t_\mathrm{Alf}$, (II) evolution towards 2D force balance until $\approx 25\; t_\mathrm{Alf}$, (III) the instability develops until $\approx 50\; t_\mathrm{Alf}$, (IV) magnetic turbulence until $\approx 150\; t_\mathrm{Alf}$.

Below we go through all the stages again and highlight the difference between the simulations.

\subsubsection{Acceleration phase}

\begin{figure*}
    \begin{minipage}{0.245\linewidth}
    \includegraphics[width=\columnwidth]{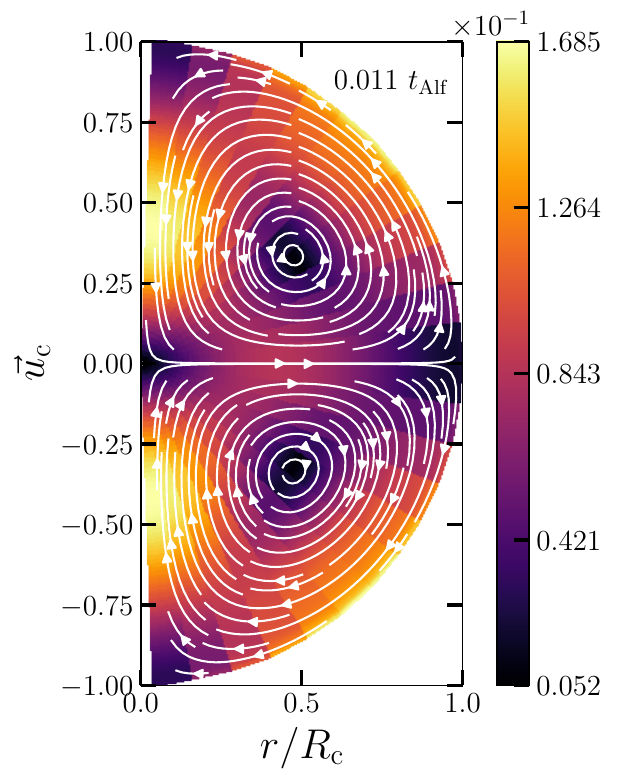}
    \end{minipage}
    \begin{minipage}{0.245\linewidth}
    \includegraphics[width=\columnwidth]{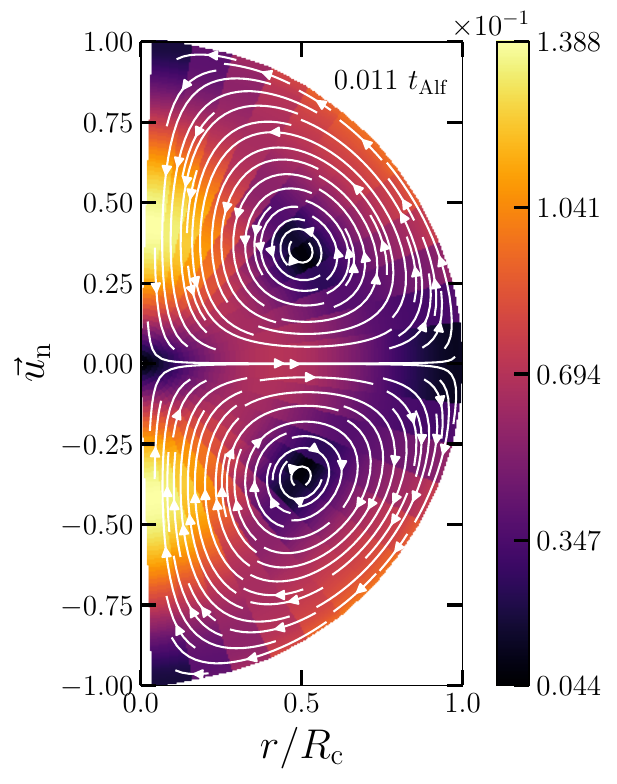}
    \end{minipage}
    \begin{minipage}{0.245\linewidth}
    \includegraphics[width=\columnwidth]{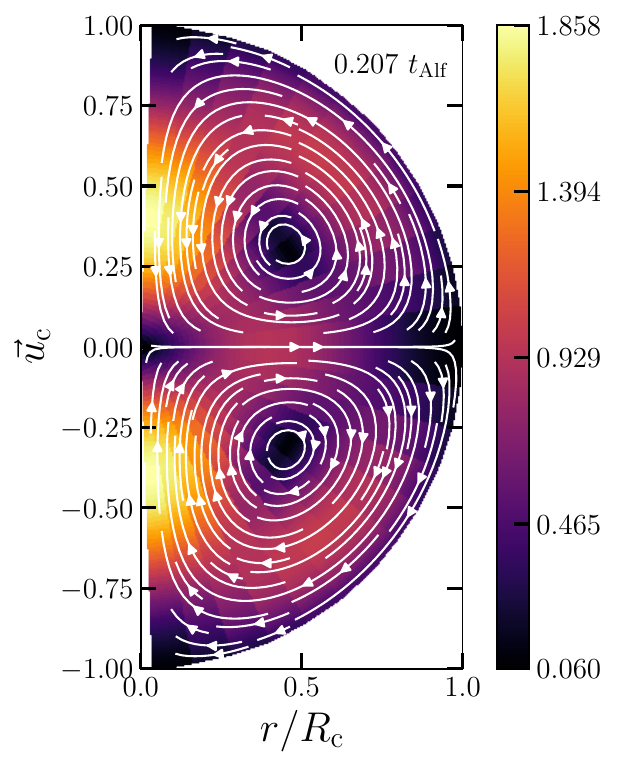}
    \end{minipage}
    \begin{minipage}{0.245\linewidth}
    \includegraphics[width=\columnwidth]{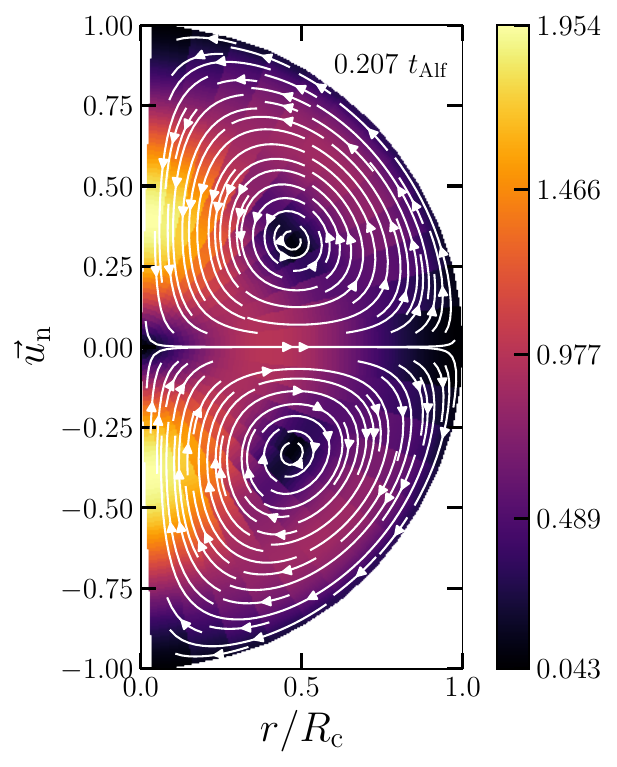}
    \end{minipage}
    \caption{Meridional cuts for charged particle and neutron velocity during the acceleration stage for two-fluid simulation H. }
    \label{f:acceleration_H}
\end{figure*}

There are a few small differences at the acceleration stage. We show the velocities during this acceleration stage in Figure~\ref{f:acceleration_H}.
The fluid velocities are significantly lower than in the case of 1-barotropic-fluid simulations, see Figure~\ref{f:rms_forces_K}. In the former case the charged particle velocity reached 3.9 and 27.4 by $0.01\; t_\mathrm{Alf}$ and $0.1\; t_\mathrm{Alf}$ respectively, while in the two-fluid case the velocities are 0.23 and 1.24 respectively. During the acceleration stage, the magnetic energy is first transferred to charged particles and then to neutrons. The number density of neutrons (and thus their mass density) is nearly an order of magnitude larger than the number density of charged particles, see Figure~\ref{fig:radial_profile}, explaining why the respective velocities are significantly smaller in the two-fluid case. 
The structure of the flow is very similar.

\begin{figure}[h]
\includegraphics[width=\columnwidth]{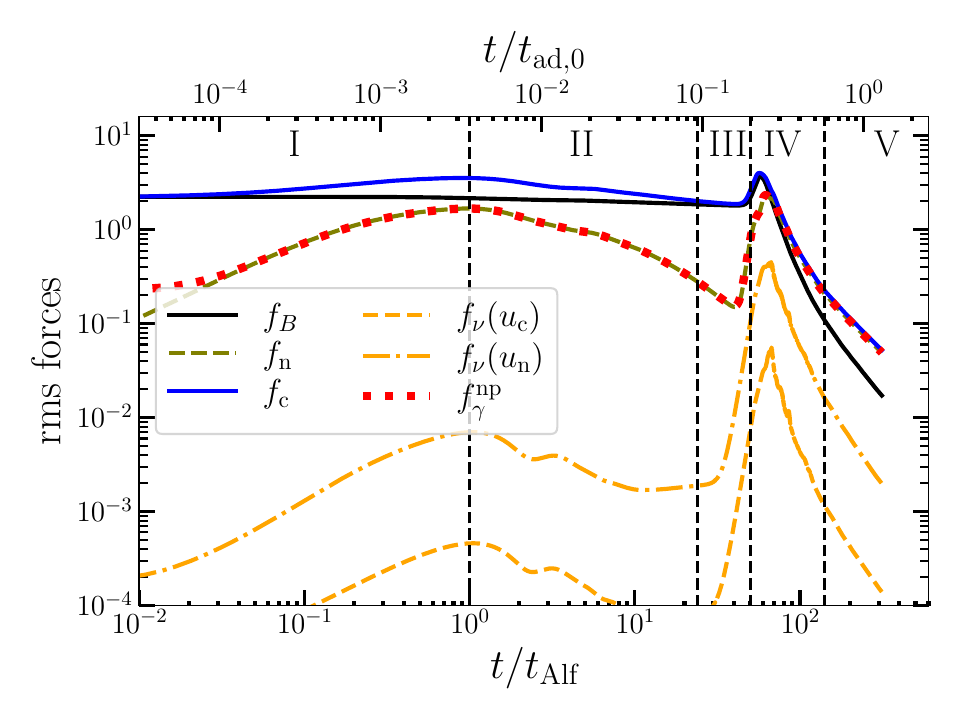}
\caption{Time evolution of root-mean square of forces in two-fluid simulation D. $f_\gamma^\mathrm{np}$ shows the rms friction force.   }
    \label{fig:rms_forces_H}
\end{figure}

\subsubsection{Evolution towards 2D force balance}

\begin{figure*}[h]
    \begin{minipage}{0.245\linewidth}
    \includegraphics[width=\columnwidth]{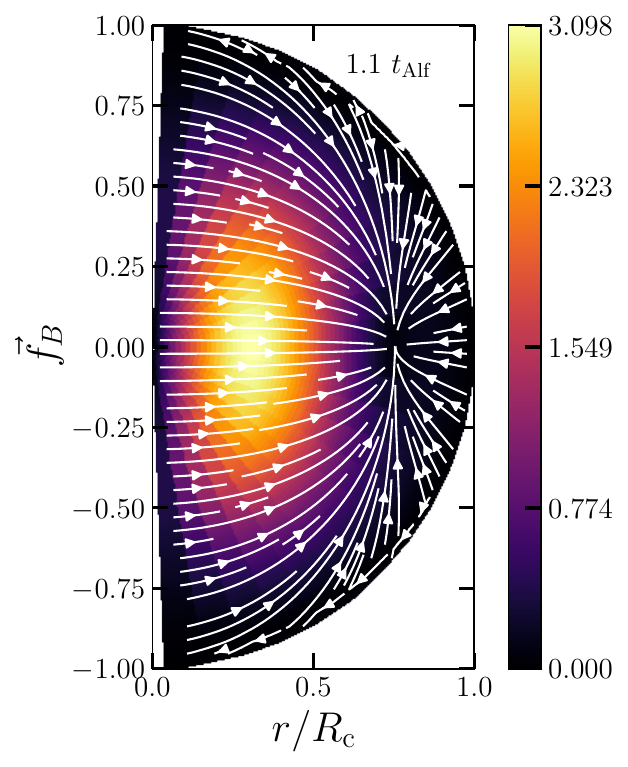}
    \end{minipage}
    \begin{minipage}{0.245\linewidth}
    \includegraphics[width=\columnwidth]{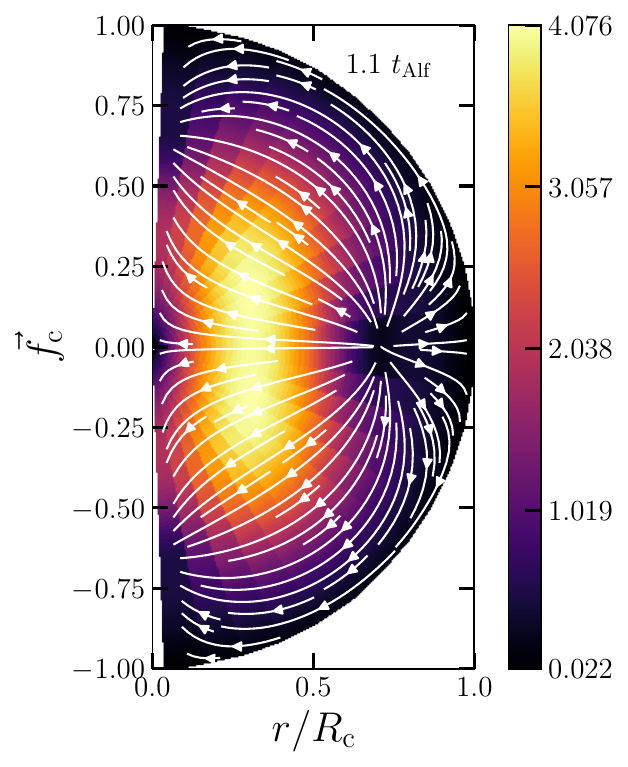}
    \end{minipage}
    \begin{minipage}{0.245\linewidth}
    \includegraphics[width=\columnwidth]{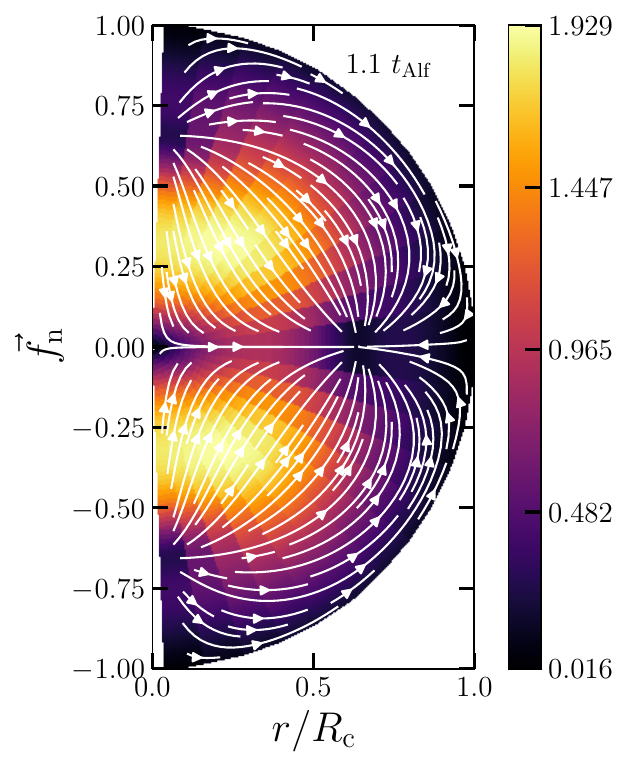}
    \end{minipage}
    \begin{minipage}{0.245\linewidth}
    \includegraphics[width=\columnwidth]{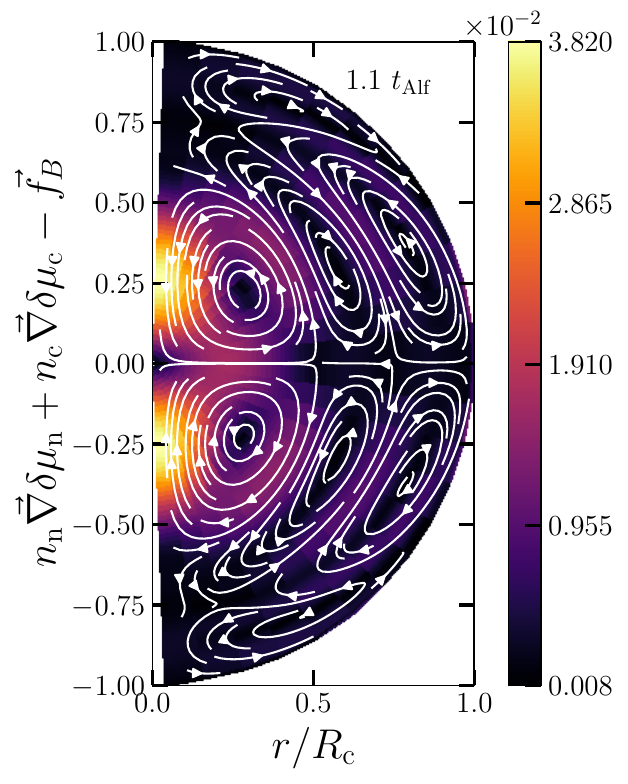}
    \end{minipage}
    \begin{minipage}{0.245\linewidth}
    \includegraphics[width=\columnwidth]{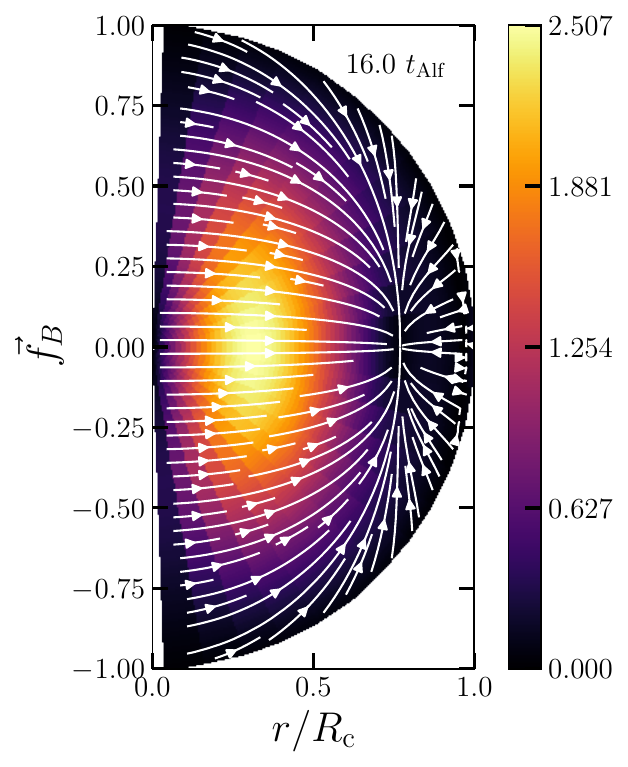}
    \end{minipage}
    \begin{minipage}{0.245\linewidth}
    \includegraphics[width=\columnwidth]{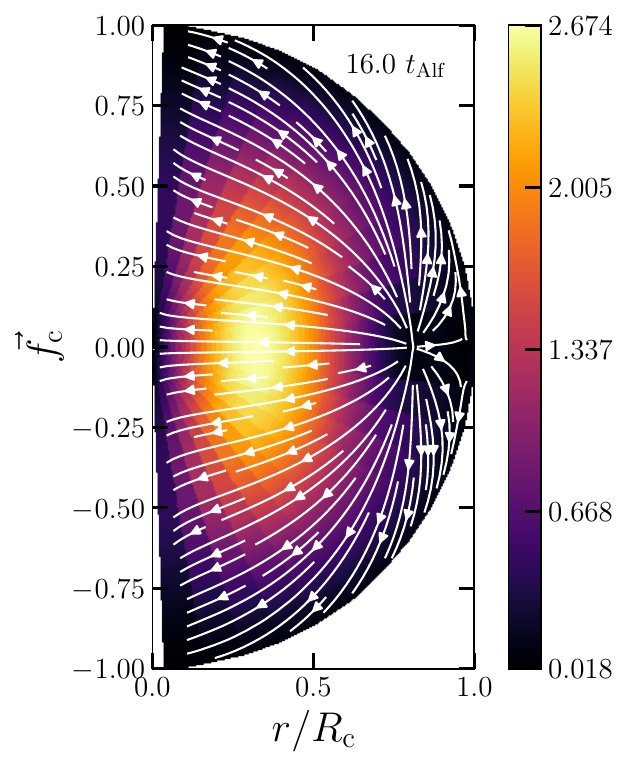}
    \end{minipage}
    \begin{minipage}{0.245\linewidth}
    \includegraphics[width=\columnwidth]{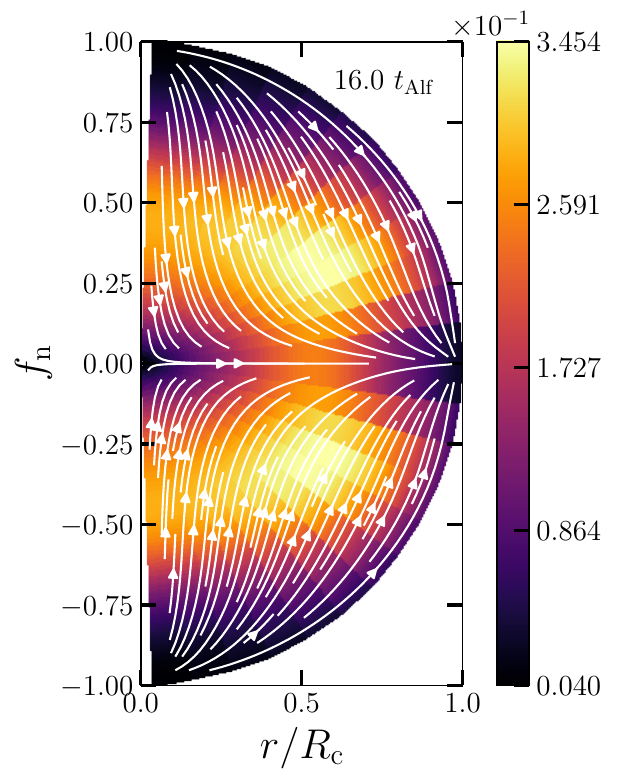}
    \end{minipage}
    \begin{minipage}{0.245\linewidth}
    \includegraphics[width=\columnwidth]{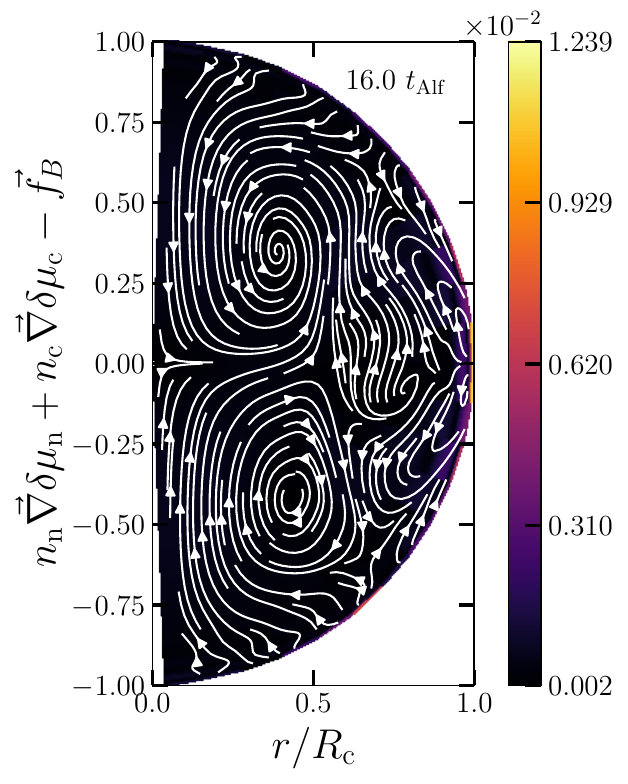}
    \end{minipage}
    \caption{Individual forces and their contribution to the force balance for two-fluid simulation D. We show the Lorentz force (first panel from the left), force of charged particles $\vec f_\mathrm{c} =-n_\mathrm{c} \vec \nabla \delta \mu_\mathrm{c}$ (second panel), force of neutrons $\vec f_\mathrm{n} =-n_\mathrm{n} \vec \nabla \delta \mu_\mathrm{n}$ (third panel), and finally the total force balance $\vec f_\mathrm{n} + \vec f_\mathrm{c} + \vec f_B$.}
    \label{f:df_H}
\end{figure*}

As we explained before, the force equilibrium in the two-fluid case includes chemical potential perturbations for both charged particles and neutrons.
We show the relative contribution of individual components via their rms values in Figure~\ref{fig:rms_forces_H} as well as with individual snapshots in Figure~\ref{f:df_H}. Stage II begins with 3 forces of the same order of magnitude and ends with $\vec f_\mathrm{c} \approx \vec f_B $, while $\vec f_\mathrm{n}$ becomes significantly smaller.

It is interesting to note that the Lorentz force is balanced better in two-fluid simulations at the beginning of stage II in comparison to stage II of the 1-barotropic-fluid simulations. Thus, a similar force balance is only reached by $\approx 6\; t_\mathrm{Alf}$ in simulation A. The system does indeed evolve in the direction of force balance where the viscous forces play a much smaller role by the end of stage II, as seen in Figure~\ref{fig:rms_forces_H}.


It is worth noting, however, that this force balance differs significantly from the one established at the end of stage I.
At that point, the Lorentz force was balanced by two pressure contributions, arising from the charged and neutral fluid components. While the neutral force was approximately half the magnitude of the charged-particle force at the end of stage I, by the end of stage II it had decreased to about one tenth of it. This suggests that the system evolves towards the Grad-Shafranov equilibrium, characterized by zero relative --ambipolar-- velocity, which in equilibrium satisfies $\vec{u}_c -\vec{u}_n \propto \grad\delta\mu_n$ (see equation~\ref{eq:un}). Hence, at this stage, neutrons reach diffusive equilibrium 
\begin{equation}
    \vec{\nabla} \delta \mu_n \approx 0,
\end{equation}
and the poloidal component of the Lorentz force is balanced solely by the pressure force arising from the charged particles 
\begin{equation}
 \vec{f}^{\mathrm{\,pol}}_B \approx - n_\mathrm{c} \vec \nabla \delta \mu_\mathrm{c}.  
\end{equation}
Thus, our results are consistent with those reported by \cite{Castillo2020MNRAS,castillo2025AA,moraga2025magnetothermal}, who found the same equilibrium.
It is worth noting that, as the system approaches this equilibrium, the charged fluid component becomes increasingly more barotropic and can be expected to behave similarly to the 1-barotropic-fluid.


\subsubsection{ Instability}

\begin{figure*}
    \begin{minipage}{0.49\linewidth}
    \includegraphics[width=\columnwidth]{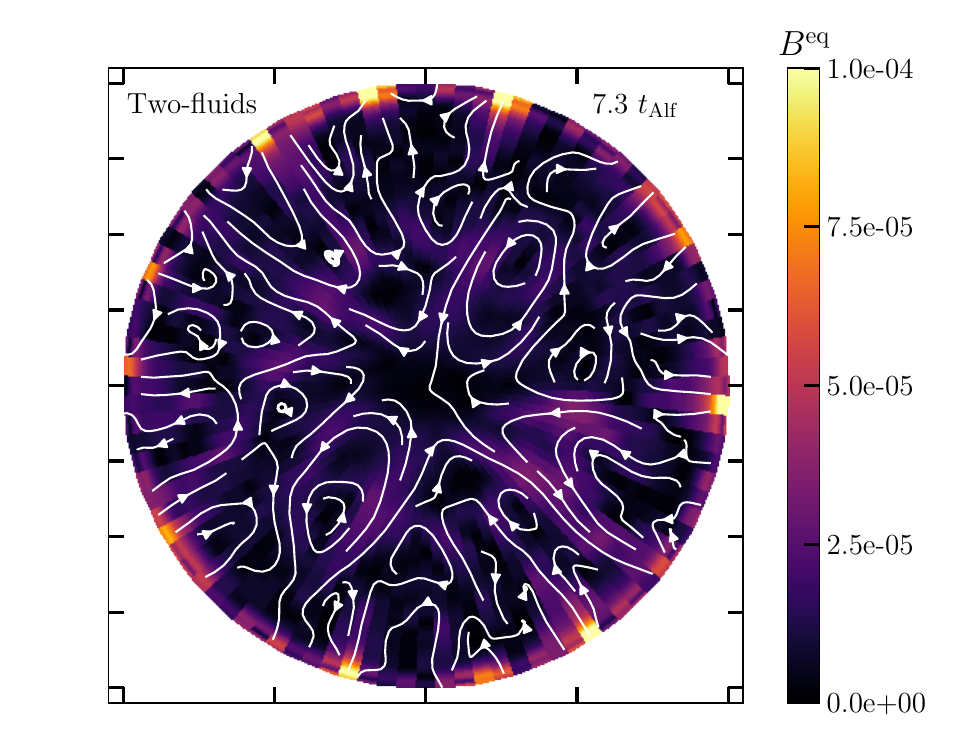}
    \end{minipage}
    \begin{minipage}{0.49\linewidth}
    \includegraphics[width=\columnwidth]{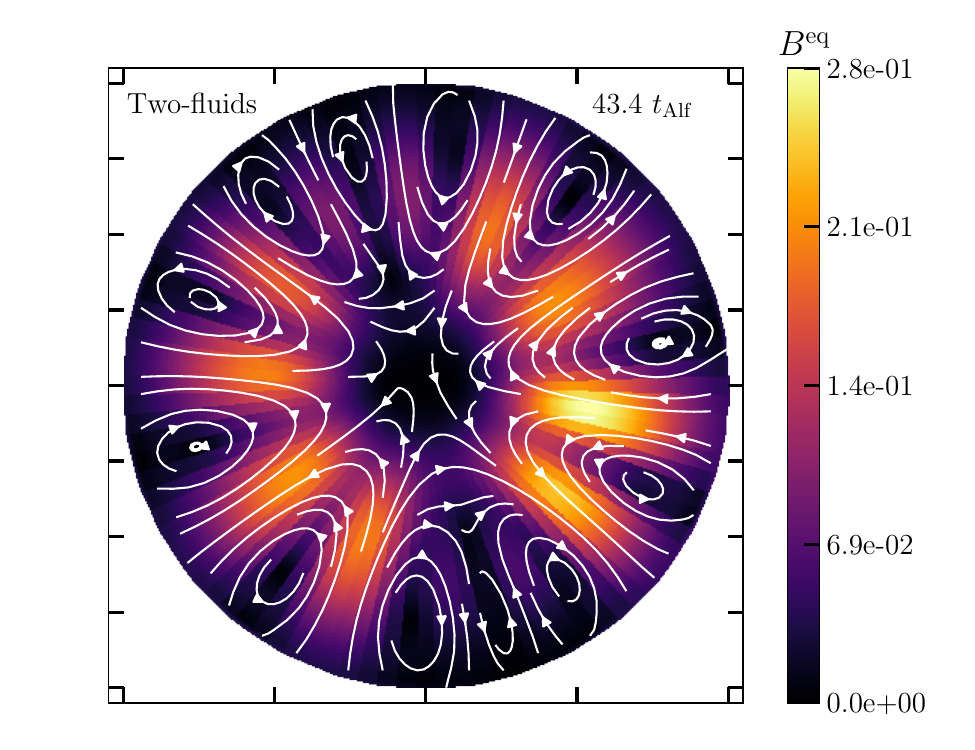}
    \end{minipage}
    \caption{Equatorial cut for two-fluid simulation. We show magnetic field in the plane. }
    \label{f:B_eq_H}
\end{figure*}

Similarly to 1-barotropic-fluid simulations, we see a development of instability in our two-fluid simulations. We show the equatorial cuts for magnetic field in Figure~\ref{f:B_eq_H}. We show these slices at different times in comparison to the 1-barotropic-fluid case because the development of instability takes longer. 

Nevertheless, the development of this instability is analogous to the 1-barotropic-fluid case. The non-axisymmetric part of $B_\phi$ is initially represented by noise at a level $10^{-4}\; B_0$. A new highly regular structure emerges from this noise and grows exponentially on a longer timescale, reaching values comparable to the initial $B_r$ and $B_\theta$ components. We can fit the growth rate with a $2t_\mathrm{Alf}$ timescale.

To check whether the instability only appears in three dimensions, we also run shorter axially symmetric simulations. Their results are shown in grey in Figure~\ref{f:energies_H}. In these simulations both kinetic energies of charged and neutral fluids decay on much longer timescales than required for instability to develop. 


\begin{figure}
    \begin{minipage}{0.49\linewidth}
    \includegraphics[width=\columnwidth]{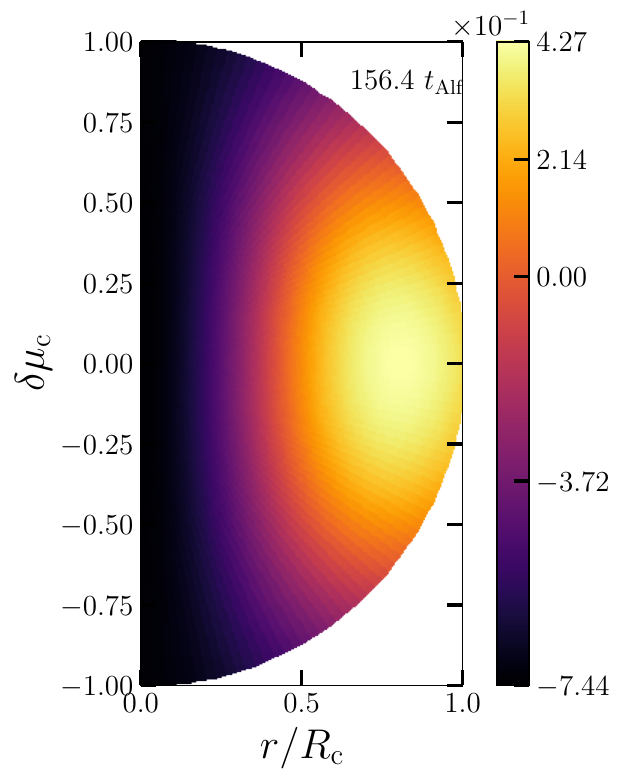}
    \end{minipage}
    \begin{minipage}{0.49\linewidth}
    \includegraphics[width=\columnwidth]{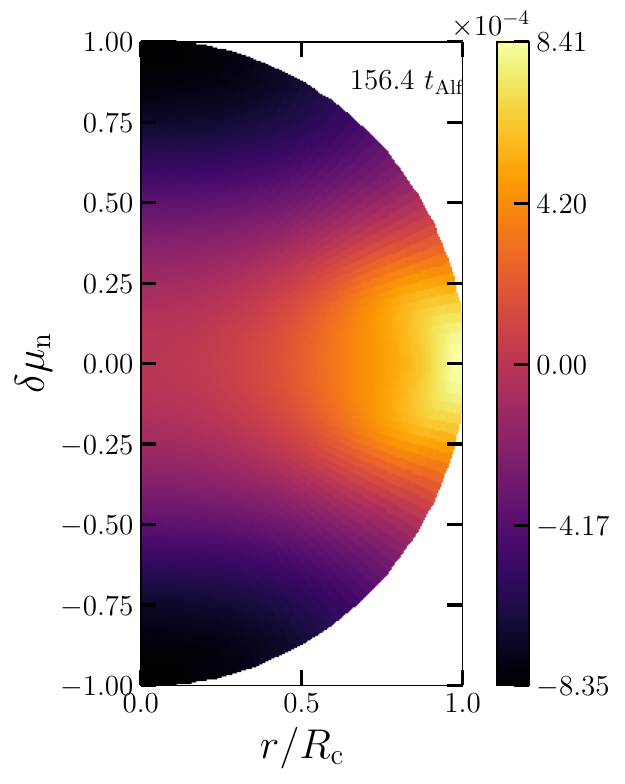}
    \end{minipage}
    \begin{minipage}{0.49\linewidth}
    \includegraphics[width=\columnwidth]{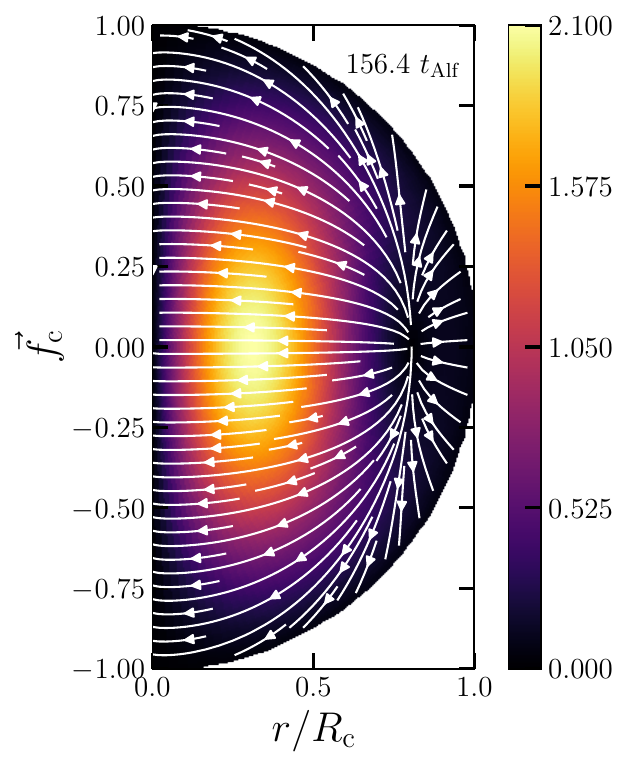}
    \end{minipage}
    \begin{minipage}{0.49\linewidth}
    \includegraphics[width=\columnwidth]{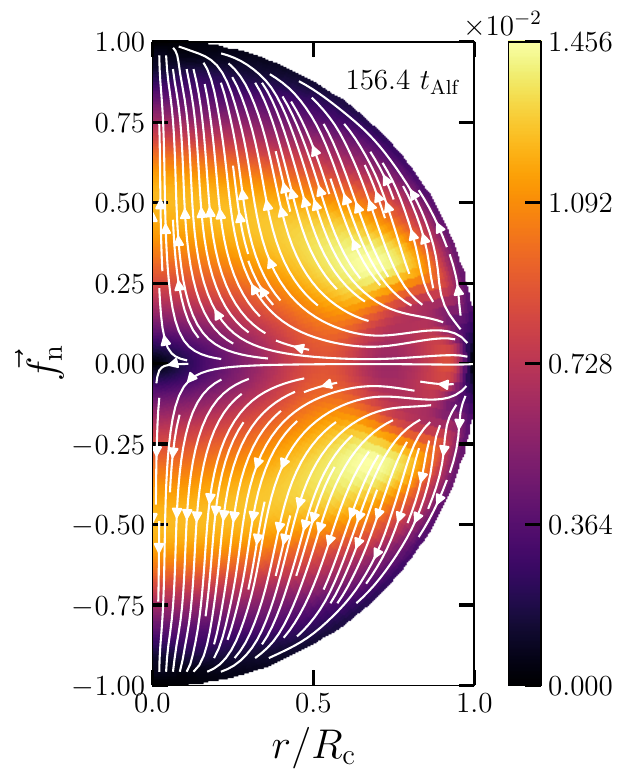}
    \end{minipage}
    \caption{Results of two-fluid axially symmetric simulation at $\approx 156\; t_\mathrm{Alf}$.}
    \label{f:axial_symmetry_H}
\end{figure}

We also examine the magnetic energy spectra at different times in Figure~\ref{f:magnetic_energy_H}. In comparison with simple 1-barotropic-fluid simulations, we see many more modes growing at $25\;t_\mathrm{Alf}$, including $m=4$, $m=6$ and $m=8$. The two modes $m=4$ and $m=6$ are the largest, with roughly similar amplitudes, reflecting the complicated field structures visible in Figure~\ref{f:B_eq_H}.

This instability can be also characterized via analysis of the velocity field. We show streamlines for $\vec u_\mathrm{c}$ in Figure~\ref{fig:uc_un_structure}. The streamlines for $\vec u_\mathrm{n}$ are very similar due to the small difference between these velocities. Around $51\;t_\mathrm{Alf}$ when the instability is saturated we see the velocity field is composed of individual circulations on spherical surfaces. This can also be seen in Figure~\ref{fig:Br_surface_H}. 

\begin{figure}
\plotone{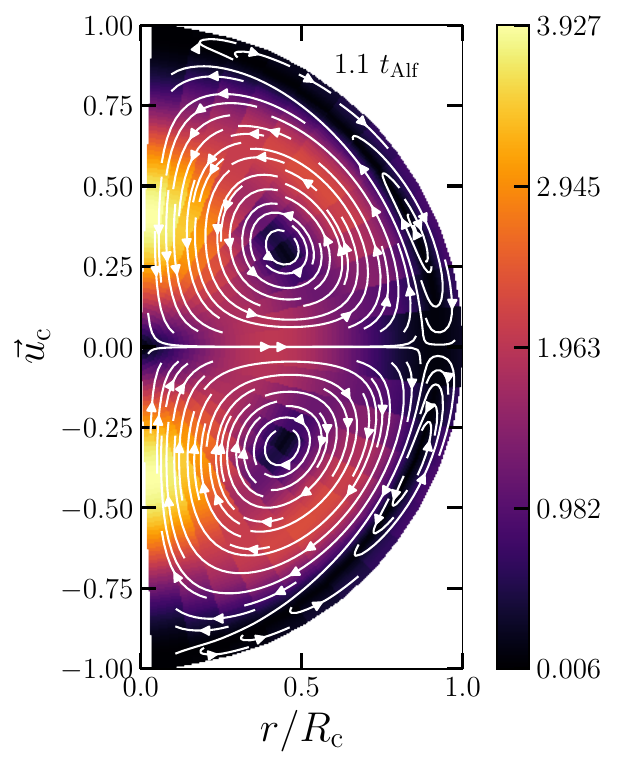}
\caption{Evolution of fluid velocities during the key stages of simulation D. First image shows charged fluid velocity at the end of stage I. The complete figure set (15 images) is available in the online journal.}
\label{f:vel_H}
\end{figure}

\begin{figure}
    \begin{minipage}{0.99\linewidth}
    \includegraphics[width=\columnwidth]{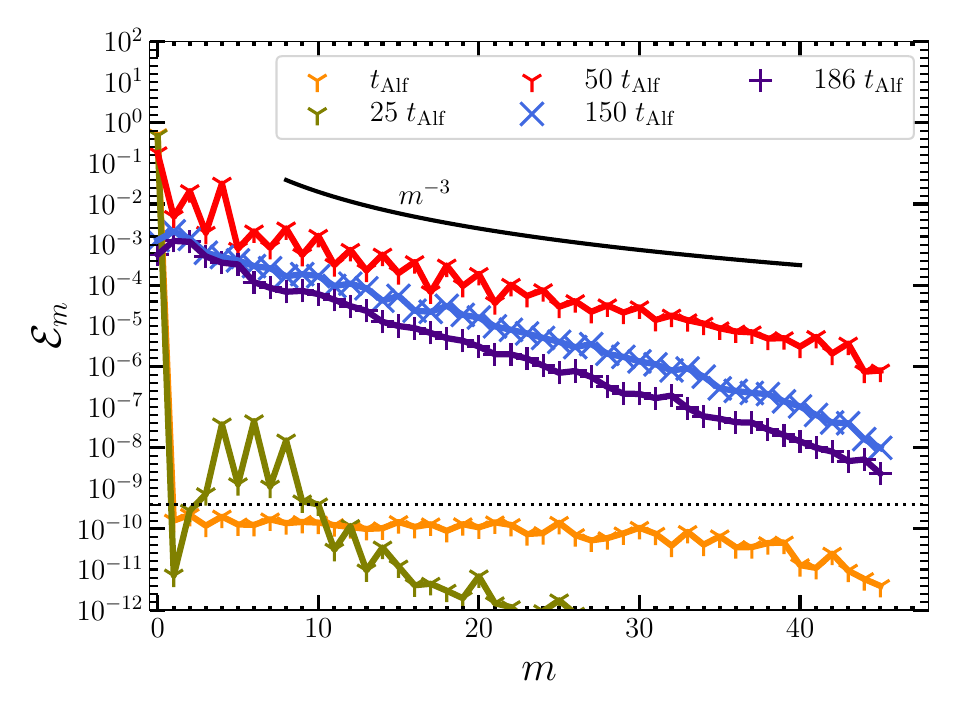}
    \end{minipage}
    \begin{minipage}{0.99\linewidth}
    \includegraphics[width=\columnwidth]{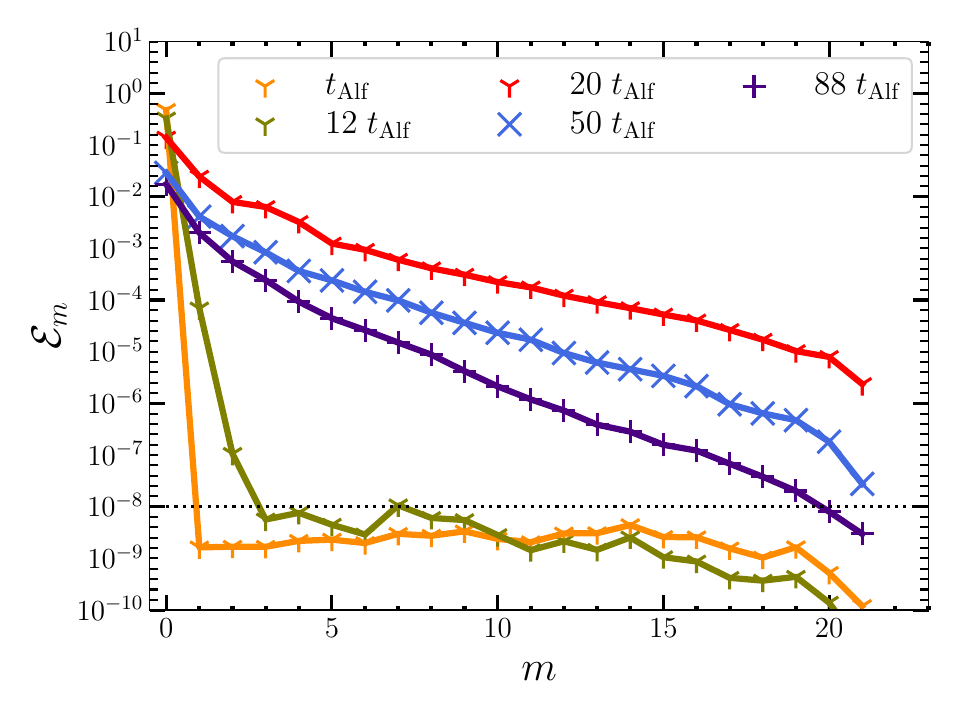}
    \end{minipage}
    \caption{Magnetic energy spectra at different stages of simulation for the two-fluid simulation. Top panel: simulation D; lower panel: simulation F.}
    \label{f:magnetic_energy_H}
\end{figure}

\begin{figure*}
    \begin{minipage}{0.32\linewidth}
    \includegraphics[width=\columnwidth]{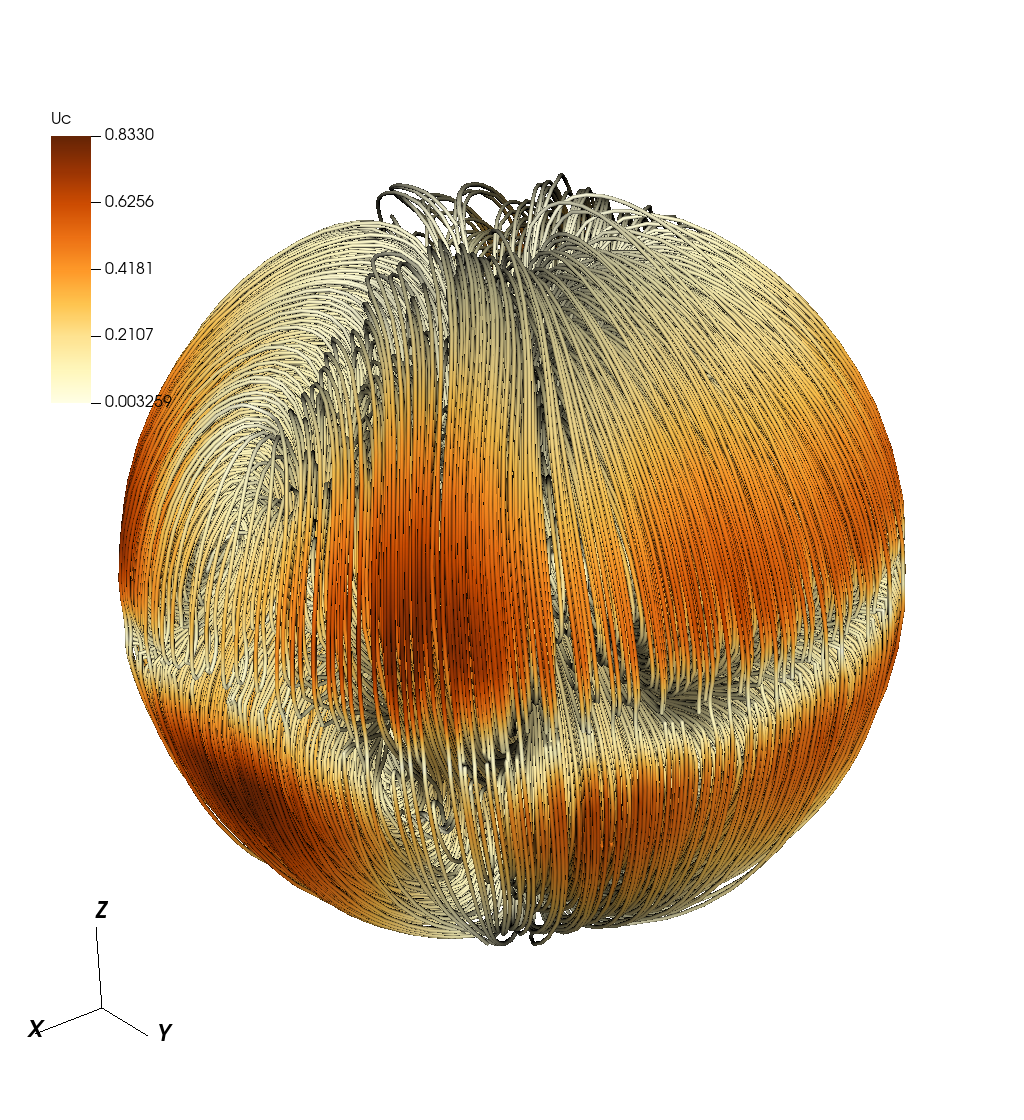}
    \end{minipage}
    \begin{minipage}{0.32\linewidth}
    \includegraphics[width=\columnwidth]{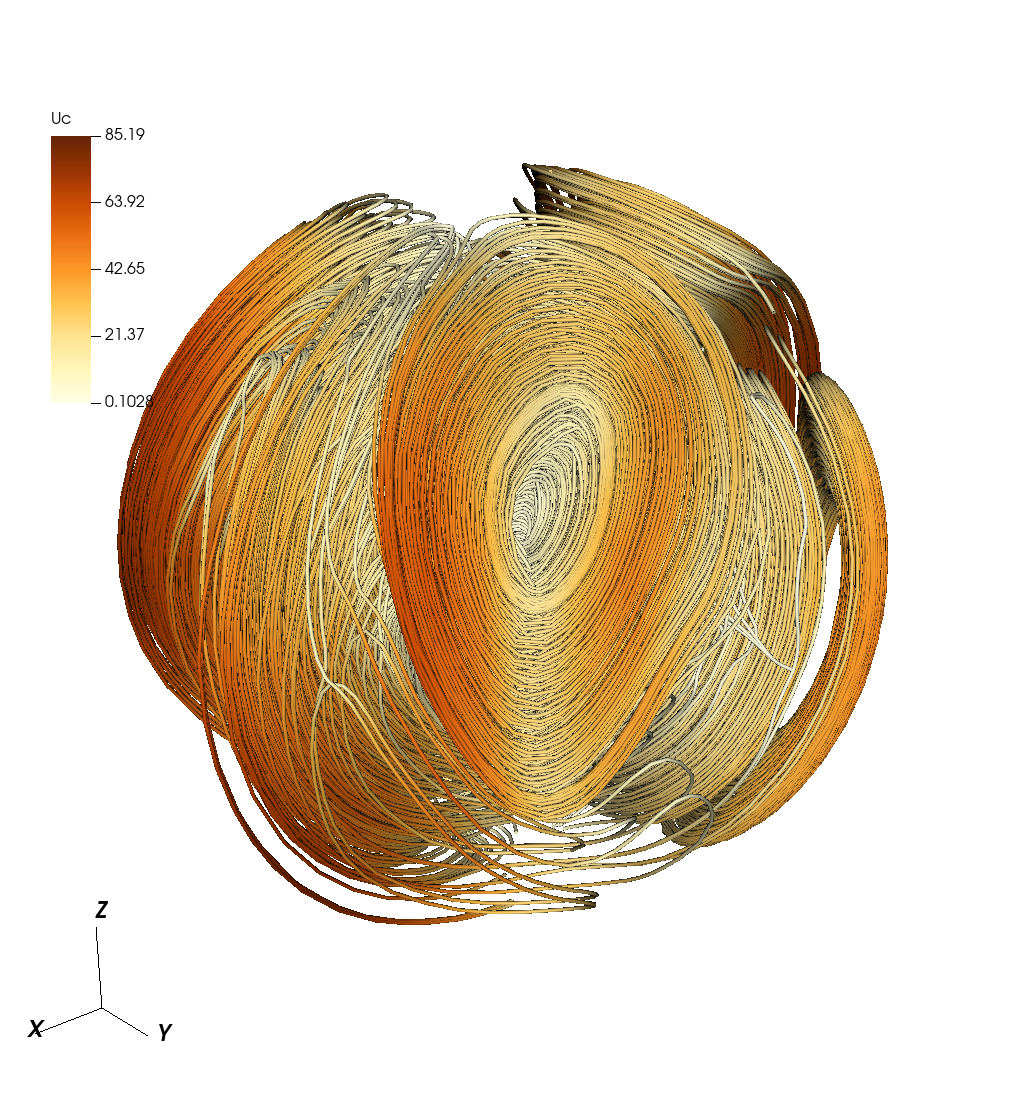}
    \end{minipage}
    \begin{minipage}{0.32\linewidth}
    \includegraphics[width=\columnwidth]{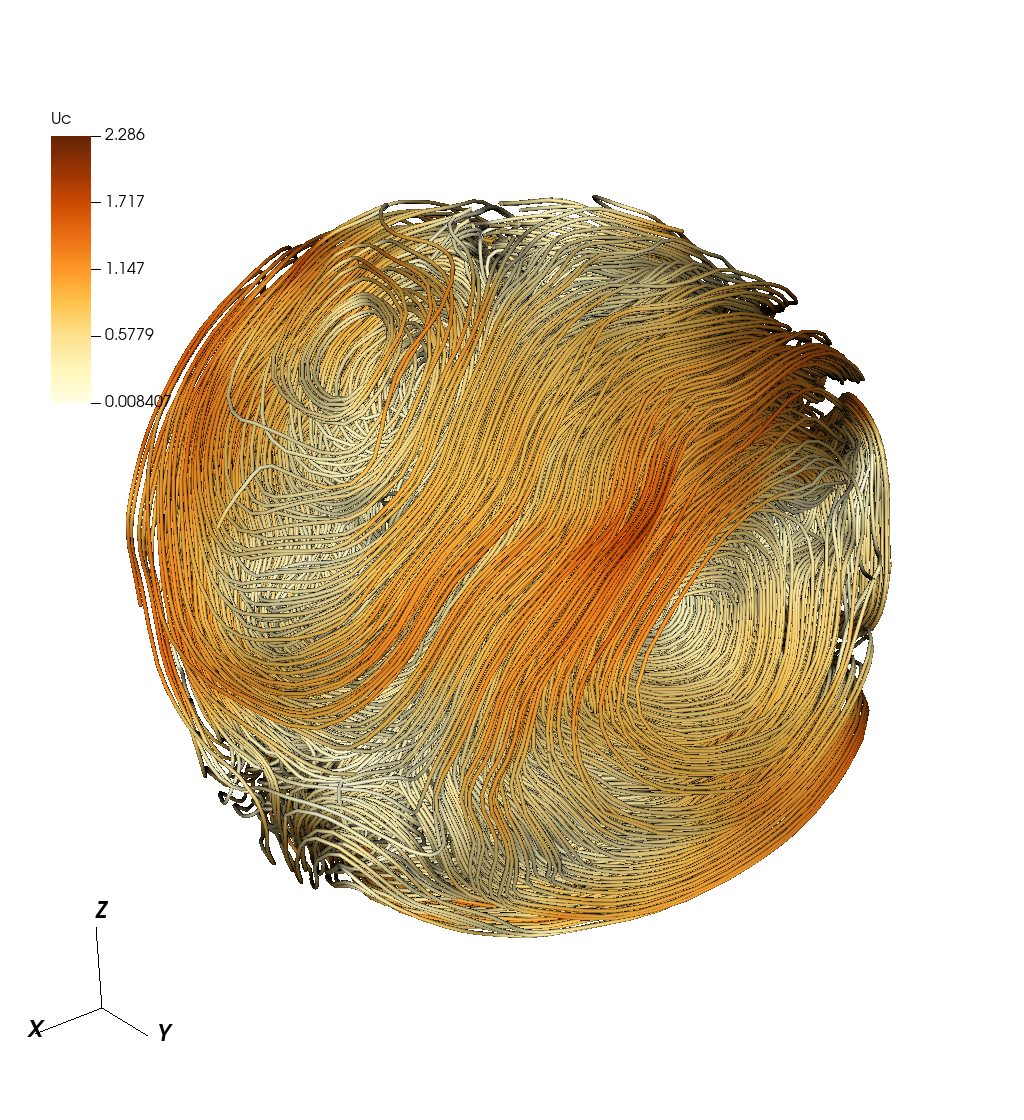}
    \end{minipage}
    \caption{Streamlines for $\vec u_\mathrm{c}$ for key moments of simulation D. The first panel shows $24.7\; t_\mathrm{Alf}$, the second $50.9\; t_\mathrm{Alf}$, and the third $186.8\; t_\mathrm{Alf}$. A video is also available at \url{https://doi.org/10.5281/zenodo.17552305}} 
    \label{fig:uc_un_structure}
\end{figure*}

The fluid velocities have a small radial component, repeating the same circulation structure at all radial distances.
We also show the meridional cuts for all velocities in Figure~\ref{f:vel_H}. Again from these figures it is possible to see that both charged and neutral fluid velocities are non-radial. At the end of stage I, the mean radial velocity $\langle|u_r|\rangle = 0.77$ is comparable to the mean orthogonal velocity $\langle u_\perp\rangle = \langle\sqrt{u_\theta^2 + u_\phi^2}\rangle = 0.76$. At the end of stage II, the mean absolute radial velocity $\langle|u_r|\rangle = 0.04$ becomes less than half the orthogonal velocity $\langle u_\perp\rangle = 0.1$. At the end of stage III, the mean radial velocity $\langle|u_r|\rangle= 1.5$ is more than an order of magnitude smaller than the orthogonal velocity $\langle u_\perp\rangle = 18.8$. Later on, at the end of stage IV, the mean radial velocity $\langle|u_r|\rangle = 0.13$ is just five times smaller than the orthogonal velocity $\langle u_\perp\rangle = 0.72$.

\figsetstart
\figsetnum{24}
\figsettitle{Evolution of fluid velocities during the key stages of simulation D.}

\figsetgrpstart

\figsetgrpnum{24.1}
\figsetplot{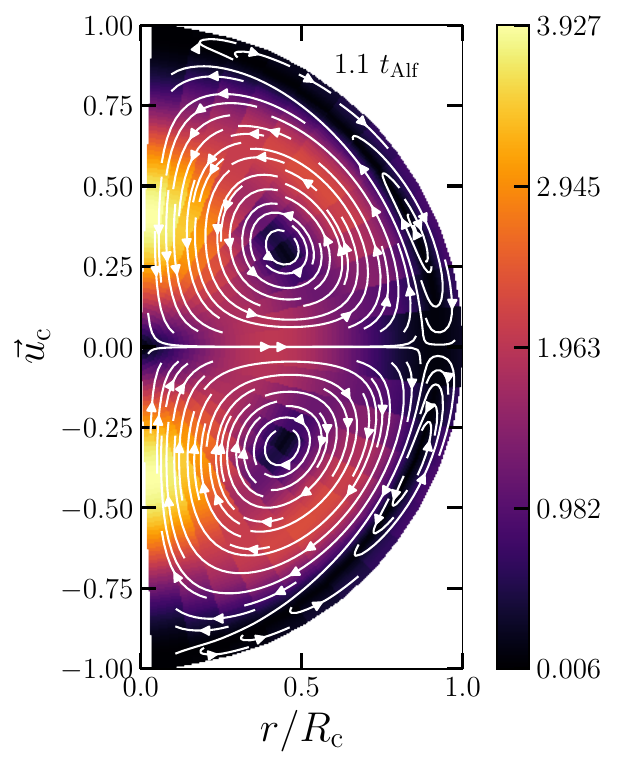}
\figsetgrpnote{Charged fluid velocity at the end of stage I}
\figsetgrpend

\figsetgrpstart
\figsetgrpnum{24.2}
\figsetgrpnote{Neutral fluid velocity at the end of stage I}
\figsetplot{figures/un_H2.pdf}

\figsetgrpend

\figsetgrpstart
\figsetgrpnum{24.3}
\figsetgrpnote{Ambipolar velocity at the end of stage I}
\figsetplot{figures/ad_H2.pdf}
\figsetgrpend

\figsetgrpnum{24.4}
\figsetgrpnote{Charged fluid velocity at the end of stage II}
\figsetplot{figures/uc_H21.pdf}
\figsetgrpend

\figsetgrpstart
\figsetgrpnum{24.5}
\figsetgrpnote{Neutral fluid velocity at the end of stage II}
\figsetplot{figures/un_H21.pdf}
\figsetgrpend

\figsetgrpstart
\figsetgrpnum{24.6}
\figsetgrpnote{Ambipolar velocity at the end of stage II}
\figsetplot{figures/ad_H21.pdf}
\figsetgrpend

\figsetgrpnum{24.7}
\figsetgrpnote{Charged fluid velocity at the end of stage III}
\figsetplot{figures/uc_H42.pdf}
\figsetgrpend

\figsetgrpstart
\figsetgrpnum{24.8}
\figsetgrpnote{Neutral fluid velocity at the end of stage III}
\figsetplot{figures/un_H42.pdf}
\figsetgrpend

\figsetgrpstart
\figsetgrpnum{24.9}
\figsetgrpnote{Ambipolar velocity at the end of stage III}
\figsetplot{figures/ad_H42.pdf}
\figsetgrpend

\figsetgrpnum{24.10}
\figsetgrpnote{Charged fluid velocity at the end of stage IV}
\figsetplot{figures/uc_H94.pdf}
\figsetgrpend

\figsetgrpstart
\figsetgrpnum{24.11}
\figsetgrpnote{Neutral fluid velocity at the end of stage IV}
\figsetplot{figures/un_H94.pdf}
\figsetgrpend

\figsetgrpstart
\figsetgrpnum{24.12}
\figsetgrpnote{Ambipolar velocity at the end of stage IV}
\figsetplot{figures/ad_H94.pdf}
\figsetgrpend

\figsetgrpnum{24.13}
\figsetgrpnote{Charged fluid velocity at the end of simulations}
\figsetplot{figures/uc_H450.pdf}
\figsetgrpend

\figsetgrpstart
\figsetgrpnum{24.14}
\figsetgrpnote{Neutral fluid velocity at the end of simulations}
\figsetplot{figures/un_H450.pdf}
\figsetgrpend

\figsetgrpstart
\figsetgrpnum{24.15}
\figsetgrpnote{Ambipolar velocity at the end of simulations}
\figsetplot{figures/ad_H450.pdf}
\figsetgrpend

\figsetend

\subsubsection{Magnetic Turbulence}

The field structure changes significantly around $50\;t_\mathrm{Alf}$. During stage III, the $B_\phi$ component grows exponentially, but does not affect the other components. At this moment, $B_\phi$ becomes comparable in strength to $B_r$ and $B_\theta$, and starts affecting the dynamics. Similarly, the energy in this component reaches a maximum at $50\; t_\mathrm{Alf}$ (see the right panel of Figure~\ref{f:energies_H}) and decays afterwards. 

The field structure at this stage becomes turbulent, i.e. the energy is no longer described using a limited number of spherical harmonics, but is represented instead with a broad power-law spectrum, see Figure~\ref{f:magnetic_energy_H}. The energy cascade seems to be steeper than $\mathcal {E}_m\propto m^{-3}$. Energy is transferred through this cascade from $m=0$ (initial field) to smaller $m$. 

\begin{figure*}
    \begin{minipage}{0.49\linewidth}
    \includegraphics[width=\columnwidth]{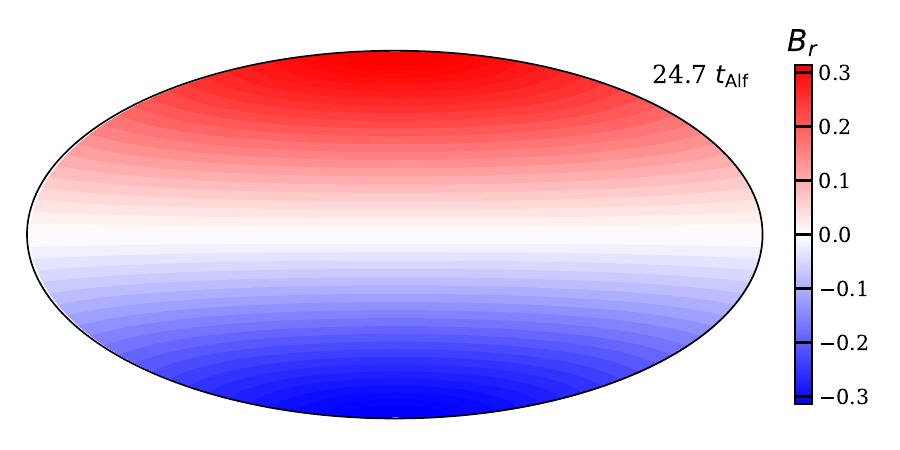}
    \end{minipage}
    \begin{minipage}{0.49\linewidth}
    \includegraphics[width=\columnwidth]{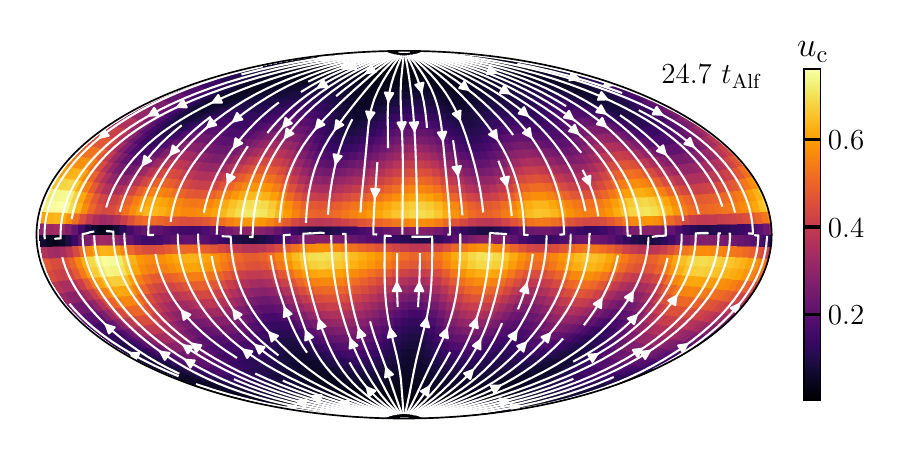}
    \end{minipage}
    \begin{minipage}{0.49\linewidth}
    \includegraphics[width=\columnwidth]{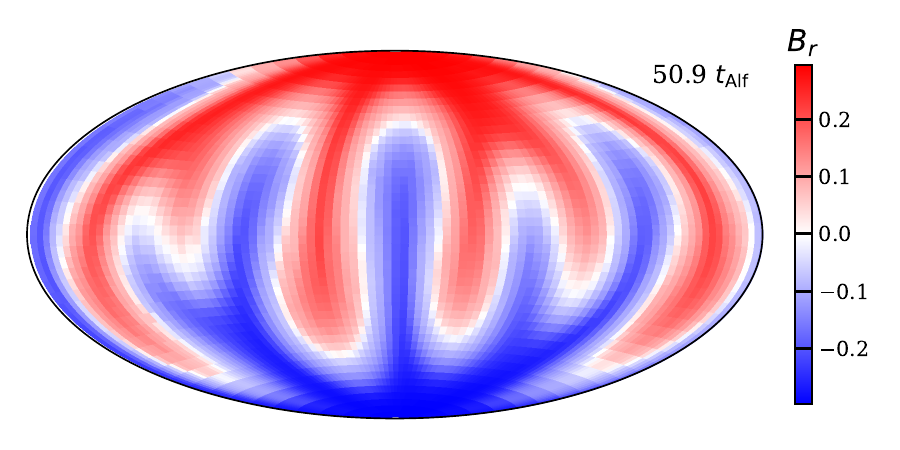}
    \end{minipage}
    \begin{minipage}{0.49\linewidth}
    \includegraphics[width=\columnwidth]{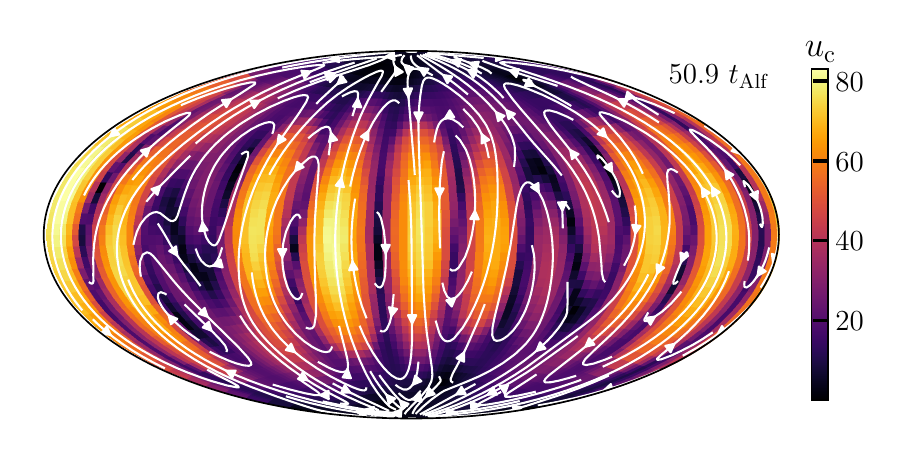}
    \end{minipage}
    \begin{minipage}{0.49\linewidth}
    \includegraphics[width=\columnwidth]{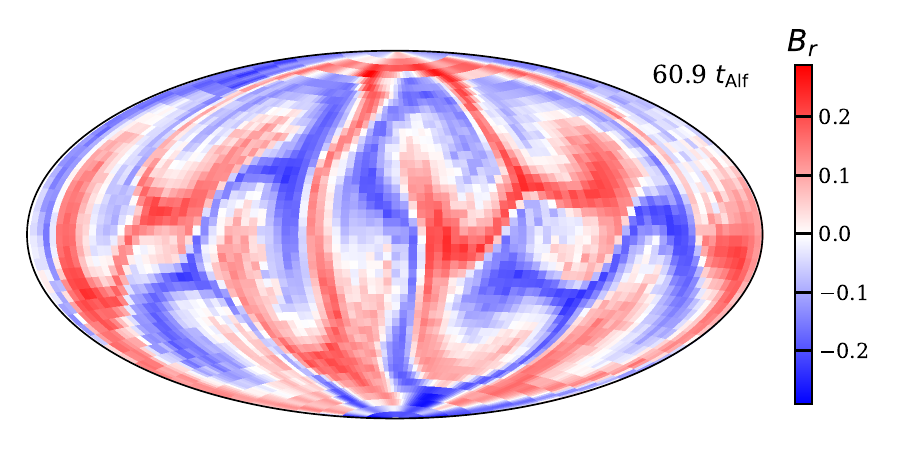}
    \end{minipage}
    \begin{minipage}{0.49\linewidth}
    \includegraphics[width=\columnwidth]{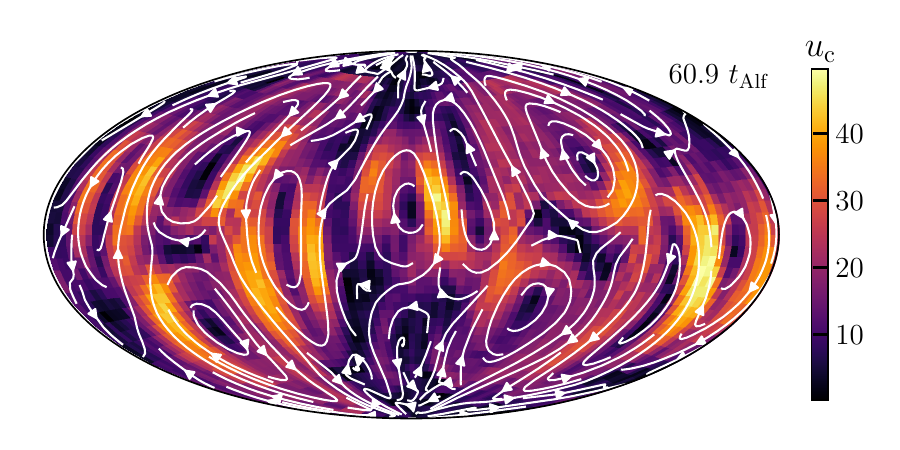}
    \end{minipage}
    \begin{minipage}{0.49\linewidth}
    \includegraphics[width=\columnwidth]{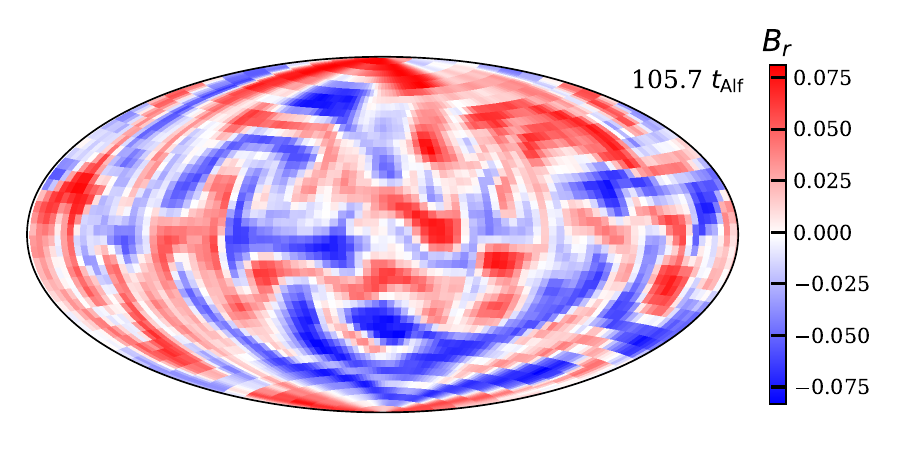}
    \end{minipage}
    \begin{minipage}{0.49\linewidth}
    \includegraphics[width=\columnwidth]{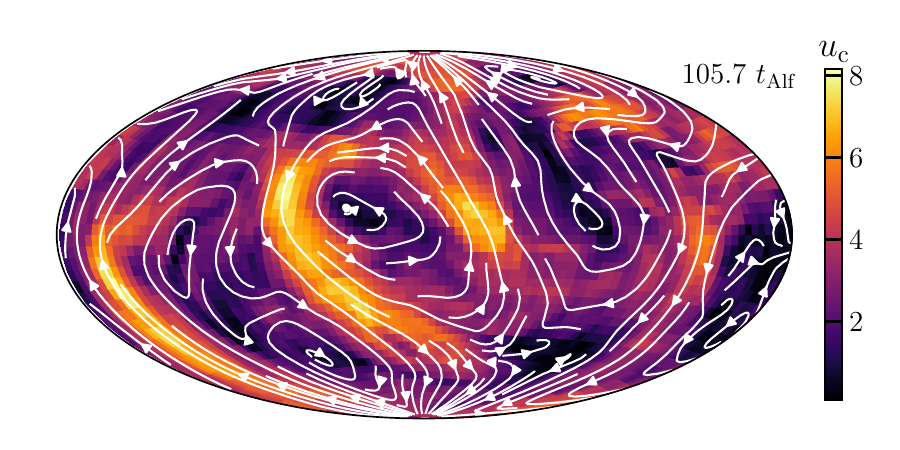}
    \end{minipage}
    \begin{minipage}{0.49\linewidth}
    \includegraphics[width=\columnwidth]{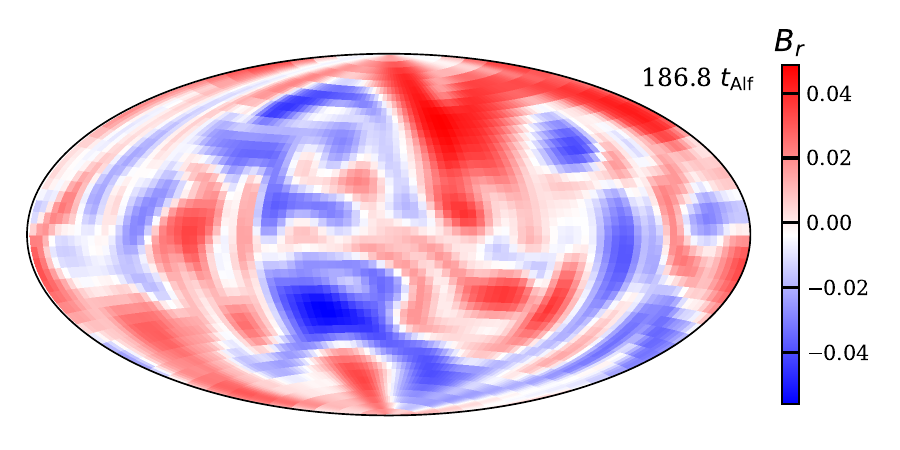}
    \end{minipage}
    \begin{minipage}{0.49\linewidth}
    \includegraphics[width=\columnwidth]{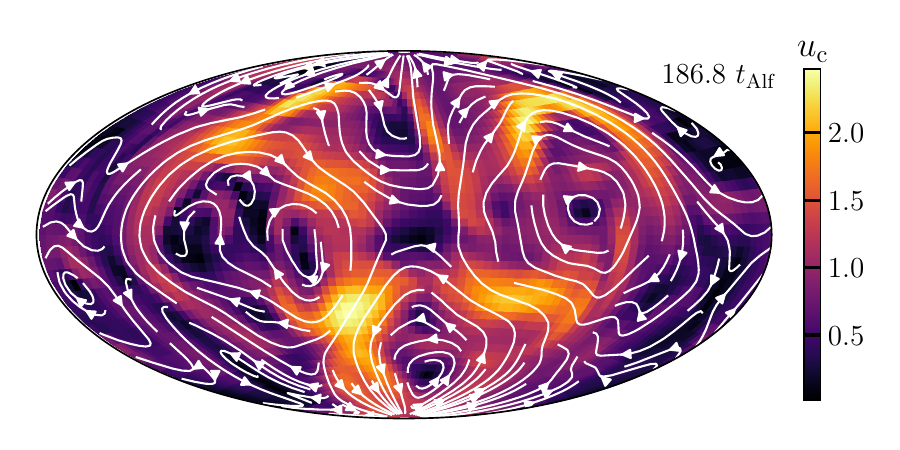}
    \end{minipage}
    \caption{Time evolution of surface radial magnetic field (left panel) as well as near-surface charged fluid velocity $\vec u_\mathrm{c} = (u_\mathrm{c}^\phi, u_\mathrm{c}^\theta)$ (right panel) in two-fluid simulation E. }
    \label{fig:Br_surface_H}
\end{figure*}

\begin{figure*}
\begin{minipage}{0.49\linewidth}
\includegraphics[width=\columnwidth]{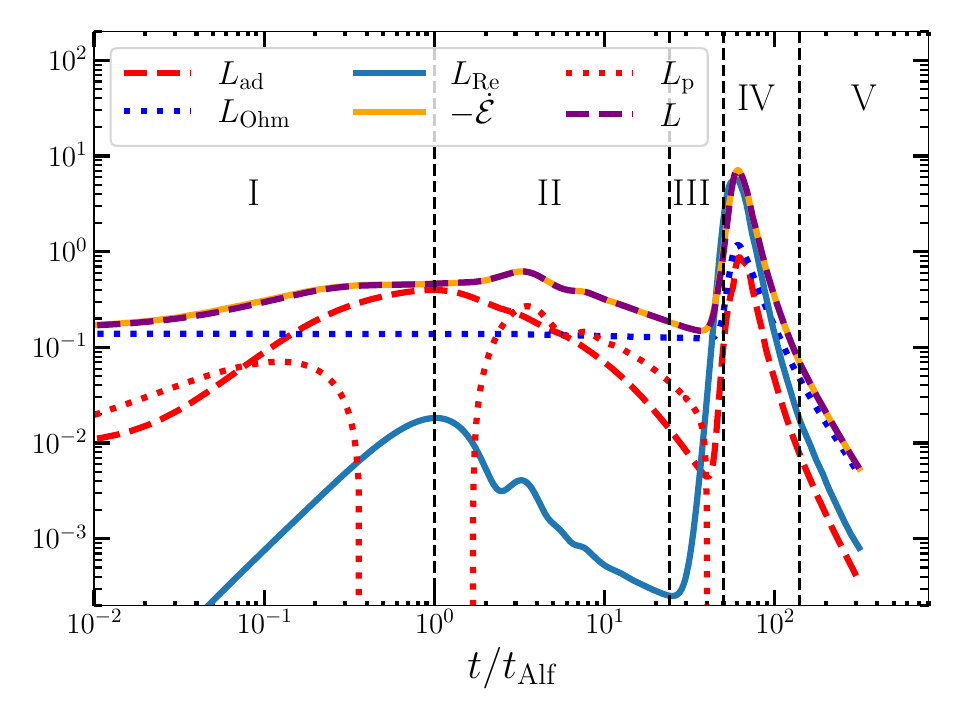}
\end{minipage}
\begin{minipage}{0.49\linewidth}
\includegraphics[width=\columnwidth]{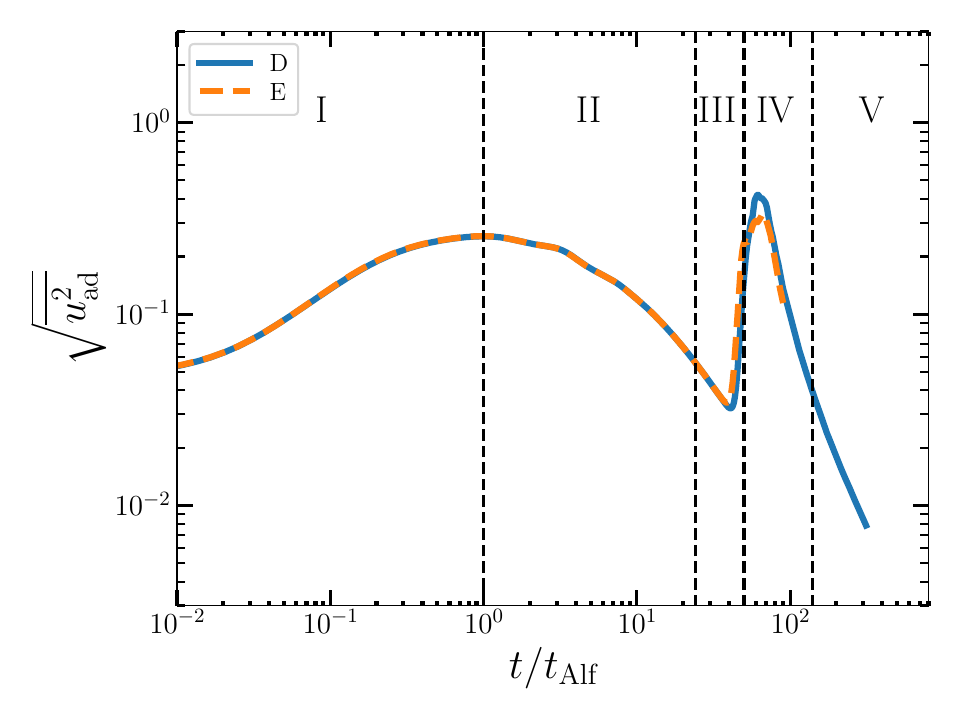}
\end{minipage}
\caption{Left panel: evolution of energy fluxes for two-fluid simulation D. Right panel: evolution of root-mean square of ambipolar velocity for two-fluid simulations D and E.}
    \label{fig:energy_H}
\end{figure*}

We show the energy evolution in Figure~\ref{fig:energy_H}. The transition from stage III to stage IV in this figure is marked by a sudden increase in energy dissipation via Ohmic and viscous decay, similarly to 1-barotropic-fluid simulations. While in a realistic NS the viscosity is significantly smaller, we would still expect a similar transition to occur at much smaller scales. Essentially, the magnetic field structure becomes small-scale, see e.g. Figure~\ref{fig:Br_surface_H}. Similarly, the fluid motion is also evolving towards small scales. When the spatial extent of these circulations approaches one comparable to a viscous limit, fluid viscosity becomes the dominant effect to dissipate energy. Via this dissipative term, the magnetic field is restructured. As soon as the magnetic field gets rid of excessive energy which prevents its stability, the viscous dissipation reduces. After this, the magnetic field decays mostly due to resistivity.

\begin{figure}
\includegraphics[width=\columnwidth]{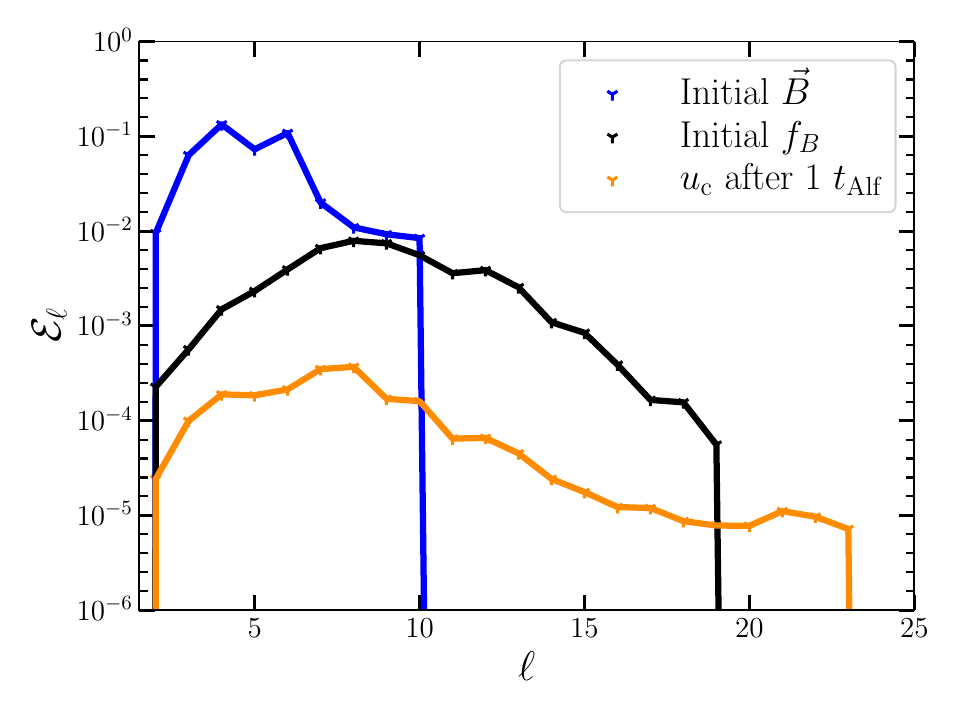}
\caption{Spectra of the additional two-fluid simulation.
}
    \label{fig:cascade}
\end{figure}

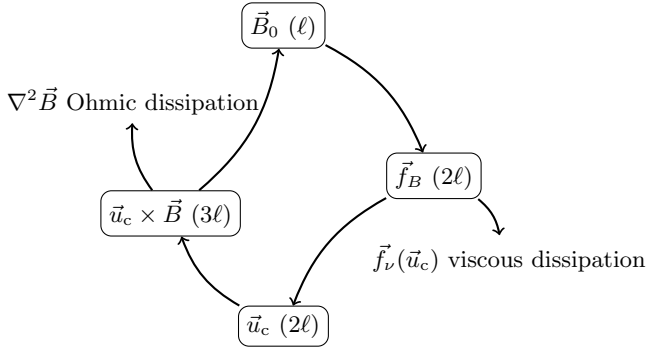
\begin{figure}
\begin{tikzpicture}
\node (a1) [rectangle,rounded corners, draw ] at (0,0) {$\vec B_0$ ($\ell$)};
\node (a2) [rectangle,rounded corners, draw] at (2,-2) {$\vec f_B$ ($2\ell$)};
\node (a3) [rectangle,rounded corners, draw] at (0,-4) {$\vec u_\mathrm{c}$ ($2\ell$)};
\node (a4) [rectangle,rounded corners, draw] at (-1.5,-2.5) {$\vec u_\mathrm{c}\times \vec B$ ($3\ell$)};
\node (a5) [rectangle, rounded corners] at (3,-3.1) {$\vec{f}_\nu (\vec u_\mathrm{c})$ viscous dissipation};
\node (a6) [rectangle, rounded corners] at (-2,-1) {$\nabla^2 \vec B$ Ohmic dissipation};
\path [->, thick] (a1)  edge [bend left=20]  (a2);
\path [->, thick] (a2)  edge [bend right=20] (a3);
\path [->, thick] (a2)  edge [bend left=20] (a5);
\path [->, thick] (a3)  edge [bend left=20] (a4);
\path [->, thick] (a4)  edge [bend right=20] (a1);
\path [->, thick] (a4)  edge [bend left=20] (a6);
\end{tikzpicture}
\caption{Formation of magnetic cascade and turbulence. An initial magnetic field limited to $\ell$ harmonics only creates Lorentz force with $\approx 2\ell$ structures, which translates into a velocity field of similar complexity. This velocity then further adds complexity to the field via the induction equation up to $3\ell$. Ohmic and viscous dissipations set limits to this increase in complexity at their respective timescales suppressing high-order harmonics.  }\label{f:cascade_formation}
\end{figure}

In usual MHD simulations, turbulence arises due to inertial forces in the fluid. In our system we removed $(\vec u_\mathrm{c} \cdot \vec \nabla) \vec u_\mathrm{c}$ because the inertial terms are expected to be small. Instead, we notice that the Lorentz force is highly non-linear and thus could lead to the development of turbulence. In order to check our hypothesis, we took the last snapshot from the two-fluid simulation, filtered all small-scale harmonics above 10 (both spherical and radial) from the magnetic field and increased the strength of the remaining configuration by 30 to cancel its previous decay. We evolved this configuration for $30\;t_\mathrm{Alf}$ and examined its spectra at different stages. We plot the spectra of the magnetic field, Lorentz force and velocity at the beginning of this new simulation in Figure~\ref{fig:cascade}. As expected from our setup, the field spectrum cuts off at $\ell = 10$. The Lorentz force spectrum produced by this field extends until $\ell = 19$ due to the non-linearity of the Lorentz force, with $(\curl\vec{B})\times\vec{B}$ producing nearly twice as many harmonics. The spectral peak is shifted from $\ell=4$ and $\ell = 6$ to $\ell=8-9$. Thus, the charged fluid is forced with a very broad angular spectrum, which naturally translates to a broad spectrum of fluid velocities after a single $t_\mathrm{Alf}$. After a few more Alfv\'en times small-scale structures propagated from the velocity field back to the magnetic field. We further illustrate this cycle in Figure~\ref{f:cascade_formation}.

\begin{figure*}
\begin{minipage}{0.49\linewidth}
\includegraphics[width=\columnwidth]{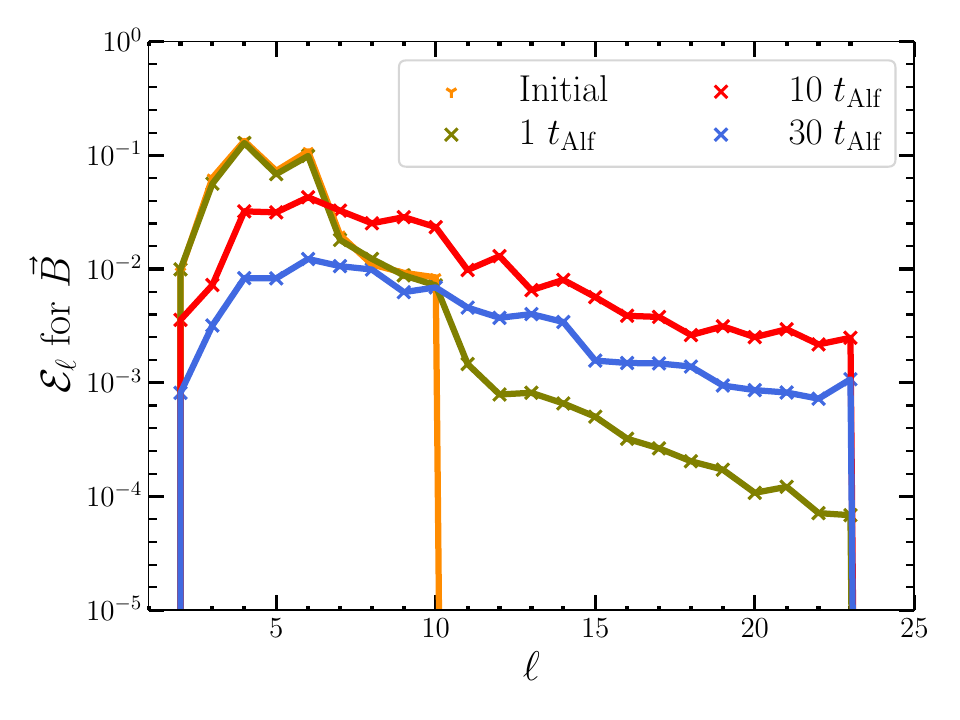}
\end{minipage}
\begin{minipage}{0.49\linewidth}
\includegraphics[width=\columnwidth]{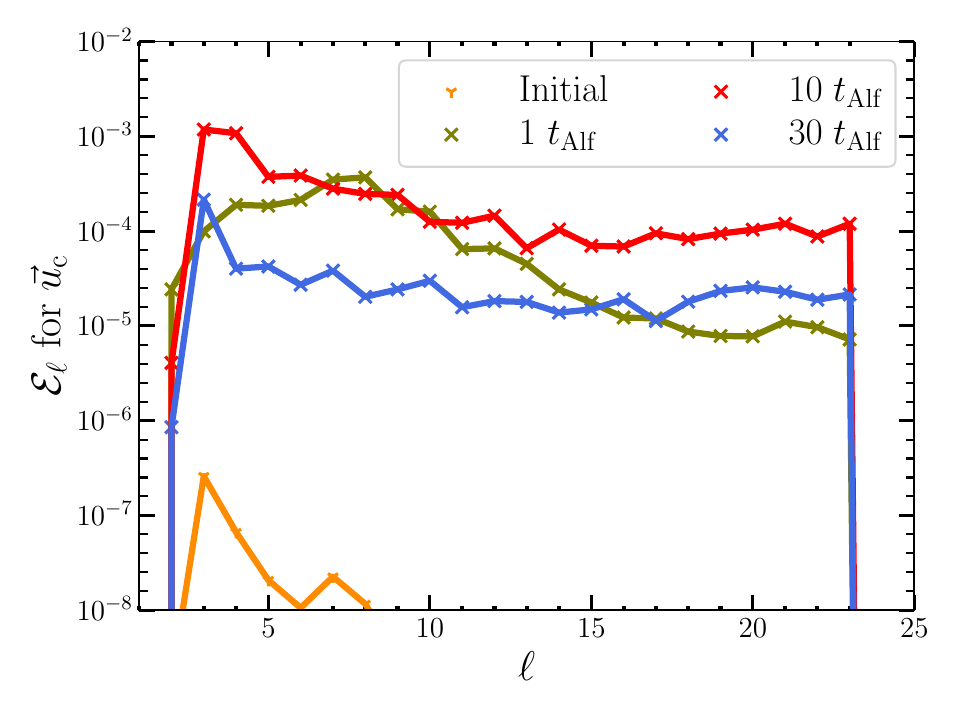}
\end{minipage}
\caption{Magnetic and kinetic energy spectra for the short additional simulation.  }
    \label{fig:cascade_detailed}
\end{figure*}

We show the energy spectra of later stages for this short additional simulation in Figure~\ref{fig:cascade_detailed}. Indeed the kinetic spectrum is relatively flat after $10\;t_\mathrm{Alf}$, so the fluid viscosity is indeed required to remove energy from the system. In terms of pseudo-spectral methods, viscosity is necessary to avoid pile-up of energy at smallest scales. The magnetic energy spectra are less flat, but also extend towards small scales. 

Summarizing, during stages IV and V the evolution proceeds as follows. Energy is transferred from magnetic field to small-scale velocities via the Lorentz force. A fraction of this energy is dissipated via the fluid viscosity. Another fraction is returned back to the field via the magnetic induction. Through this process, the magnetic field also becomes small-scale and loses a fraction of its energy via Ohmic losses. While the magnetic field is strong, the transfer of energy from magnetic field to charged particles and neutrons is more efficient, so the total energy is mostly lost via the viscous interactions. When the magnetic field and its Lorentz force becomes weaker, a lot of small-scale structure is developed in the field itself, and it starts losing energy efficiently due to Ohmic losses. 


\subsubsection{Resistive decay and stable configuration}

Similarly to our 1-barotropic-fluid simulations, after a period of intense turbulence a new magnetic field structure emerges. This field structure has no similarity to the initial configuration. Instead, it consists of small-scale fields including both toroidal and poloidal components. The further evolution of this magnetic field structure is primarily driven by Ohmic decay which is caused by weaker magnetic field dominated by small-scale structures.

It is interesting to note we have a new force equilibrium at this stage. We see from Figure~\ref{fig:rms_forces_H} that the rms Lorentz force is at least one order of magnitude smaller at the beginning of stage V than it used to be at stage II. Second, the force balance at the end of stage I was dominated by charged particles, with a noticeable contribution of neutrons. At the end of stage II, the force balance is completely dominated by charged particles, i.~e. it is close to the GS-equilibrium. Instead, at the beginning of stage V, the Lorenz force is in delicate balance with both charged particles and neutrons, which contribute nearly equally. Furthermore, the rms values of $\vec{f}_\mathrm{c}$ and $\vec{f}_\mathrm{n}$ are almost identical, implying that $\vec f_\mathrm{c} \approx -\vec f_\mathrm{n}$. In other words, the fluid forces acting on the charged particles and neutrons approximately balance each other and are larger than the Lorentz force $\vec f_B$. This behavior, first observed by \cite{Moraga2024MNRAS}, is a characteristic feature of the force balance equation in the two-fluid model (Eq.~\ref{eq:two_fluid_force_balance}). \cite{Moraga2024MNRAS} explicitly demonstrated that 
\begin{equation}
   |\vec f_\mathrm{c}| \sim |\vec f_\mathrm{n}|\sim(\ell_\mathrm{c}/\ell_B)|\vec f_{B}|,
\end{equation}
where $\ell_{B}$ is the typical length scale of the magnetic field (a fraction of the core radius), and 
\begin{equation}\label{eq:lc}
    \ell_\mathrm{c}\equiv -[d\ln(n_\mathrm{c}/n_\mathrm{n})/dr]^{-1}.
\end{equation}
Thus, this force balance is an indirect consequence of the stratification of the background stellar model. While $\vec{f}_B$ varies over a lengthscale $\ell_B$, the density ratio $n_\mathrm{c}/n_\mathrm{n}$ varies over a larger lengthscale $\ell_\mathrm{c} \sim R_{\mathrm{c}}$. Consequently, the individual fluid forces are larger than the Lorentz force.  

Since the decay is driven by Ohmic losses with a weak feedback from the advection term, the evolution at this stage proceeds on a fraction of an Ohmic timescale. Thus, the configuration could be treated as stable on Alfv\'en and ambipolar diffusion timescales. 

It is worth discussing the force balance and origin of fluid velocities at stage V in three-dimensional simulations. The stage seems puzzling because the viscosity seems to play an important role in stabilizing the fluid, unlike the two-dimensional case.
To better understand this stage, we first examine the force balance. 

\begin{figure*}
    \begin{minipage}{0.24\linewidth}
    \includegraphics[width=\columnwidth]{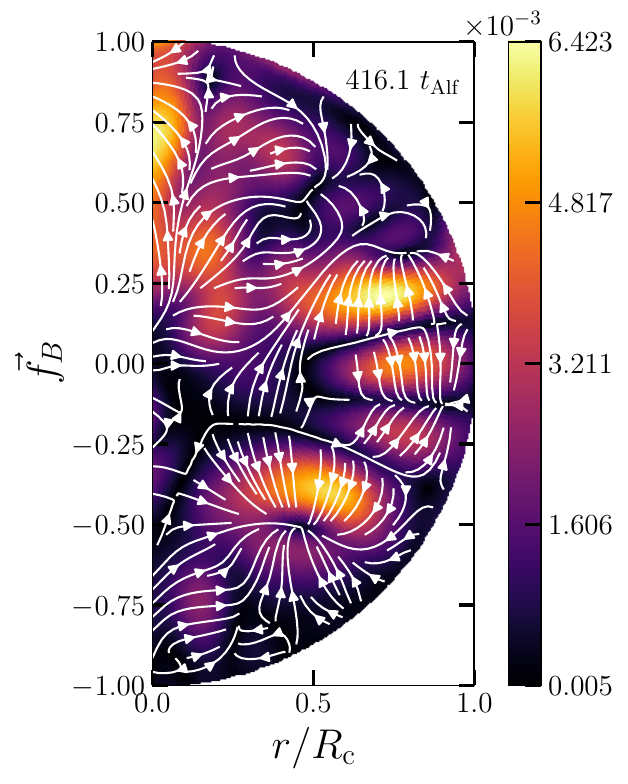}
    \end{minipage}
    \begin{minipage}{0.24\linewidth}
    \includegraphics[width=\columnwidth]{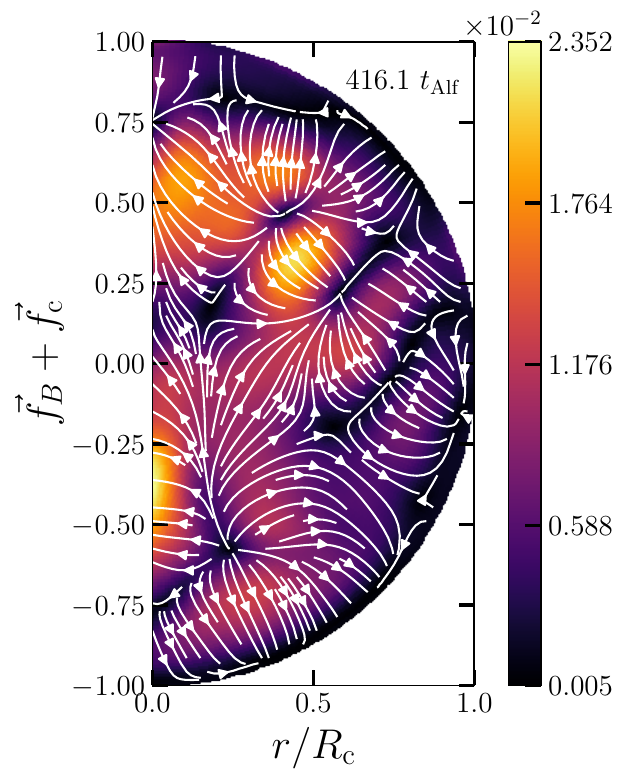}
    \end{minipage}
    \begin{minipage}{0.24\linewidth}
    \includegraphics[width=\columnwidth]{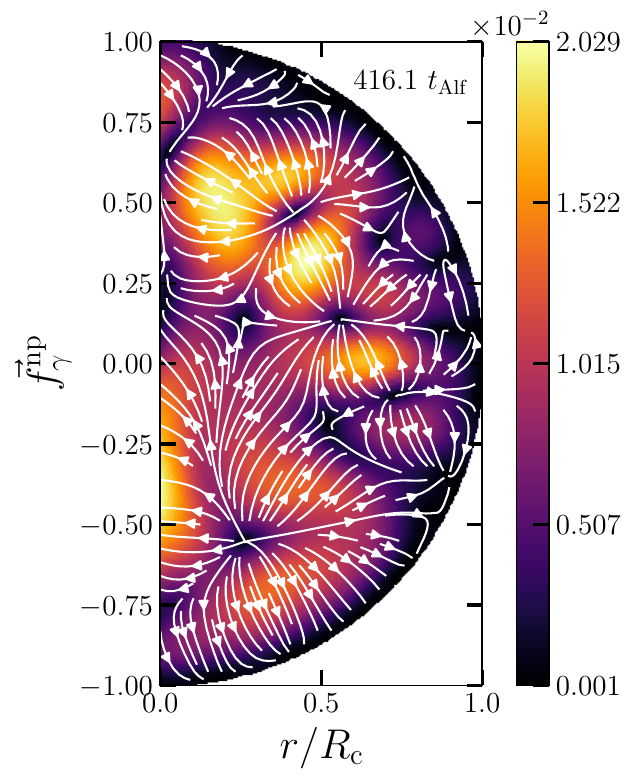}
    \end{minipage}
    \begin{minipage}{0.24\linewidth}
    \includegraphics[width=\columnwidth]{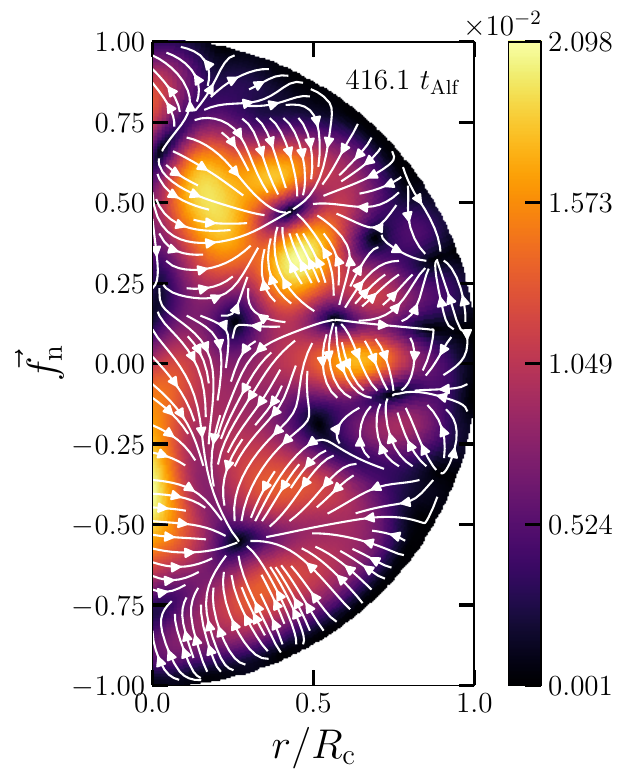}
    \end{minipage}
    \begin{minipage}{0.24\linewidth}
    \includegraphics[width=\columnwidth]{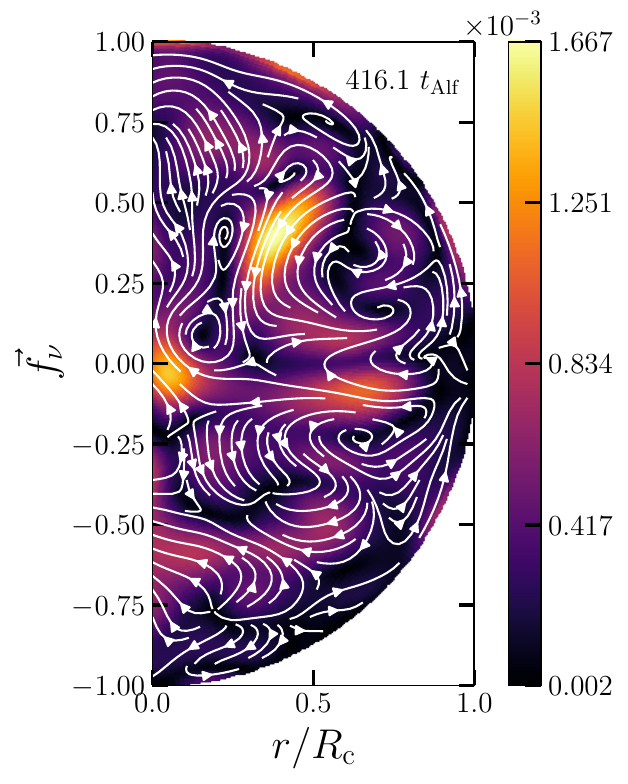}
    \end{minipage}
    \begin{minipage}{0.24\linewidth}
    \includegraphics[width=\columnwidth]{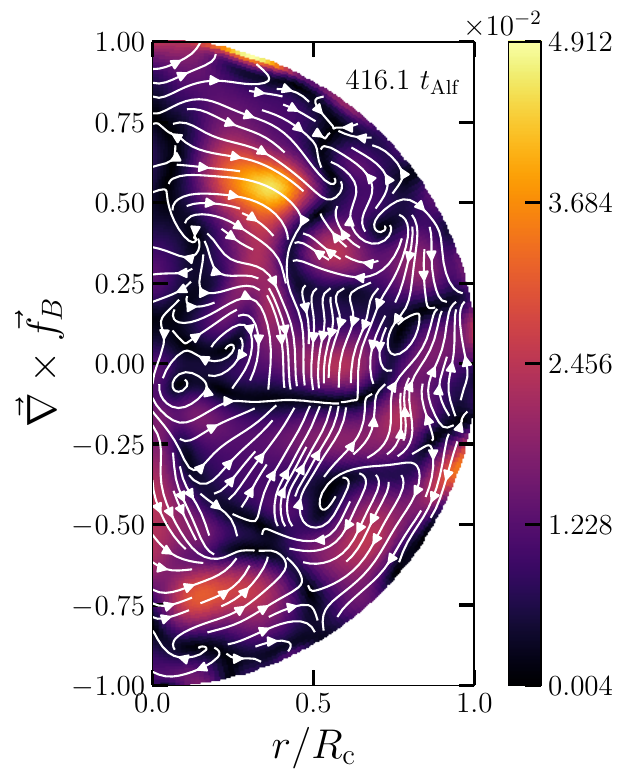}
    \end{minipage}
    \begin{minipage}{0.24\linewidth}
    \includegraphics[width=\columnwidth]{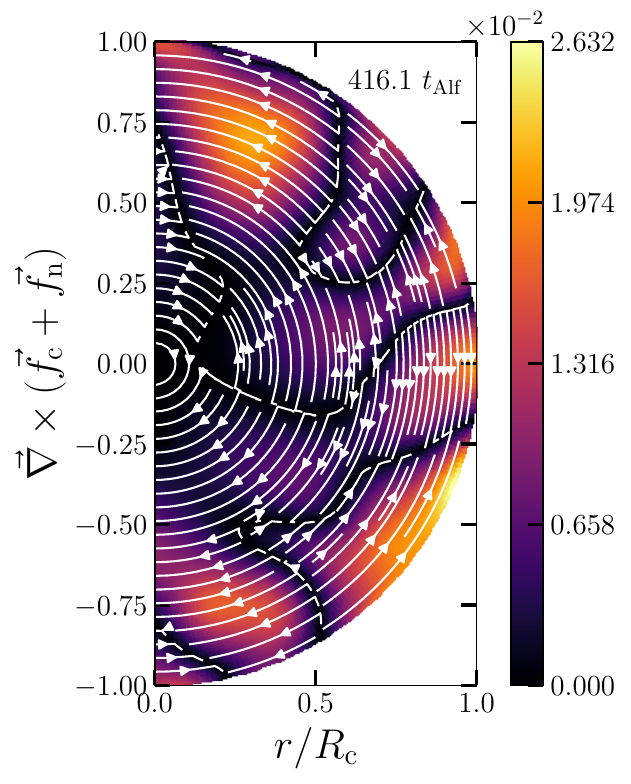}
    \end{minipage}
    \begin{minipage}{0.24\linewidth}
    \includegraphics[width=\columnwidth]{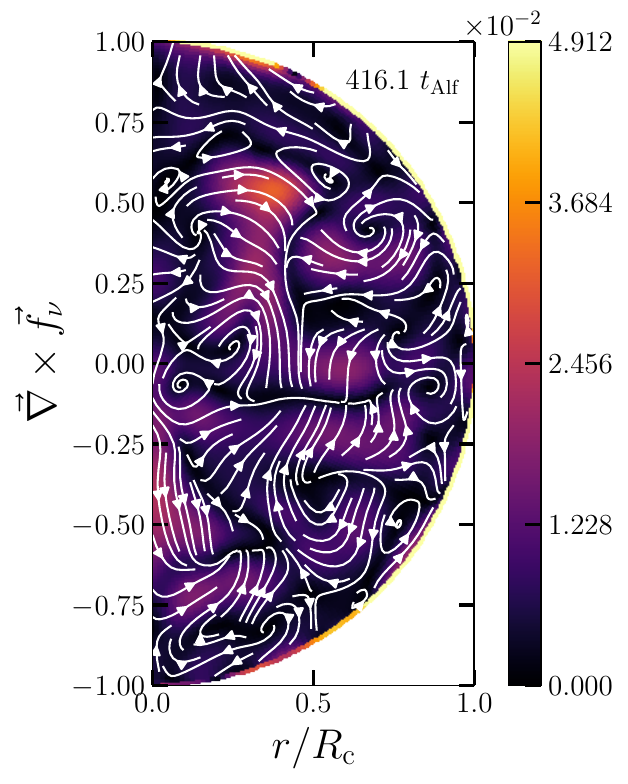}
    \end{minipage}
    \caption{Force balance for later stages of simulation D. Top panels: contribution of individual terms: (1) Lorentz force; (2) difference between Lorentz force and force caused by chemical potential perturbations of charged fluid; (3) collisional friction force; (4) force caused by chemical potential perturbations of neutrons only $\vec f_\mathrm{n}$; Bottom panel: (1) viscous forces; (2) curl of Lorentz force; (3) curl of forces caused by chemical potential perturbations; (4) curl of the viscous forces. }
    \label{fig:force_balance_late_h}
\end{figure*}

At the level of individual equations, we can notice that the collisional friction force $\vec f_\gamma^\mathrm{np} = \gamma_\mathrm{np} n_\mathrm{n} n_\mathrm{c} (\vec u_\mathrm{n} - \vec u_\mathrm{c})$ keeps playing an important role at stage V. We show the rms of this force in Figure~\ref{fig:rms_forces_H}. We have also examined individual forces in meridional cuts, see Figure~\ref{fig:force_balance_late_h}. In these figures, we see that $\vec f_\gamma^\mathrm{np}\approx -\vec f_\mathrm{n} \approx \vec f_B + \vec  f_\mathrm{c}$.
The Lorentz force is still mostly balanced by $\vec f_\mathrm{c} + \vec f_\mathrm{n}$, where each individual force is relatively large. Thus, during this phase, the system evolves through a broader range of consecutive quasi-equilibrium states, characterized by $\vec f_\gamma^\mathrm{np}\approx -\vec f_\mathrm{n} \approx \vec f_B + \vec  f_\mathrm{c}$. These quasi-equilibrium states are broader and less constrained than the GS equilibrium, where the additional condition $\vec{f}_B= \vec{f}_{c}$ must also be satisfied. Moreover, they appear to be more stable in three dimensions.

The viscous force $\vec f_\nu$ is not negligible and reaches up to $\approx 25$~\% of the Lorentz force. The viscous force does not peak at the boundary where it increases to match the fluid motion to the boundary conditions. Instead, it peaks deep in the core volume. 
So, the role of viscosity here is different.
It is worth reminding that the Navier-Stokes equations can be understood in terms of vorticity, similarly to the analysis we have done for the acceleration stage; see eq. (\ref{eq:vorticity}). 

Both chemical imbalances have the same mathematical structure $\vec f_\mathrm{c} + \vec f_\mathrm{n} = - n_\mathrm{n} \vec \nabla \delta \mu_\mathrm{n} - n_\mathrm{c} \vec \nabla \delta \mu_\mathrm{c}$. 
If we compute the curl of this expression (i.e. analyse vorticity), we obtain:
\begin{equation}
\curl (\vec f_\mathrm{c} + \vec f_\mathrm{n}) = \frac{d n_\mathrm{n} (r)}{d r} \hat r \times \vec \nabla \delta \mu_\mathrm{n}  + \frac{d n_\mathrm{c} (r)}{d r} \hat r \times \vec \nabla \delta \mu_\mathrm{c}  .
\end{equation}
This form of vector field has no $r$-component, so due to its mathematical structure it cannot balance the $r$-component of the curl of the Lorentz force in three dimensions. In axial symmetry, toroidal motions can easily adjust the magnetic field lines in such a way that $f_{B,\phi}=0$, which implies that $\hat r \cdot \curl \vec f_\mathrm{B}=0$. In full three dimensions, this appears not to happen, as can be seen 
in the bottom panel of Figure~\ref{fig:force_balance_late_h}, which shows a significant mismatch between $\vec\nabla\times\vec f_B$ and $\vec\nabla\times(\vec f_n+\vec f_c)$. 
The Navier-Stokes equation for fluid motion still needs to be satisfied precisely everywhere in the volume, and only the viscous force can 
compensate for this mismatch. 

We run an additional simulation G with higher numerical resolution and twice lower viscosity. In this simulation we find that viscosity plays a very similar role at stage V. We look at time $308\; t_\mathrm{Alf}$. At this stage rms $f_\nu = 7.3\times 10^{-4}$ in simulation D and rms $f_\nu = 5.8 \times 10^{-4}$ in simulation G. If it is linearly dependent on viscosity we would expect it to be twice smaller in this case. Values for the curl of Lorentz force and curl of viscous terms are very similar. We see rms $\curl \vec f_B = 0.02725$ for simulation D and rms $\curl \vec f_B = 0.02769$ for simulation G. Similarly, rms $\curl \vec f_\nu = 0.02283$ in D and rms $\curl \vec f_\nu = 0.0231$ in G. So, the viscosity still balances the same fraction of the Lorentz force. 

Despite its robustness against small variation of viscosity, this force balance is probably an artifact of our model. If we try to balance a significant fraction $\xi\approx 0.1$ of Lorentz force with viscous force in a realistic neutron star, we obtain:
\begin{equation}
\xi \frac{B^2}{R_c} = \frac{n_\mathrm{c} m_\mathrm{p} \nu u_\mathrm{c} }{l_u^2}    
\end{equation}
If we substitute physical values in this equation, this would give us unrealistically high charged fluid velocities (up to 10 km/s) which have spatial scale $l_u = 1$ cm or smaller to balance Lorentz force. Such motions will be efficiently damped by fluid viscosity. This alternative seems implausible if magnetic field survives for Myrs. Another alternative is to balance much weaker magnetic field or a negligible fraction of the Lorentz force $\xi \ll 0.1$. To model it more realistically, we need to modify our equations penalizing development of $\hat r \cdot \vec \nabla \times \vec f_B$ component. This would require a modification to our equation set. Nevertheless, it seems that our set of equations models evolution reliably until the turbulence is fully developed.

Still, it might be the case that in the long-term evolution $\hat r \cdot \curl \vec f_\mathrm{B}$ decays and the magnetic field evolves towards an equilibrium where viscosity plays no role. This needs to be checked in the future in longer simulations.

It is interesting to note here that despite the importance of the collisional friction force, the ambipolar velocity decays over time, as shown in the right panel of Figure~\ref{fig:energy_H}. Similarly, we can see from Figure~\ref{f:energies_H} that kinetic energies of charged and neutral fluids are decaying, so the velocities of both fluids are decaying. 
We can see it in Figure~\ref{fig:Br_surface_H} that velocities decay from $\mathrm{max}|u_c|\approx 4$ at 115~$t_\mathrm{Alf}$ to $\mathrm{max}|u_c|\approx 0.2$ by 558~$t_\mathrm{Alf}$. For evolution to proceed on Alfv\'en timescale we expect velocities to be $\approx 1/0.003 \approx 300$. If we rescale the Alfv\'en timescale due to decay of magnetic field they still should be $\approx 30$.   

\subsubsection{Role of Am}

\begin{figure}
\begin{minipage}{0.99\linewidth}
\includegraphics[width=\columnwidth]{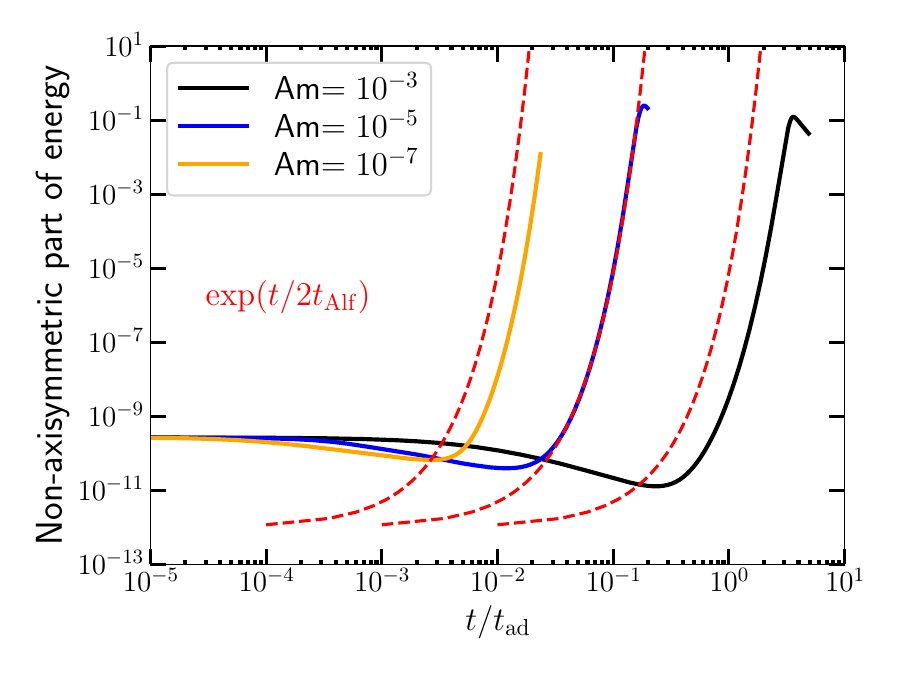}
\end{minipage}
\begin{minipage}{0.99\linewidth}
\includegraphics[width=\columnwidth]{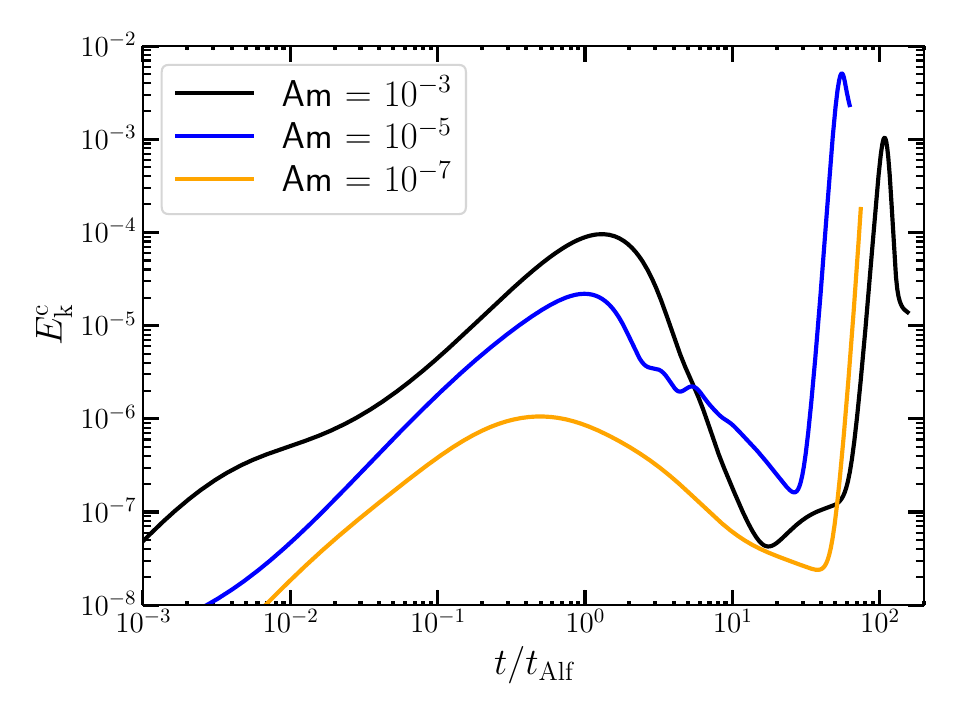}
\end{minipage}
\caption{Dependence of energy evolution of parameter Am. Top panel: growth of non-axisymmetric part of magnetic energy for different choices of parameter Am. Bottom panel: total kinetic energy of charged fluid. We show two-fluid simulations only.}
\label{f:Am}
\end{figure}

The Am value simply differentiates between the Alfv\'en and ambipolar diffusion timescales. Setting it at the value $10^{-5}$ yields $t_\mathrm{ad,0}\approx 316\; t_\mathrm{Alf}$. We vary this value by setting it to $\mathrm{Am} = 10^{-3}$, which means $t_\mathrm{ad,0}\approx 31.6\; t_\mathrm{Alf}$, and to $\mathrm{Am} = 10^{-7}$ which means $t_\mathrm{ad,0}\approx 3162\; t_\mathrm{Alf}$. We show the results in Figure~\ref{f:Am}. We compare the growth in the non-axisymmetric part of magnetic energy with exponential growth. Only the simulation with Am$=10^{-5}$ follows the line $\exp \left(t / (2 t_\mathrm{Alf}) \right)$ precisely. Other lines can be made more similar to an exponential if the numerical factor is slightly varied. For example, growth in the non-axisymmetric part of magnetic energy for $\mathrm{Am}=10^{-3}$ coincides with  $\exp \left(t / (4 t_\mathrm{Alf}) \right)$, while for Am$=10^{-7}$ we need a factor of 3, i.e. $\exp \left(t / (3 t_\mathrm{Alf}) \right)$. We relate this to the fact that the choice of Am affects the total viscosity of simulations. We have to adapt the Reynolds number to keep a similar ratio of Am/Re in the viscosity term. 

However, it is important to notice that changes in Am over four orders of magnitude do result in respective changes of two orders of magnitude in the instability growth timescale (in units of $t_\mathrm{ad,0}$). We therefore convincingly prove that the instability grows on a timescale $\sim t_\mathrm{Alf}$ and not $\sim t_\mathrm{ad,0}$.

\subsection{Toroidal magnetic field}
\label{s:toroidal}

\begin{figure*}
    \begin{minipage}{0.49\linewidth}
    \includegraphics[width=\columnwidth]{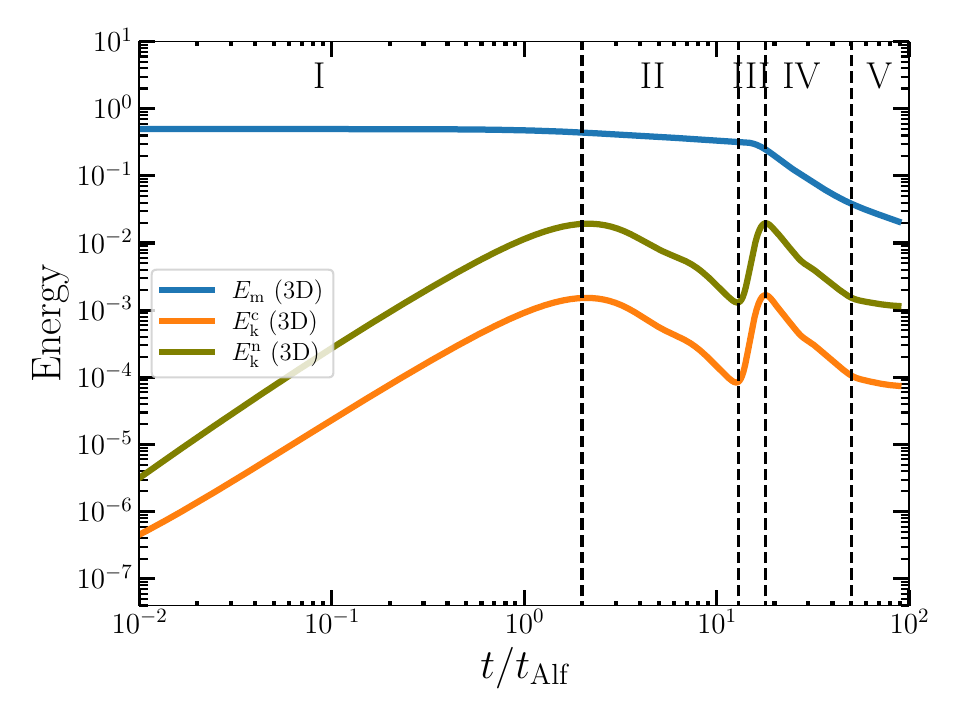}
    \end{minipage}
    \begin{minipage}{0.49\linewidth}
    \includegraphics[width=\columnwidth]{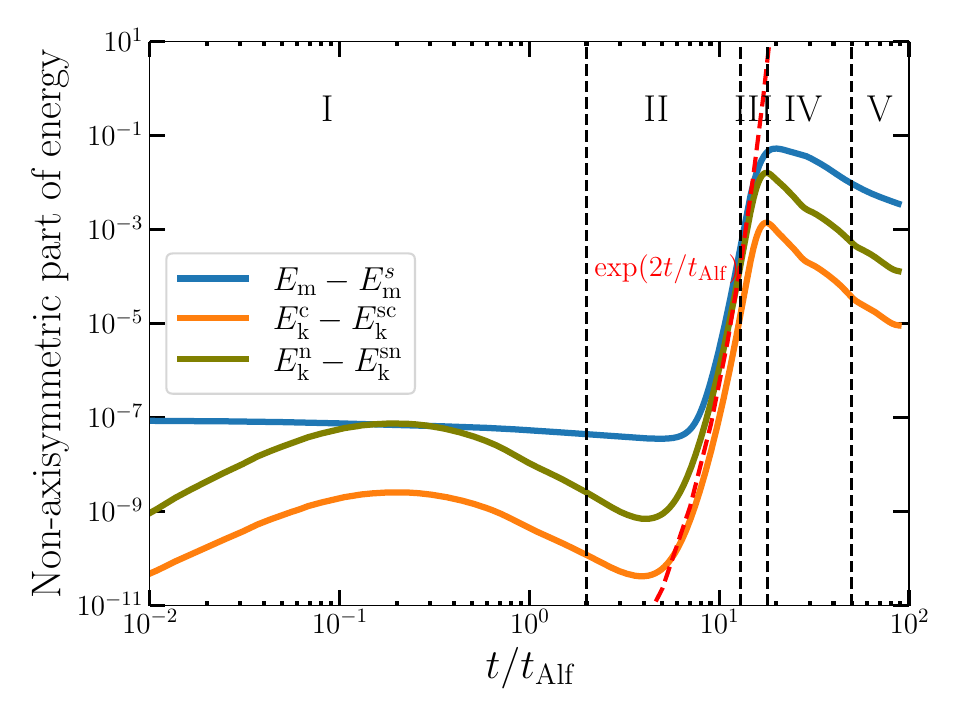}
    \end{minipage}
    \caption{Evolution of kinetic and magnetic energies for two-fluid MHD simulation F. Left panel: total magnetic and kinetic energies. Right panel: non-axisymmetric part of magnetic and kinetic energies. We mark the following stages: I is the acceleration stage, II is evolution towards 2D force balance, III is  instability, IV is magnetic turbulence, V is resistive decay. The red dashed line in the right panel shows exponential growth with timescale $2\;t_\mathrm{Alf}$.  }
    \label{f:energies_F}
\end{figure*}

We also run two-fluid simulation F starting with an axially symmetric toroidal magnetic field. The energy evolution closely follows the one we discussed already for 1-barotropic-fluid and two-fluid simulations A-E. We show the energy evolution in Figure~\ref{f:energies_F}. 

Here we list the main changes in comparison to poloidal magnetic field evolution. The acceleration stage takes even longer, so charged and neutral fluids reach equilibrium only at $2\; t_\mathrm{Alf}$. The kinetic energy of charged fluid at the end of stage I is comparable to the end of stage III. The  instability grows on a shorter timescale of $0.5\;t_\mathrm{Alf}$. Otherwise, the evolution of simulation F proceeds similarly to the poloidal magnetic field configuration.

\begin{figure}
    \begin{minipage}{0.49\linewidth}
    \includegraphics[width=\columnwidth]{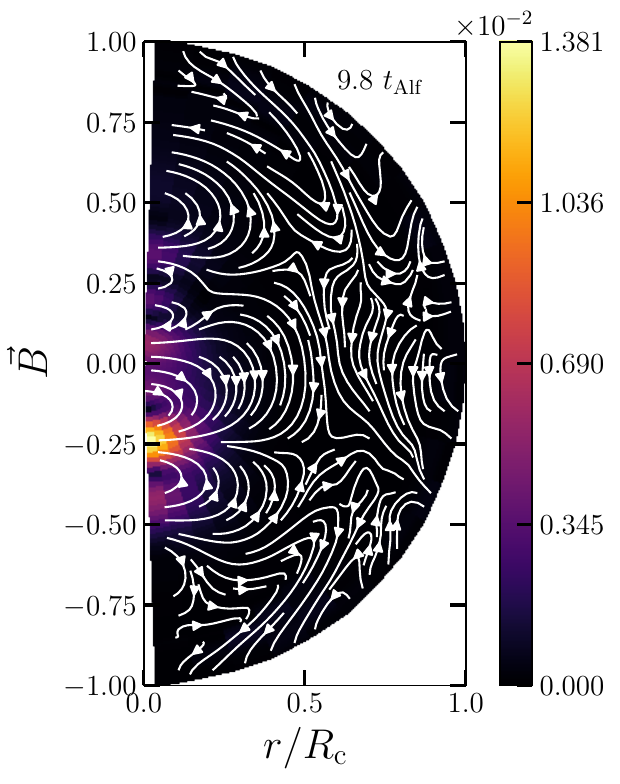}
    \end{minipage}
    \begin{minipage}{0.49\linewidth}
    \includegraphics[width=\columnwidth]{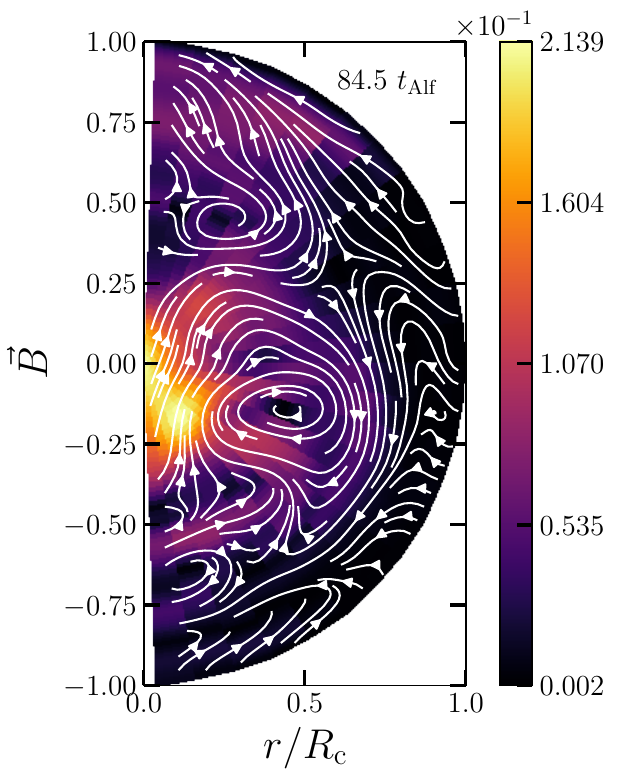}
    \end{minipage}
    \caption{Poloidal magnetic field grown as a result of instability in simulation F. }
    \label{f:B_merid_F}
\end{figure}

It is interesting to see how the instability grows in the case of the toroidal magnetic field. To illustrate this, we show the meridional cut of magnetic fields in Figure~\ref{f:B_merid_F}. The small-scale field grows near the north-south axis closer to the NS center. The final state of the simulation does not look too different from what we found in the case of initial poloidal magnetic field. Similarly to that case, the magnetic field is quite turbulent. We also show the magnetic energy spectra at different stages of simulation F in lower panel of Figure~\ref{f:magnetic_energy_H}. As expected, the dominant growing mode is $m=1$.

\begin{figure}
    \includegraphics[width=\columnwidth]{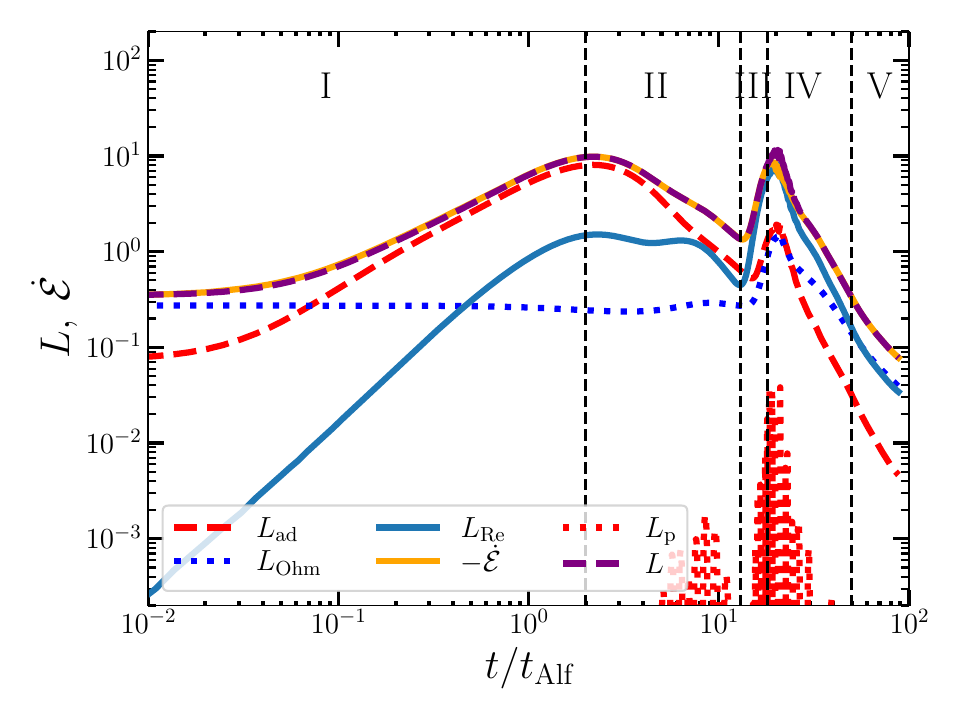}
    \caption{Evolution of energy fluxes for two-fluid simulation F. }
    \label{f:energy_F}
\end{figure}

We show the evolution of the total energy in Figure~\ref{f:energy_F}. It is also very similar to the case of an initial poloidal magnetic field. During the turbulence stage, the viscous dissipation increases significantly and it decays later on to a level comparable to Ohmic dissipation.

\section{Discussion}

\subsection{Comparison with earlier works}
\label{s:previous}

It is interesting to compare the results found here with the earlier analysis by \cite{IgoshevHollerbach2023MNRAS}. This comparison is not straightforward because some of the assumptions differ significantly. In this research we studied two approximations: 1-barotropic-fluid and two-fluids. \cite{IgoshevHollerbach2023MNRAS} studied 
ambipolar diffusion in the so-called single-fluid approximation, in which neutrons are fixed and not moving. Due to ambipolar diffusion, the single-fluid approximation has an additional dissipative term in comparison to the 1-barotropic-fluid case. This term is directly proportional to the charged particle (ambipolar) velocity in this simplified scenario. Additionally, \cite{IgoshevHollerbach2023MNRAS} neglected the shortest timescales of the problem, assuming that charged particle velocities adapt to the Lorentz force instantaneously, with $t_\mathrm{Alf} = 0$. While this approximation is fully justified because the Alfv\'en timescale is many orders of magnitude shorter than the relevant timescale of ambipolar diffusion, this approximation led to incorrect conclusions about the timescale of instability development. \cite{Igoshev2023MNRAS} concluded that the non-axisymmetric instability they found develops on $0.2\;t_\mathrm{ad,0}$, while in this work we see that this instability is general and develops much earlier on timescales comparable to $t_\mathrm{Alf}$. Here we also double-check with additional axially symmetric simulations that the instability only develops in three dimensions. The detailed development of the instability also differs. Here we found a selection of the following azimuthal harmonics: $m=4$, $m=6$ and $m=8$. In earlier work by \cite{IgoshevHollerbach2023MNRAS} the dominant harmonics included $m=10$, $m=12$ and $m=14$.

Because we describe fluids in this work using the Navier-Stokes equations, we had to include fluid viscosities and set proper boundary conditions. The fluid viscosity of neutron star matter is very small and boundary layers forming at the crust-core interface cannot be resolved. In this work we follow familiar planetary and stellar simplifications such as usage of the anelastic approximation, and increased viscosities significantly. Beside forming the boundary layers, viscosity also constrains fluid velocities everywhere in the volume, see e.g. \cite{Livermore2016NatSR}. 

Moreover, \cite{Igoshev2023MNRAS} had to cut the realistic radial profile for charged particle number density in its steepest part because the numerical scheme could not handle it otherwise. In this research, we introduced the anelastic approximation which is well-suited to deal with steep gradients similar to ones which develop in self-gravitating bodies, e.g. in planetary atmospheres.

Another comparison of our 1-barotropic-fluid simulations is possible with the fully-compressible simulations by \cite{Becerra2022MNRAS}, which follow similar patterns to our stages II and III identified above. They relaxed their initial conditions before the start of simulations, so the first stage is excluded completely. Their simulations with no toroidal magnetic field show some oscillations until $\approx 10\; t_\mathrm{Alf}$, and then the total magnetic energy decays by approximately a factor of two, reaching a new equilibrium soon after $10\; t_\mathrm{Alf}$, see their figure 2. In our case, the total magnetic energy drops from $\approx 0.5$ to $\approx 0.35$ during this stage. Naturally, this comparison is not perfect because of the following aspects: (1) our simulations follow the anelastic approximation, (2) the equation of state is different, and (3) the initial configuration is different.

\subsection{Dimensionless numbers and their role in shaping the flow}
\label{s:dimensionless}


\subsubsection{Reynolds number}
We estimate the Reynolds number inside a NS as
\begin{equation}
\mathrm{Re} = \frac{u l}{\nu} = 10^{-3}-10^{-4}  ,
\end{equation}
using a typical charged particle velocity of $10^{-6}$~--~$10^{-7}$~cm~s$^{-1}$ and a viscosity estimate of
$\nu = \mu/ \rho =\mathrm{(10^{18}~g~cm^{-1}~s^{-1}) / (10^{15}~g~cm^{-3}) = 10^3~cm^2~s^{-1}}$, using the dynamic viscosity $\mu$ from \citet{CutlerLindblom1987ApJ} at $10^8$~K, combined with density $\rho = 10^{15}$~g~cm$^{-3}$.  

While normally such a small Reynolds number is easy to model and indicates laminar flow, this number is less relevant for NS cores, since inertial forces are tiny in comparison to magnetic forces. Nevertheless, it does still indicate that we are working in the laminar regime and normal fluid turbulence is not expected.

In our numerical experiments we note that the size of hydrodynamic boundary layer scales as
\begin{equation}
h \sim 0.003 R_\mathrm{c} \sqrt{\frac{\mathrm{Am}}{\mathrm{Re}}}  .  
\end{equation}

\subsubsection{Magnetic Reynolds number}

We estimate the magnetic Reynolds number as
\begin{equation}
\mathrm{Rm} = \frac{ul}{\eta} \sim 10^4 - 10^{9}  .  
\end{equation}
We estimated the core conductivity as $10^{26}$~--~$10^{29}$~s$^{-1}$ \citep{PonsVignao2019LRCA}, which translates to diffusivity $c^2/(4\pi\sigma) = \mathrm{(10^{-6}-10^{-9})~cm^2~s^{-1}}$. 

The size of the magnetic boundary layer can be estimated as
\begin{equation}
h_\mathrm{m} \sim \frac{R_\mathrm{c}}{\sqrt{\mathrm{Rm}}} = 30 - 10^4 \; \mathrm{cm}    .
\end{equation}

We can also compute magnetic Prandtl number:
\begin{equation}
\mathrm{Pm} = \frac{\mathrm{Rm}}{\mathrm{Re}}    
\end{equation}
In neutron star core, we expect $\mathrm{Pm}\approx 10^7$~--~$10^{12}$. In our simulations, we typically have $\mathrm{Pm} = 50 \gg 1$ i.e. we perform our simulations in the right regime of high magnetic Prandtl number.

\subsection{Future work}
\label{s:future}

This article represents a first step in novel modeling of neutron star cores, and as such it neglects multiple important effects, including the star's rotation, its crust and cooling, superfluidity and superconductivity, as well as the effects of general relativity. We neglect these effects to make the problem tractable. Some of these effects are relatively straightforward to add to future modeling, such as rotation and cooling. Other effects, such as general relativity, are likely less relevant for magnetic field evolution in NSs anyway. The most difficult effects to tackle are neutron superfluidity and proton superconductivity in NS cores. Even for these effects the techniques developed here could prove to be very useful. For example, \cite{Fuentes2024ApJ} model superfluidity using a relatively similar two-fluid system of equations in \texttt{Dedalus}. Thus, we hope to address some of these effects in future work.

The interaction between Hall evolution in the crust and ambipolar diffusion in the core also seems to be an important effect (see \citealt{Skiathas2024MNRAS}). This is therefore another possible future direction of research.
Finally, magnetic fields could have a very complicated structure in neutron star cores, as was recently demonstrated in detailed dynamo simulations \citep{Raynaud2020SciA,ReboulSalze2022AA,Barrere2022AA,Barrere2023MNRAS}. These complicated structures are expected to be stable on Alfv\'en timescales. We plan to model evolution of some of these magnetic field configurations including mixture of poloidal and toroidal magnetic fields in our future work.

\section{Conclusions}
\label{s:conclusion}

We performed the first three-dimensional numerical simulations for ambipolar diffusion in the two-fluid approximation. The NS core was modeled as non-Cooper-paired (normal) matter consisting of neutrons, protons, and electrons, surrounded by a non-conducting medium.
The non-conducting medium was imposed by using the current-free boundary condition on the magnetic field outside the core. We ran our simulations in spherical geometry and started with simple, nearly axially symmetric initial conditions for the magnetic field. In our simulations we neglected the NS thermal evolution. Our equations were written in the anelastic approximation.
In order to better understand our results we additionally ran simple 1-barotropic-fluid simulations in the same geometry.

We found:
\begin{itemize}
    \item The magnetic field evolution proceeds via five distinct stages in both cases, 1-barotropic-fluid and two-fluid. The stages are as follows: (I) acceleration of fluids, (II) evolution towards 2D force balance, (III)  instability, (IV) magnetic turbulence, and (V) Ohmic decay.
    \item The main difference between axially symmetric (2D) simulations and full 3D simulations is the development of an instability. Two-dimensional simulations continue to slowly evolve toward a Grad-Shafranov equilibrium even at tens of $t_\mathrm{Alf}$, while the 3D results develop an instability at $6\; t_\mathrm{Alf}$ (1-barotropic-fluid) and $25\; t_\mathrm{Alf}$ (two-fluid). In the axially symmetric simulations, the fluid velocities slowly decay due to viscosity. 
    \item In the case of a poloidal magnetic field, the instability leads to the growth of $B_\phi$ with leading azimuthal wavenumbers $m=2$ and $m=4$ for 1-barotropic-fluid. It is similar to the results of \cite{Braithwaite2009MNRAS}, who found a growth of a mixture of modes with $m\geq 2$ with $m=2$ as the most unstable mode for poloidal magnetic field. For two fluids, the leading harmonics are $m=4$ and $m=6$.
    \item The instability is seen in the fluid velocities as circulations on spherical shells. In the two-fluids case, the radial motion of the fluid is strongly suppressed, but both charged and neutral fluids can move together with large $u_\phi$ and $u_\theta$ components of velocity. Such a motion is prohibited in two spatial dimensions where both $u_\theta$ and $u_\phi$ components of the velocity are available, but they do not depend on the $\phi$ coordinate.  
    \item Overall, the main influence of the two-fluid evolution is on longer timescales and consists in the emergence of more complicated patterns. The timescale for this instability is $t_\mathrm{Alf}$ (1-barotropic-fluid) and $2t_\mathrm{Alf}$ (two-fluids).
    \item Due to the non-linear Lorentz force, the magnetic energy is redistributed from large scales to small scales, driving the fluids. Until the cascade is completely shaped, the leading energy loss mechanism is via the fluid viscosity. After small-scale magnetic field structure is formed, it decays due to Ohmic losses. 
    \item At the last stage V, our model shows a force balance where a significant fraction of Lorentz force is balanced by viscous forces. This balance is probably unreachable in a real neutron star. Further modifications to our model are required to reach a more plausible steady state at stage V. Nevertheless, the model is extremely useful to study stability under influence of ambipolar diffusion for different initial configuration and their evolution until the turbulence stage. 
    \item In our model,  the magnetic field configuration evolves towards a state which is stable on Alfv\'en timescales. Its evolution is mostly driven by Ohmic decay.
    \item In our simulations,  the magnetic field decays nearly one order of magnitude in the process.  
    Future work is required to study more realistic magnetic field configurations, including both poloidal and toroidal components.
\end{itemize}

\begin{acknowledgments}

A.I. was supported by STFC grant no.\ ST/W000873/1 while at Leeds, and by the Royal Society University Research Fellowship URF\textbackslash R1\textbackslash 241531 in Newcastle. A.I. thanks Dr Toby Wood and Dr Aur\'elie Astoul for useful discussions. A.I. is grateful to the \texttt{Dedalus} developers for answering multiple questions. N.M. is grateful for support by ANID doctoral fellowship 21210909, and thanks the University of Leeds for their hospitality during two extended visits. A.I. and N.M. thank Francisco Castillo for helping with data visualization.  This research made use of the Rocket High Performance Computing service at Newcastle University. This work was partially performed using the DiRAC Data Intensive service at Leicester, operated by the University of Leicester IT Services, which forms part of the STFC DiRAC HPC Facility (www.dirac.ac.uk). The equipment was funded by BEIS capital funding via STFC capital grants ST/K000373/1 and ST/R002363/1 and STFC DiRAC Operations grant ST/R001014/1. DiRAC is part of the National e-Infrastructure.

\end{acknowledgments}



\software{Dedalus \citep{Vasil2019JCPX}, Numpy \citep{numpy}, Matplotlib \citep{matplotlib}
, VisIt \citep{visit_code}}



\bibliographystyle{aasjournalv7}
\bibliography{two_fluids}{} 

@ARTICLE{IgoshevHollerbach2023MNRAS,
       author = {{Igoshev}, Andrei P. and {Hollerbach}, Rainer},
        title = "{Three-dimensional numerical simulations of ambipolar diffusion in NS cores in the one-fluid approximation: instability of poloidal magnetic field}",
      journal = {\mnras},
         year = 2023,
        month = jan,
       volume = {518},
       number = {1},
        pages = {821-846},
          doi = {10.1093/mnras/stac3126},
archivePrefix = {arXiv},
       eprint = {2210.10869},
 primaryClass = {astro-ph.HE},
       adsurl = {https://ui.adsabs.harvard.edu/abs/2023MNRAS.518..821I}
}

@ARTICLE{Igoshev2021Univ,
       author = {{Igoshev}, Andrei P. and {Popov}, Sergei B. and {Hollerbach}, Rainer},
        title = "{Evolution of Neutron Star Magnetic Fields}",
      journal = {Universe},
         year = 2021,
        month = sep,
       volume = {7},
       number = {9},
          eid = {351},
        pages = {351},
          doi = {10.3390/universe7090351},
archivePrefix = {arXiv},
       eprint = {2109.05584},
 primaryClass = {astro-ph.HE},
       adsurl = {https://ui.adsabs.harvard.edu/abs/2021Univ....7..351I}
}

@ARTICLE{GoldreichReisenegger1992ApJ,
       author = {{Goldreich}, Peter and {Reisenegger}, Andreas},
        title = "{Magnetic Field Decay in Isolated Neutron Stars}",
      journal = {\apj},
         year = 1992,
        month = aug,
       volume = {395},
        pages = {250},
          doi = {10.1086/171646},
       adsurl = {https://ui.adsabs.harvard.edu/abs/1992ApJ...395..250G}
}

@ARTICLE{LanderJones2012MNRAS,
       author = {{Lander}, S.~K. and {Jones}, D.~I.},
        title = "{Are there any stable magnetic fields in barotropic stars?}",
      journal = {\mnras},
         year = 2012,
        month = jul,
       volume = {424},
       number = {1},
        pages = {482-494},
          doi = {10.1111/j.1365-2966.2012.21213.x},
archivePrefix = {arXiv},
       eprint = {1202.2339},
 primaryClass = {astro-ph.SR},
       adsurl = {https://ui.adsabs.harvard.edu/abs/2012MNRAS.424..482L}
}

@ARTICLE{CutlerLindblom1987ApJ,
       author = {{Cutler}, Curt and {Lindblom}, Lee},
        title = "{The Effect of Viscosity on Neutron Star Oscillations}",
      journal = {\apj},
         year = 1987,
        month = mar,
       volume = {314},
        pages = {234},
          doi = {10.1086/165052},
       adsurl = {https://ui.adsabs.harvard.edu/abs/1987ApJ...314..234C}
}

@ARTICLE{PonsVignao2019LRCA,
       author = {{Pons}, Jos{\'e} A. and {Vigan{\`o}}, Daniele},
        title = "{Magnetic, thermal and rotational evolution of isolated neutron stars}",
      journal = {Living Reviews in Computational Astrophysics},
         year = 2019,
        month = dec,
       volume = {5},
       number = {1},
          eid = {3},
        pages = {3},
          doi = {10.1007/s41115-019-0006-7},
archivePrefix = {arXiv},
       eprint = {1911.03095},
 primaryClass = {astro-ph.HE},
       adsurl = {https://ui.adsabs.harvard.edu/abs/2019LRCA....5....3P}
}

@ARTICLE{Tayler1973MNRAS,
       author = {{Tayler}, R.~J.},
        title = "{The adiabatic stability of stars containing magnetic fields-I.Toroidal fields}",
      journal = {\mnras},
         year = 1973,
        month = jan,
       volume = {161},
        pages = {365},
          doi = {10.1093/mnras/161.4.365},
       adsurl = {https://ui.adsabs.harvard.edu/abs/1973MNRAS.161..365T}
}

@ARTICLE{MarkeyTayler1973MNRAS,
       author = {{Markey}, P. and {Tayler}, R.~J.},
        title = "{The adiabatic stability of stars containing magnetic fields. II. Poloidal fields}",
      journal = {\mnras},
         year = 1973,
        month = mar,
       volume = {163},
        pages = {77-91},
          doi = {10.1093/mnras/163.1.77},
       adsurl = {https://ui.adsabs.harvard.edu/abs/1973MNRAS.163...77M}
}

@ARTICLE{OfengeimGusakov2018PhRvD,
       author = {{Ofengeim}, D.~D. and {Gusakov}, M.~E.},
        title = "{Fast magnetic field evolution in neutron stars: The key role of magnetically induced fluid motions in the core}",
      journal = {\prd},
         year = 2018,
        month = aug,
       volume = {98},
       number = {4},
          eid = {043007},
        pages = {043007},
          doi = {10.1103/PhysRevD.98.043007},
archivePrefix = {arXiv},
       eprint = {1805.03956},
 primaryClass = {astro-ph.HE},
       adsurl = {https://ui.adsabs.harvard.edu/abs/2018PhRvD..98d3007O}
}

@ARTICLE{Castillo2017MNRAS,
       author = {{Castillo}, F. and {Reisenegger}, A. and {Valdivia}, J.~A.},
        title = "{Magnetic field evolution and equilibrium configurations in neutron star cores: the effect of ambipolar diffusion}",
      journal = {\mnras},
         year = 2017,
        month = oct,
       volume = {471},
       number = {1},
        pages = {507-522},
          doi = {10.1093/mnras/stx1604},
archivePrefix = {arXiv},
       eprint = {1705.10020},
 primaryClass = {astro-ph.HE},
       adsurl = {https://ui.adsabs.harvard.edu/abs/2017MNRAS.471..507C}
}

@ARTICLE{Becerra2022random,
       author = {{Becerra}, Laura and {Reisenegger}, Andreas and {Valdivia}, Juan Alejandro and {Gusakov}, Mikhail E.},
        title = "{Evolution of random initial magnetic fields in stably stratified and barotropic stars}",
      journal = {\mnras},
         year = 2022,
        month = mar,
       volume = {511},
       number = {1},
        pages = {732-745},
          doi = {10.1093/mnras/stac102},
archivePrefix = {arXiv},
       eprint = {2111.10673},
 primaryClass = {astro-ph.SR},
       adsurl = {https://ui.adsabs.harvard.edu/abs/2022MNRAS.511..732B}
}

@ARTICLE{Becerra2022MNRAS,
       author = {{Becerra}, Laura and {Reisenegger}, Andreas and {Valdivia}, Juan Alejandro and {Gusakov}, Mikhail},
        title = "{Stability of axially symmetric magnetic fields in stars}",
      journal = {\mnras},
         year = 2022,
        month = nov,
       volume = {517},
       number = {1},
        pages = {560-568},
          doi = {10.1093/mnras/stac2704},
archivePrefix = {arXiv},
       eprint = {2209.01042},
 primaryClass = {astro-ph.SR},
       adsurl = {https://ui.adsabs.harvard.edu/abs/2022MNRAS.517..560B}
}

@ARTICLE{Gourgouliatos2022Symm,
       author = {{Gourgouliatos}, Konstantinos N. and {De Grandis}, Davide and {Igoshev}, Andrei},
        title = "{Magnetic Field Evolution in Neutron Star Crusts: Beyond the Hall Effect}",
      journal = {Symmetry},
         year = 2022,
        month = jan,
       volume = {14},
       number = {1},
          eid = {130},
        pages = {130},
          doi = {10.3390/sym14010130},
archivePrefix = {arXiv},
       eprint = {2201.08345},
 primaryClass = {astro-ph.HE},
       adsurl = {https://ui.adsabs.harvard.edu/abs/2022Symm...14..130G}
}

@ARTICLE{Pons2019LRCA,
       author = {{Pons}, Jos{\'e} A. and {Vigan{\`o}}, Daniele},
        title = "{Magnetic, thermal and rotational evolution of isolated neutron stars}",
      journal = {Living Reviews in Computational Astrophysics},
         year = 2019,
        month = dec,
       volume = {5},
       number = {1},
          eid = {3},
        pages = {3},
          doi = {10.1007/s41115-019-0006-7},
archivePrefix = {arXiv},
       eprint = {1911.03095},
 primaryClass = {astro-ph.HE},
       adsurl = {https://ui.adsabs.harvard.edu/abs/2019LRCA....5....3P}
}

@ARTICLE{Igoshev2023MNRAS,
       author = {{Igoshev}, Andrei P. and {Hollerbach}, Rainer and {Wood}, Toby},
        title = "{Three-dimensional magnetothermal evolution of off-centred dipole magnetic field configurations in neutron stars}",
      journal = {\mnras},
         year = 2023,
        month = nov,
       volume = {525},
       number = {3},
        pages = {3354-3375},
          doi = {10.1093/mnras/stad2404},
archivePrefix = {arXiv},
       eprint = {2308.09132},
 primaryClass = {astro-ph.HE},
       adsurl = {https://ui.adsabs.harvard.edu/abs/2023MNRAS.525.3354I}
}

@ARTICLE{Jones2011Icar,
       author = {{Jones}, C.~A. and {Boronski}, P. and {Brun}, A.~S. and {Glatzmaier}, G.~A. and {Gastine}, T. and {Miesch}, M.~S. and {Wicht}, J.},
        title = "{Anelastic convection-driven dynamo benchmarks}",
      journal = {\icarus},
         year = 2011,
        month = nov,
       volume = {216},
       number = {1},
        pages = {120-135},
          doi = {10.1016/j.icarus.2011.08.014},
       adsurl = {https://ui.adsabs.harvard.edu/abs/2011Icar..216..120J}
}

@ARTICLE{Moss2022PhRvF,
       author = {{Moss}, J.~B. and {Wood}, T.~S. and {Bushby}, P.~J.},
        title = "{Validity of sound-proof approximations for magnetic buoyancy}",
      journal = {Physical Review Fluids},
         year = 2022,
        month = oct,
       volume = {7},
       number = {10},
          eid = {103701},
        pages = {103701},
          doi = {10.1103/PhysRevFluids.7.103701},
archivePrefix = {arXiv},
       eprint = {2209.13315},
 primaryClass = {physics.flu-dyn},
       adsurl = {https://ui.adsabs.harvard.edu/abs/2022PhRvF...7j3701M}
}

@ARTICLE{Wilczynski2022JFM,
       author = {{Wilczy{\'n}ski}, Fryderyk and {Hughes}, David W. and {Kersal{\'e}}, Evy},
        title = "{Magnetic buoyancy instability and the anelastic approximation: regime of validity and relationship with compressible and Boussinesq descriptions}",
      journal = {Journal of Fluid Mechanics},
         year = 2022,
        month = jul,
       volume = {942},
          eid = {A46},
        pages = {A46},
          doi = {10.1017/jfm.2022.325},
       adsurl = {https://ui.adsabs.harvard.edu/abs/2022JFM...942A..46W}
}

@ARTICLE{Burns2020PhRvR,
       author = {{Burns}, Keaton J. and {Vasil}, Geoffrey M. and {Oishi}, Jeffrey S. and {Lecoanet}, Daniel and {Brown}, Benjamin P.},
        title = "{Dedalus: A flexible framework for numerical simulations with spectral methods}",
      journal = {Physical Review Research},
         year = 2020,
        month = apr,
       volume = {2},
       number = {2},
          eid = {023068},
        pages = {023068},
          doi = {10.1103/PhysRevResearch.2.023068},
archivePrefix = {arXiv},
       eprint = {1905.10388},
 primaryClass = {astro-ph.IM},
       adsurl = {https://ui.adsabs.harvard.edu/abs/2020PhRvR...2b3068B}
}

@ARTICLE{Vasil2019JCPX,
       author = {{Vasil}, Geoff and {Lecoanet}, Daniel and {Burns}, Keaton and {Oishi}, Jeff and {Brown}, Ben},
        title = "{Tensor calculus in spherical coordinates using Jacobi polynomials. Part-I: Mathematical analysis and derivations}",
      journal = {Journal of Computational Physics: X},
         year = 2019,
        month = sep,
       volume = {3},
          eid = {100013},
        pages = {100013},
          doi = {10.1016/j.jcpx.2019.100013},
archivePrefix = {arXiv},
       eprint = {1804.10320},
 primaryClass = {math.NA},
       adsurl = {https://ui.adsabs.harvard.edu/abs/2019JCPX....300013V}
}

@article{Lecoanet2019,
title = {Tensor calculus in spherical coordinates using Jacobi polynomials. Part-II: Implementation and examples},
journal = {Journal of Computational Physics: X},
volume = {3},
pages = {100012},
year = {2019},
issn = {2590-0552},
doi = {https://doi.org/10.1016/j.jcpx.2019.100012},
url = {https://www.sciencedirect.com/science/article/pii/S2590055219300289},
author = {Daniel Lecoanet and Geoffrey M. Vasil and Keaton J. Burns and Benjamin P. Brown and Jeffrey S. Oishi}
}

@ARTICLE{Moraga2024MNRAS,
       author = {{Moraga}, Nicol{\'a}s A. and {Castillo}, Francisco and {Reisenegger}, Andreas and {Valdivia}, Juan A. and {Gusakov}, Mikhail E.},
        title = "{Magnetothermal evolution in the cores of adolescent neutron stars: The Grad-Shafranov equilibrium is never reached in the 'strong-coupling' regime}",
      journal = {\mnras},
         year = 2024,
        month = jan,
       volume = {527},
       number = {3},
        pages = {9431-9449},
          doi = {10.1093/mnras/stad3787},
archivePrefix = {arXiv},
       eprint = {2309.14182},
 primaryClass = {astro-ph.HE},
       adsurl = {https://ui.adsabs.harvard.edu/abs/2024MNRAS.527.9431M}
}

@ARTICLE{Passamonti2017MNRAS,
       author = {{Passamonti}, Andrea and {Akg{\"u}n}, Taner and {Pons}, Jos{\'e} A. and {Miralles}, Juan A.},
        title = "{The relevance of ambipolar diffusion for neutron star evolution}",
      journal = {\mnras},
         year = 2017,
        month = mar,
       volume = {465},
       number = {3},
        pages = {3416-3428},
          doi = {10.1093/mnras/stw2936},
archivePrefix = {arXiv},
       eprint = {1608.00001},
 primaryClass = {astro-ph.HE},
       adsurl = {https://ui.adsabs.harvard.edu/abs/2017MNRAS.465.3416P}
}

@ARTICLE{Castillo2020MNRAS,
       author = {{Castillo}, F. and {Reisenegger}, A. and {Valdivia}, J.~A.},
        title = "{Two-fluid simulations of the magnetic field evolution in neutron star cores in the weak-coupling regime}",
      journal = {\mnras},
         year = 2020,
        month = oct,
       volume = {498},
       number = {2},
        pages = {3000-3012},
          doi = {10.1093/mnras/staa2543},
archivePrefix = {arXiv},
       eprint = {2006.13186},
 primaryClass = {astro-ph.HE},
       adsurl = {https://ui.adsabs.harvard.edu/abs/2020MNRAS.498.3000C}
}

@ARTICLE{Livermore2016NatSR,
       author = {{Livermore}, Philip W. and {Bailey}, Lewis M. and {Hollerbach}, Rainer},
        title = "{A comparison of no-slip, stress-free and inviscid models of rapidly rotating fluid in a spherical shell}",
      journal = {Scientific Reports},
         year = 2016,
        month = mar,
       volume = {6},
          eid = {22812},
        pages = {22812},
          doi = {10.1038/srep22812},
       adsurl = {https://ui.adsabs.harvard.edu/abs/2016NatSR...622812L}
}

@ARTICLE{Fuentes2024ApJ,
       author = {{Fuentes}, J.~R. and {Graber}, Vanessa},
        title = "{Superfluid Spin-up: Three-dimensional Simulations of Post-glitch Dynamics in Neutron Star Cores}",
      journal = {\apj},
         year = 2024,
        month = oct,
       volume = {974},
       number = {2},
          eid = {300},
        pages = {300},
          doi = {10.3847/1538-4357/ad77d5},
archivePrefix = {arXiv},
       eprint = {2407.18810},
 primaryClass = {astro-ph.HE},
       adsurl = {https://ui.adsabs.harvard.edu/abs/2024ApJ...974..300F}
}

@ARTICLE{Raynaud2020SciA,
       author = {{Raynaud}, Rapha{\"e}l and {Guilet}, J{\'e}r{\^o}me and {Janka}, Hans-Thomas and {Gastine}, Thomas},
        title = "{Magnetar formation through a convective dynamo in protoneutron stars}",
      journal = {Science Advances},
         year = 2020,
        month = mar,
       volume = {6},
       number = {11},
        pages = {eaay2732},
          doi = {10.1126/sciadv.aay2732},
archivePrefix = {arXiv},
       eprint = {2003.06662},
 primaryClass = {astro-ph.HE},
       adsurl = {https://ui.adsabs.harvard.edu/abs/2020SciA....6.2732R}
}

@ARTICLE{ReboulSalze2022AA,
       author = {{Reboul-Salze}, A. and {Guilet}, J. and {Raynaud}, R. and {Bugli}, M.},
        title = "{MRI-driven {\ensuremath{\alpha}}{\ensuremath{\Omega}} dynamos in protoneutron stars}",
      journal = {\aap},
         year = 2022,
        month = nov,
       volume = {667},
          eid = {A94},
        pages = {A94},
          doi = {10.1051/0004-6361/202142368},
archivePrefix = {arXiv},
       eprint = {2111.02148},
 primaryClass = {astro-ph.HE},
       adsurl = {https://ui.adsabs.harvard.edu/abs/2022A&A...667A..94R}
}

@ARTICLE{Barrere2022AA,
       author = {{Barr{\`e}re}, P. and {Guilet}, J. and {Reboul-Salze}, A. and {Raynaud}, R. and {Janka}, H. -T.},
        title = "{A new scenario for magnetar formation: Tayler-Spruit dynamo in a proto-neutron star spun up by fallback}",
      journal = {\aap},
         year = 2022,
        month = dec,
       volume = {668},
          eid = {A79},
        pages = {A79},
          doi = {10.1051/0004-6361/202244172},
archivePrefix = {arXiv},
       eprint = {2206.01269},
 primaryClass = {astro-ph.HE},
       adsurl = {https://ui.adsabs.harvard.edu/abs/2022A&A...668A..79B}
}

@ARTICLE{Barrere2023MNRAS,
       author = {{Barr{\`e}re}, Paul and {Guilet}, J{\'e}r{\^o}me and {Raynaud}, Rapha{\"e}l and {Reboul-Salze}, Alexis},
        title = "{Numerical simulations of the Tayler-Spruit dynamo in proto-magnetars}",
      journal = {\mnras},
         year = 2023,
        month = nov,
       volume = {526},
       number = {1},
        pages = {L88-L93},
          doi = {10.1093/mnrasl/slad120},
archivePrefix = {arXiv},
       eprint = {2306.12296},
 primaryClass = {astro-ph.HE},
       adsurl = {https://ui.adsabs.harvard.edu/abs/2023MNRAS.526L..88B}
}

@ARTICLE{Skiathas2024MNRAS,
       author = {{Skiathas}, Dimitrios and {Gourgouliatos}, Konstantinos N.},
        title = "{Combined magnetic field evolution in neutron star cores and crusts: ambipolar diffusion, Hall effect, and Ohmic dissipation}",
      journal = {\mnras},
         year = 2024,
        month = mar,
       volume = {528},
       number = {3},
        pages = {5178-5188},
          doi = {10.1093/mnras/stae190},
archivePrefix = {arXiv},
       eprint = {2401.08979},
 primaryClass = {astro-ph.HE},
       adsurl = {https://ui.adsabs.harvard.edu/abs/2024MNRAS.528.5178S}
}

@ARTICLE{shafranov66,
       author = {{Shafranov}, V.~D.},
        title = "{Plasma Equilibrium in a Magnetic Field}",
      journal = {Reviews of Plasma Physics},
         year = 1966,
        month = jan,
       volume = {2},
        pages = {103},
       adsurl = {https://ui.adsabs.harvard.edu/abs/1966RvPP....2..103S}
}

@article{gradrubin54,
title = "Hydromagnetic equilibria and force-free fields",
journal = "Journal of Nuclear Energy (1954)",
volume = "7",
number = "3",
pages = "284 - 285",
year = "1958",
issn = "0891-3919",
doi = "https://doi.org/10.1016/0891-3919(58)90139-6",
url = "http://www.sciencedirect.com/science/article/pii/0891391958901396",
author = "Harold Grad and Hanan Rubin"
}

@article{Armaza_2015,
	doi = {10.1088/0004-637x/802/2/121},
	url = {https://doi.org/10.1088\%2F0004-637x\%2F802\%2F2\%2F121},
	year = 2015,
	month = {apr},
	publisher = {{IOP} Publishing},
	volume = {802},
	number = {2},
	pages = {121},
	author = {Crist{\'{o}}bal Armaza and Andreas Reisenegger and Juan Alejandro Valdivia},
	title = {{ON} {MAGNETIC} {EQUILIBRIA} {IN} {BAROTROPIC} {STARS}},
	journal = {ApJ}
}

@article{migdal1959superfluidity,
  title={Superfluidity and the moments of inertia of nuclei},
  author={Migdal, AB},
  journal={Nuclear Physics},
  volume={13},
  number={5},
  pages={655--674},
  year={1959},
  publisher={Elsevier}
}

@article{Heiselberg_1999,
	doi = {10.1086/312321},
	url = {https://doi.org/10.1086/312321},
	year = 1999,
	month = {nov},
	publisher = {AAS},
	volume = {525},
	number = {1},
	pages = {L45--L48},
	author = {H. Heiselberg and M. Hjorth-Jensen},
	title = {Phase Transitions in Neutron Stars and Maximum Masses},
	journal = {ApJ}
}

@article{glampedakis2011magnetohydrodynamics,
  title={Magnetohydrodynamics of superfluid and superconducting neutron star cores},
  author={Glampedakis, Kostas and Andersson, Nils and Samuelsson, Lars},
  journal={MNRAS},
  volume={410},
  number={2},
  pages={805--829},
  year={2011},
  publisher={Blackwell Publishing Ltd Oxford, UK}
}

@article{reisenegger2009A&A,
  title={Stable magnetic equilibria and their evolution in the upper main sequence, white dwarfs, and neutron stars},
  author={Reisenegger, Andreas},
  journal={A \& A},
  volume={499},
  number={2},
  pages={557--566},
  year={2009},
  publisher={EDP Sciences},
  doi = {10.1051/0004-6361/200810895},
 url = {https://www.aanda.org/articles/aa/abs/2009/20/aa10895-08/aa10895-08.html}
}

@Article{numpy,
 title         = {Array programming with {NumPy}},
 author        = {Charles R. Harris and K. Jarrod Millman and St{\'{e}}fan J.
                 van der Walt and Ralf Gommers and Pauli Virtanen and David
                 Cournapeau and Eric Wieser and Julian Taylor and Sebastian
                 Berg and Nathaniel J. Smith and Robert Kern and Matti Picus
                 and Stephan Hoyer and Marten H. van Kerkwijk and Matthew
                 Brett and Allan Haldane and Jaime Fern{\'{a}}ndez del
                 R{\'{i}}o and Mark Wiebe and Pearu Peterson and Pierre
                 G{\'{e}}rard-Marchant and Kevin Sheppard and Tyler Reddy and
                 Warren Weckesser and Hameer Abbasi and Christoph Gohlke and
                 Travis E. Oliphant},
 year          = {2020},
 month         = sep,
 journal       = {Nature},
 volume        = {585},
 number        = {7825},
 pages         = {357--362},
 doi           = {10.1038/s41586-020-2649-2},
 publisher     = {Springer Science and Business Media {LLC}},
 url           = {https://doi.org/10.1038/s41586-020-2649-2}
}

@Article{matplotlib,
  Author    = {Hunter, J. D.},
  Title     = {Matplotlib: A 2D graphics environment},
  Journal   = {Computing in Science \& Engineering},
  Volume    = {9},
  Number    = {3},
  Pages     = {90--95},
  publisher = {IEEE COMPUTER SOC},
  doi       = {10.1109/MCSE.2007.55},
  year      = 2007
}

@ARTICLE{castillo2025AA,
       author = {{Castillo}, F. and {Moraga}, N.~A. and {Gusakov}, M.~E. and {Valdivia}, J.~A. and {Reisenegger}, A.},
        title = "{Validating and improving two-fluid simulations of the magnetic field evolution in neutron star cores}",
      journal = {arXiv e-prints},
         year = 2025,
        month = mar,
          eid = {arXiv:2503.11530},
        pages = {arXiv:2503.11530},
          doi = {10.48550/arXiv.2503.11530},
archivePrefix = {arXiv},
       eprint = {2503.11530},
 primaryClass = {astro-ph.HE},
       adsurl = {https://ui.adsabs.harvard.edu/abs/2025arXiv250311530C}
}

@article{moraga2025magnetothermal,
  title = {Magnetothermal evolution of neutron star cores in the weak-coupling regime: Implications of ambipolar diffusion for the quiescent x-ray luminosity of magnetars},
  author = {Moraga, N. A. and Castillo, F. and Ofengeim, D. D. and Reisenegger, A. and Valdivia, J. A. and Gusakov, M. E. and Kantor, E. M. and Potekhin, A. Y.},
  journal = {Phys. Rev. D},
  volume = {112},
  issue = {8},
  pages = {083022},
  numpages = {25},
  year = {2025},
  month = {Oct},
  publisher = {American Physical Society},
  doi = {10.1103/16ny-kw3h},
  url = {https://link.aps.org/doi/10.1103/16ny-kw3h}
}

@ARTICLE{Mitchell2015MNRAS,
       author = {{Mitchell}, J.~P. and {Braithwaite}, J. and {Reisenegger}, A. and {Spruit}, H. and {Valdivia}, J.~A. and {Langer}, N.},
        title = "{Instability of magnetic equilibria in barotropic stars}",
      journal = {\mnras},
         year = 2015,
        month = feb,
       volume = {447},
       number = {2},
        pages = {1213-1223},
          doi = {10.1093/mnras/stu2514},
archivePrefix = {arXiv},
       eprint = {1411.7252},
 primaryClass = {astro-ph.SR},
       adsurl = {https://ui.adsabs.harvard.edu/abs/2015MNRAS.447.1213M}
}

@misc{visit_code,
author = {Childs, Hank and Brugger, Eric and Whitlock, Brad and Meredith, Jeremy and Ahern, Sean and Pugmire, David and Biagas, Kathleen and Miller, Mark C. and Harrison, Cyrus and Weber, Gunther H. and Krishnan, Hari and Fogal, Thomas and Sanderson, Allen and Garth, Christoph and Bethel, E. Wes and Camp, David and Rubel, Oliver and Durant, Marc and Favre, Jean M. and Navratil, Paul},
doi = {10.1201/b12985},
title = {{High Performance Visualization--Enabling Extreme-Scale Scientific Insight}},
url = {https://visit.llnl.gov},
year = {2012}
}

@ARTICLE{Braithwaite2009MNRAS,
       author = {{Braithwaite}, Jonathan},
        title = "{Axisymmetric magnetic fields in stars: relative strengths of poloidal and toroidal components}",
      journal = {\mnras},
         year = 2009,
        month = aug,
       volume = {397},
       number = {2},
        pages = {763-774},
          doi = {10.1111/j.1365-2966.2008.14034.x},
archivePrefix = {arXiv},
       eprint = {0810.1049},
 primaryClass = {astro-ph},
       adsurl = {https://ui.adsabs.harvard.edu/abs/2009MNRAS.397..763B}
}

@ARTICLE{GusakovKantor2017PhRvD,
       author = {{Gusakov}, M.~E. and {Kantor}, E.~M. and {Ofengeim}, D.~D.},
        title = "{Evolution of the magnetic field in neutron stars}",
      journal = {\prd},
         year = 2017,
        month = nov,
       volume = {96},
       number = {10},
          eid = {103012},
        pages = {103012},
          doi = {10.1103/PhysRevD.96.103012},
archivePrefix = {arXiv},
       eprint = {1705.00508},
 primaryClass = {astro-ph.HE},
       adsurl = {https://ui.adsabs.harvard.edu/abs/2017PhRvD..96j3012G}
}

@ARTICLE{IakovlevShalybkov1991ApSS,
       author = {{Iakovlev}, D.~G. and {Shalybkov}, D.~A.},
        title = "{Electrical Conductivity of Neutron Star Cores in the Presence of a Magnetic Field - Part One - General Solution for a Multicompponent Fermi Liquid}",
      journal = {\apss},
         year = 1991,
        month = feb,
       volume = {176},
       number = {2},
        pages = {171-189},
          doi = {10.1007/BF00646697},
       adsurl = {https://ui.adsabs.harvard.edu/abs/1991Ap&SS.176..171I}
}

\end{document}